\documentclass{jfm}
\usepackage[section]{placeins}
\usepackage{graphicx}
\usepackage{epstopdf, epsfig}
\usepackage{float}% Include figure files
\usepackage{subfig}
\usepackage{enumitem, amsmath,bm}
\usepackage{bbold}
\usepackage{dsfont}
\usepackage{comment}
\usepackage[dvipsnames]{xcolor}
\usepackage{mathtools}
\usepackage{float}% Include figure files
\usepackage{adjustbox}
\usepackage{multirow}
\usepackage[normalem]{ulem}

\usepackage{scalerel,stackengine}
\stackMath
\newcommand\reallywidehat[1]{%
\savestack{\tmpbox}{\stretchto{%
  \scaleto{%
    \scalerel*[\widthof{\ensuremath{#1}}]{\kern.1pt\mathchar"0362\kern.1pt}%
    {\rule{0ex}{\textheight}}%WIDTH-LIMITED CIRCUMFLEX
  }{\textheight}% 
}{2.4ex}}%
\stackon[-6.9pt]{#1}{\tmpbox}%
}

\newcommand\utot{\boldsymbol{u}}
\newcommand\ubar{\overline{\boldsymbol{u}}}

\newcommand\uhat{\boldsymbol{\hat{u}}}
\newcommand\uprimehat{\boldsymbol{\hat{u^\prime}}}

\newcommand\ujphat{\widehat{u_j^\prime}}

\newcommand\ujphatc{\widehat{u_j^\prime}^{*}}

\newcommand\ujmhatc{\widehat{\overline{u}}_j^*}

\newcommand\Emean{\overline{E}} % energy
\newcommand\Esmean{\widehat{\overline{E}}} %Mean flow spectral energy

\newcommand\tke{\widehat{K}}
\newcommand\Etot{\widehat{E}}

\newcommand\Prodp{P} % Physical production
\newcommand\Dissipp{\epsilon}
\newcommand\Prods{\widehat{\Pi}} %Spectral production
\newcommand\Dissips{\widehat{D}}
\newcommand\Prodmp{P} %Physical production in the mean balance
\newcommand\Dissipmp{\overline{\epsilon}} 
\newcommand\Prodms{\widehat{\overline{\Pi}}} %Spectral prod in the mean balance
\newcommand\Dissipms{\widehat{\overline{D}}}

\newcommand\Advs{\widehat{A}}
\newcommand\Advms{\widehat{\overline{A}}}

\newcommand\Transvm{\overline{T}_{v}}
\newcommand\Transvs{\widehat{T}_{v}}
\newcommand\Transv{T_{v}}
\newcommand\Transvms{\widehat{\overline{T}}_{v}}
\newcommand\Transpm{\overline{T}_{p}}
\newcommand\Transps{\widehat{T}_{p}}
\newcommand\Transp{T_{p}}
\newcommand\Transpms{\widehat{\overline{T}}_{p}}
\newcommand\Transnlm{\overline{T}_{nl}}
\newcommand\Transnls{\widehat{T}_{nl}}

\newcommand\Transnl{T_{nl}}
\newcommand\Transnlms{\widehat{\overline{T}}_{nl}}

\newcommand\ksmax{k_{\text{rolls}}} % Scale of the most-energetic small-scale eddies / "injection" scale
\newcommand\kLS{k_{\text{LS}}} %SCales receiving an inverse transfer
\newcommand\kSS{k_{\text{SS}}} %SCales receiving an inverse transfer
\newcommand{\be}{\mathbf{e}}
\newcommand{\strm}{\text{strm}}
\newcommand{\spwise}{\text{span}}
\newcommand{\dx}{\Delta x}
\newcommand{\dz}{\Delta z}

\newcommand\resub[1]{{\color{black}#1}}

\shorttitle{Energy transfer and mean flow in transitional shear turbulence}
\shortauthor{S. Gom\'e, L.S. Tuckerman and D. Barkley}

\title{Patterns in transitional shear turbulence.\\
Part 1. Energy transfer and mean-flow interaction}

\author{S\'ebastien Gom\'e\aff{1},
  Laurette S.\ Tuckerman\aff{1},    \corresp{\email{laurette.tuckerman@espci.fr}}
 \and Dwight Barkley\aff{2}}

\affiliation{\aff{1}Laboratoire de Physique et M\'ecanique des Milieux H\'et\'erog\`enes, CNRS, ESPCI Paris, PSL Research
University, Sorbonne Universit\'e, Universit\'e Paris-Cit\'e, Paris 75005, France
\aff{2}Mathematics Institute, University of Warwick, Coventry CV4 7AL, United Kingdom}

\begin{document}

\maketitle

\begin{abstract}
Low Reynolds number turbulence in wall-bounded shear flows en route to laminar flow takes the form of spatially intermittent turbulent structures. In plane shear flows, these appear as a regular 
\resub{pattern of alternating turbulent and quasi-laminar flow.}
Both the physical and the spectral energy balance of a turbulent-laminar pattern \resub{in plane Couette flow} are computed and compared to those of uniform turbulence. 
In the patterned state, the mean flow is strongly modulated and is fuelled by two mechanisms: primarily, the nonlinear self-interaction of the mean flow (via mean advection),
and secondly, the extraction of energy from turbulent fluctuations (via negative spectral production, associated with
an energy transfer from small to large scales). 
\resub{Negative production at large scales is also found in the uniformly turbulent state. Important features of the energy budgets are surveyed as a function of $Re$ through the transition between uniform turbulence and turbulent-laminar patterns.}
\end{abstract}

\section{Introduction}

Transitional patterns in plane shear flows arise naturally from uniform turbulence at sufficiently low Reynolds number. These patterns feature a selected orientation of around $24^\circ$ when they emerge \citep[]{prigent2003long, tsukahara2005dns, ShimizuPRF2019, kashyap2020flow}.
When the Reynolds number is further reduced,
these spatio-temporally intermittent structures display important features of non-equilibrium phase transitions; both experimental and numerical studies have demonstrated their membership in
the directed percolation universality class in the case of plane Couette flow \citep[]{lemoult2016directed, chantry_universal,klotz2022phase}. 

Oblique patterns consist of turbulent regions (or bands) alternating with (quasi-) laminar gaps. An inherent feature of the coexistence of these two phases in planar shear flows is the large-scale flow along the laminar-turbulent interface. This along-band flow 
has been observed in both experimental and numerical configurations \citep[]{coles1966progress, barkley2007mean, duguet2013oblique, 
couliou2015large,
tuckerman2020patterns,  klotz2021experimental, marensi2022dynamics}, and
can be seen as a consequence of the breaking of spanwise symmetry and incompressibility \citep[]{duguet2013oblique}. 

Transitional turbulence presents a separation of scales: flow along the laminar-turbulent interface paves
%LST I still don't think that "paves" is correct here, even though it is a nice word.
the large scales, while the streaks and the rolls governed by the self-sustaining process of turbulence \citep[]{hamilton1995regeneration, waleffe1997self} are the basic ingredients of the small-scale flow.
In channel flow, the spanwise streak spacing is commonly found to be around $\lambda_{\spwise}^+ \simeq 100$ \citep[]{kim1987turbulence}, whereas it is found to be larger ($\lambda_{\spwise}^+ \simeq 132$) in plane Couette flow at low enough Reynolds number \citep[]{komminaho1996very, jimenez1998largest, tsukahara2006dns}. (The superscript $+$ indicates non-dimensionalisation by wall variables, e.g. $\lambda^+= \lambda u_\tau / \nu$, where $\nu$ is the kinematic viscosity and $u_\tau$ is the wall-shear velocity. Subscripts $\strm$ and $\spwise$ respectively denote streamwise and spanwise directions.)
%In channel flow, the spanwise streak spacing is commonly found to be around $100$ wall units \citep[]{kim1987turbulence}, whereas it is found to be larger ($\simeq 132$) in plane Couette flow at low-enough Reynolds number \citep[]{komminaho1996very, jimenez1998largest, tsukahara2006dns} %(both measurements use the half-gap $h$ for wall-unit renormalisation).

%
%YD: \lambda^+ = 100 undiscussed for $Re_\tau$ large enough! for PCF check Tsukahara's published date (2006?) 
In contrast, the wavelength of the large-scale patterns is much larger than that of the rolls and streaks, with a ratio on the order of 20 in patterned plane Couette flow.
This scale separation is visible in the spectral analysis presented by several authors. We mention \citet[]{tsukahara2005dns} in channel flow, \citet[]{tuckerman2011patterns, duguet2013oblique} in Couette flow and \citet[]{ishida2017turbulent} in annular pipe flow. 
However, the exact contribution of the rolls and streaks in energising the large-scale patterns has never been thoroughly investigated.

In pipe flow, the energy distribution within turbulent structures was measured in the classic experiments of Wygnanski {\em et al.} \citep{wygnanski1973transition, wygnanski1975transition} and later in numerical simulations by \citet[]{song2017speed}. For localised turbulent structures known as puffs, turbulent production $P$ at the upstream side of a puff is larger than turbulent dissipation $\epsilon$, whereas at the downstream side, dissipation dominates production, as it does throughout regions of quasi-laminar flow in general. No local balance between $P$ and $\epsilon$ is found within the puff. In contrast, in expanding or retracting turbulent zones, known as slugs, the flow in the turbulent core is locally in equilibrium, with production balancing dissipation ($P \simeq \epsilon$). Theoretical efforts to model turbulent-laminar structures in pipe flow are based on these properties of the turbulent production and dissipation \citep{barkley2011a,barkley2016theoretical}.
\resub{We will report a similar out-of-equilibrium spatial distribution of energy in transitional plane Couette flow.}

Spectral energy budgets have been extensively used to quantify energy transfers and interactions between mean flow and turbulent kinetic energy (TKE) in high Reynolds number wall-bounded flows.
This approach dates from \citet[]{lumley1964spectral}, who conjectured that energy is transferred from small to large scales in shear flows as distance from the wall increases. 
This concept of inverse energy transfer was later investigated by \citet[]{domaradzki1994energy, bolotnov2010spectral, lee2015direct, mizuno2016spectra, cho2018scale, lee2019spectral, kawata2021scale} (and references therein).
%
%However, it is only recently that the spectral energy budget has been computed at low $Re_\tau$, in particular by \citet{symon2021energy} in a turbulent channel of minimal size at $Re_\tau =180$ and in an exact coherent state of channel flow \citep[]{park_graham_2015}. 
%
\resub{However, only recently has the spectral energy budget been computed at low $Re_\tau$ by \citet{symon2021energy}, in a turbulent channel of minimal size at $Re_\tau =   u_\tau h /\nu = 180$ and for an exact coherent state of channel flow at $Re_\tau =85$ found by \citet{park_graham_2015}. ($h$ denotes the channel half-gap.)}
Currently, there is a lack of understanding of the spectral distribution of energy in transitional wall-bounded turbulence, especially regarding the role of energy transfers and triad interactions in the emergence of the large-scale flow.
%\resub{Our focus will be on low-$Re$ Couette flow in which turbulence is spatially intermittent, below $Re_\tau = u_\tau h /\nu = 66$.}

This article is devoted to the relationship between the inhomogeneous mean flow and turbulent fluctuations in transitional plane Couette flow, \resub{below $Re_\tau = u_\tau h /\nu = 66$}. These are investigated through the computation of both physical (\S \ref{sec:phys_balance}) and spectral  (\S \ref{sec:spec_balance}) energy balances
in the regime where patterns emerge from uniform turbulence.
%\SG{in a one-dimensional numerical domain oriented with the natural angle of patterns.}
%, $Re \in [380-480]$.
We will survey %various energy transfers 
the energy balance
as a function of $Re$ in \S \ref{sec:bal_Re}. 
%In \S \ref{app:y_spec}, turbulent production and nonlinear transfers are analysed at various wall-normal positions. 
\resub{Turbulent production and nonlinear transfers at various wall-normal locations are included in an appendix.}
The energy processes reported in this article will be further investigated as a function of the pattern wavelength in our companion paper \citet[Part 2]{gome2}, where we will discuss their role in wavelength selection.

%selecting the pattern wavelength.

\section{Numerical setup}
Plane Couette Flow is driven by two parallel rigid plates moving at opposite velocities $\pm U_\text{wall}$. Lengths are nondimensionalised by the half-gap $h$ between the plates, velocities by $U_\text{wall}$, and time by $h/U_\text{wall}$. The Reynolds number is defined to be $Re= U_\text{wall}h/\nu$. 
We will require one last dimensional quantity, the horizontal mean shear at the walls, which we denote by $U^\prime_\text{wall}$. 
We will use non-dimensional variables throughout.
We use the pseudospectral parallel code {\tt Channelflow} \citep{channelflow} to simulate the incompressible Navier-Stokes equations
\begin{subequations}
    \begin{align}
    \frac{\partial \utot}{\partial t}
    + \left(\utot \cdot \nabla\right) \utot
    &= -\nabla p  + \frac{1}{Re} \nabla^2 \utot,\\
    \nabla \cdot \utot &= 0.
    \end{align}
    \label{eq:NS}
\end{subequations}
%
%in a domain that is periodic in the $x$ and $z$ directions. \resub{We use $\boldsymbol{U}_b$ to denote the laminar velocity profile: $\boldsymbol{U}_b = (y,0,0)$.}

\begin{figure}
    \centering
\includegraphics[width=0.8\columnwidth]{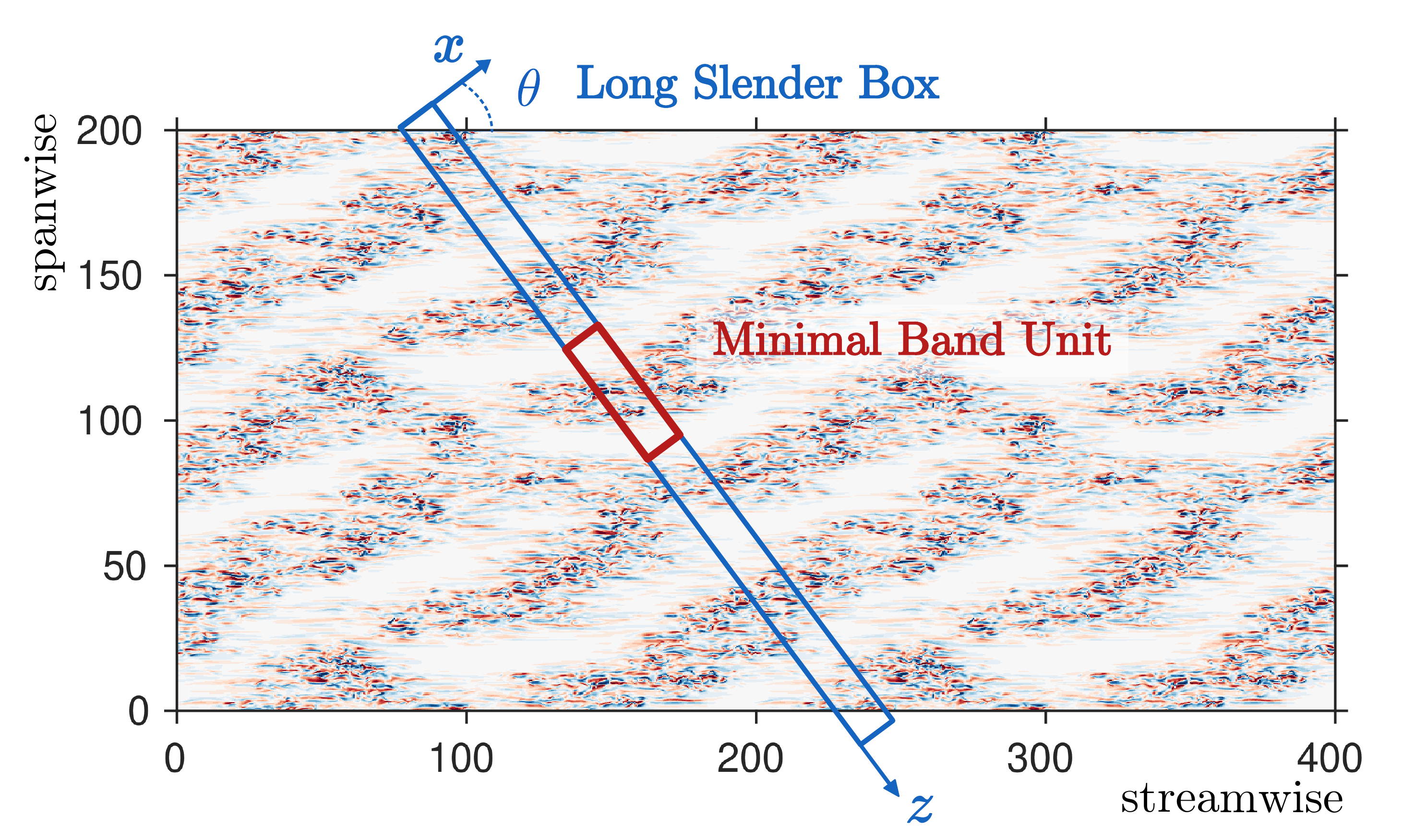}
\caption{Visualisation of the numerically simulated flow at $Re=360$ and of our numerical domains.  Colors show $y$ velocity at $y=0$ (blue: $-0.2$, white: 0, red: 0.2)  in a domain of size $L_{\strm}=400$, $L_{\spwise}=200$. Red and blue boxes show a Minimal Band Unit and a Long Slender Box, respectively. }
    \label{fig:domain}
\end{figure}

Since the bands are found to be oriented obliquely with respect to the streamwise direction, we use a periodic numerical domain which is tilted with respect to the streamwise direction of the flow, shown as the oblique rectangle in figure \ref{fig:domain}. This choice was introduced by
\citet{barkley2005computational}
and has become common in studying turbulent bands \citep[e.g.,][]{ reetz2019exact, paranjape2020oblique,tuckerman2020patterns}.
The $x$ direction is chosen to be aligned with a typical turbulent band and the $z$ direction to be orthogonal to the band.  The relationship between streamwise-spanwise coordinates and tilted band-oriented $(x,z)$ coordinates is:
\begin{subequations}
\label{tilted}
\begin{align}
\be_\strm &= \quad\cos{\theta} \, \be_x + \sin{\theta} \, \be_z \\
\be_\spwise &= -\sin{\theta } \, \be_x + \cos{\theta} \, \be_z \quad 
\end{align}
\end{subequations}
The domain is taken to be periodic in the $x$ and $z$ directions.
The usual wall-normal coordinate is denoted by $y$ and the corresponding velocity by $v$. 
%in a domain that is periodic in the $x$ and $z$ directions. 
The laminar base flow is $\boldsymbol{U}_b \equiv y \be_\strm = (U_b, 0, W_b)$.
The field visualised in figure \ref{fig:domain} (black box) is obtained by concatenating four times a field resulting from a simulation in $L_{\strm}=200$, $L_{\spwise}=100$.

The tilted box effectively reduces the dimensionality of the system by discarding large-scale variations along the short $x$ direction. This direction is considered homogeneous over large scales because it is only determined by small turbulent scales, and because the band is assumed to be infinite in $x$. The main underlying assumption is the angle of the pattern. \resub{In large non-tilted domains, turbulent bands in plane Couette flow exhibit two possible orientations (related by spanwise reflection) \citep{prigent2002large, duguet2010formation, klotz2022phase}, whereas only one orientation is permitted by our tilted box.}

%In large non-tilted domains, plane Couette flow shows two statistical orientations that equilibrate \citep{prigent2002large, duguet2010formation, klotz2022phase} whereas only one orientation is permitted by our tilted box.

In our simulations, we fix the angle $\theta = 24^{\circ}$, \resub{the number of grid points in the $y$ direction $N_y=33$}, the $x$ domain length $L_x=10$, the $x$ resolution $\dx = L_x/N_x=10/120$, and $z$ resolution $\dz = L_z/N_z = 0.08$ (similar to that used by \citet[]{tsukahara2006dns, barkley2007mean}). 
\resub{The values of $N_x$ and $N_z$ include dealiasing in the $x$ and $z$ directions.}
We will make extensive use of two numerical domains, with different domain sizes $L_z$, shown in figure \ref{fig:domain}.

\begin{enumerate}[leftmargin=*,labelindent=8mm,labelsep=3mm, itemindent=0mm, label=(\arabic*)]
    \item \textbf{Minimal Band Units}, shown as the red box in figure \ref{fig:domain}, which can accommodate a single turbulent band and associated quasi-laminar gap. This effectively restricts the flow to a perfectly periodic turbulent-laminar pattern of wavelength $\lambda=L_z$. The size $L_z$ governing the periodicity of the pattern can be modified, as is investigated in the companion paper \citet[Part 2]{gome2}.
    In the present article, the Minimal Band Unit is fixed to $L_z=40$, which corresponds to the natural spacing of bands observed experimentally and numerically. 
    \item \textbf{Long Slender Boxes}, which have a large $L_z$ direction that allows for a large number of gaps and bands in the system. The blue box in figure \ref{fig:domain} is an example of such a domain size with $L_z = 240$, but a larger size of $L_z=800$ is used 
    in this article and investigated in detail in \citet[Part 2]{gome2}.
\end{enumerate}

\begin{figure}
    \centering
\subfloat[]{  \includegraphics[width=0.5\columnwidth]{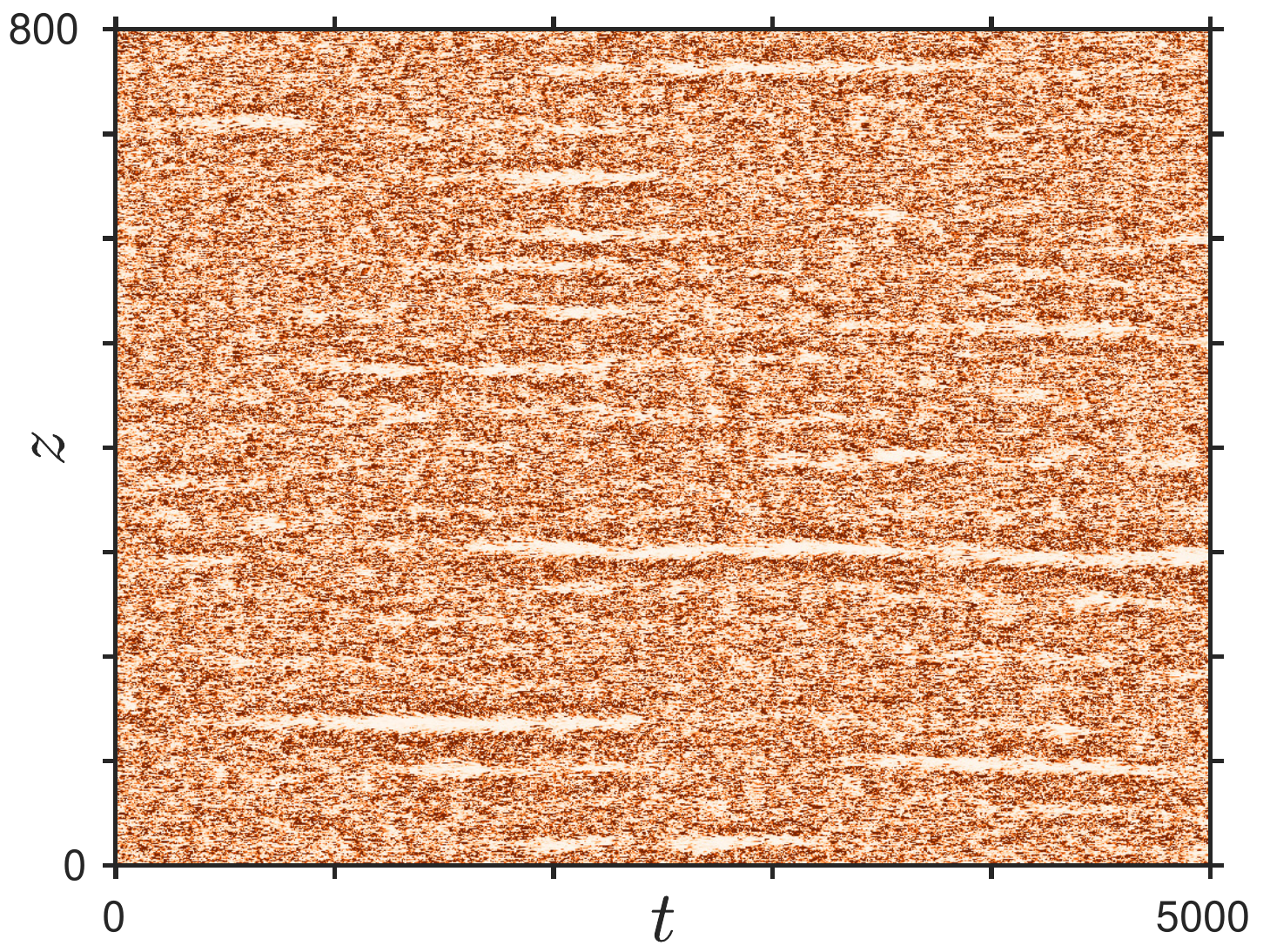}  \label{fig:probes_Lz800_R440}}
\subfloat[]{  \includegraphics[width=0.5\columnwidth]{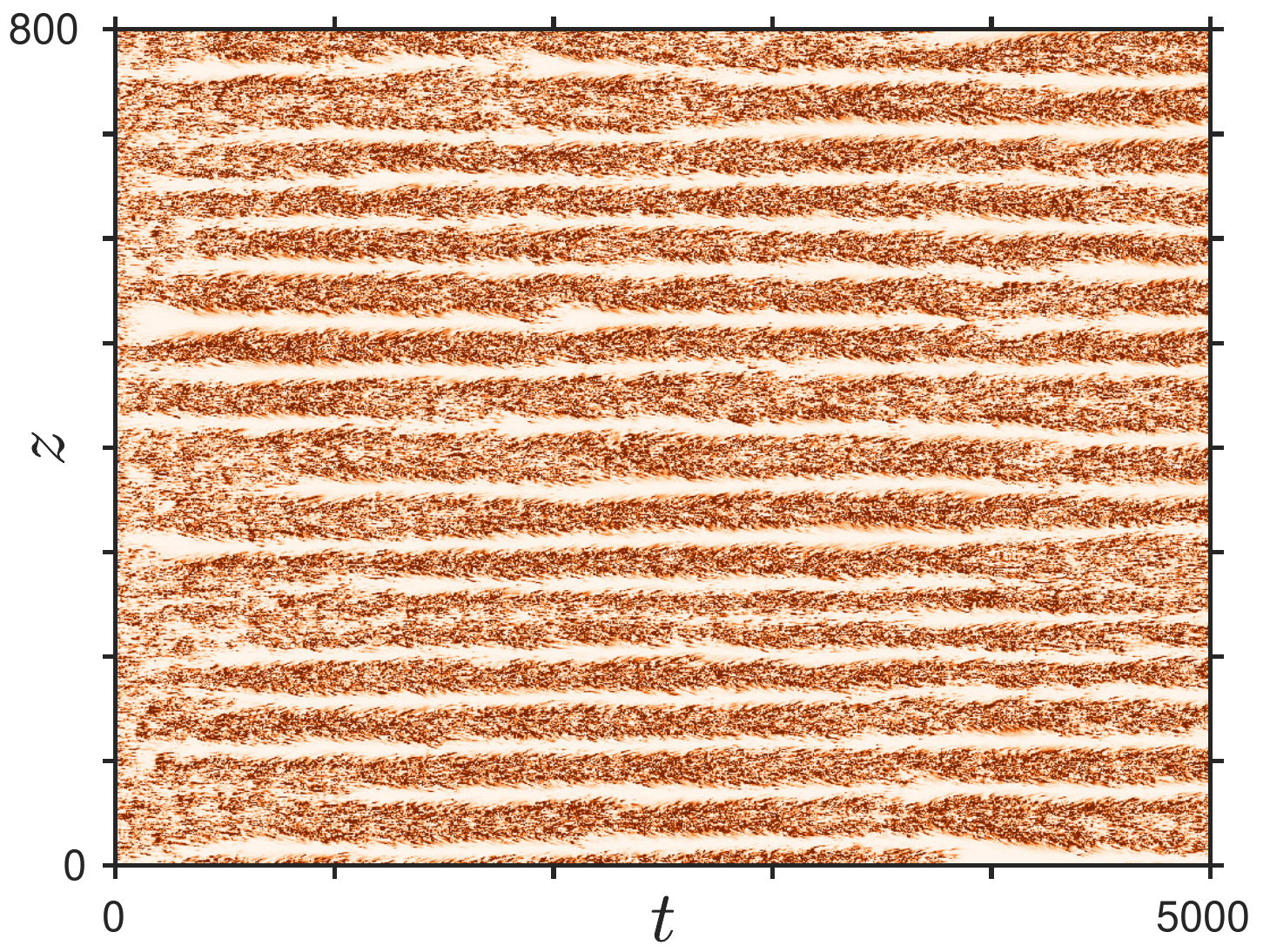} \label{fig:probes_Lz800_R380}} \\
\vspace{-1em}
\subfloat[]{  \includegraphics[width=\columnwidth]{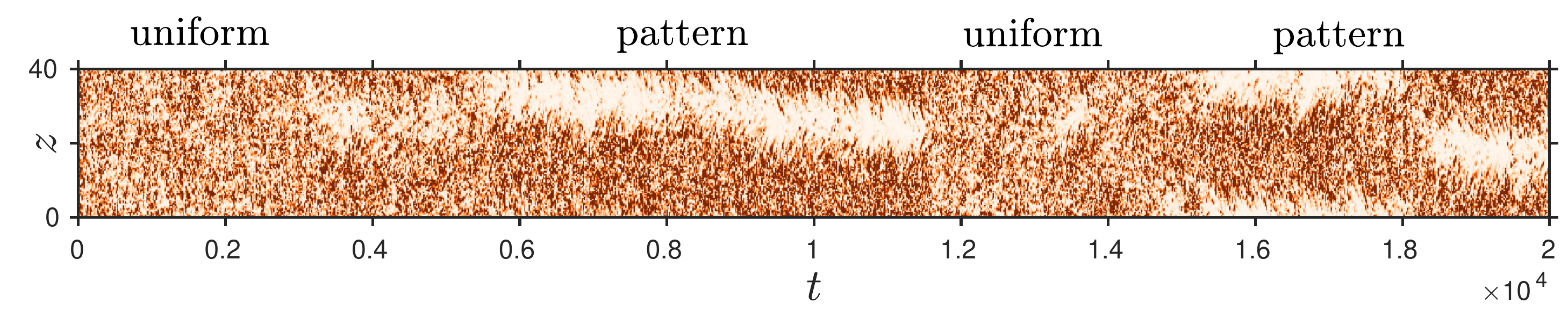} \label{fig:probes_MBU_R430}} 
\caption{Spatio-temporal  visualisations of the emergence of gaps and patterns in a Long Slender Box with $L_z=800$, for (a) $Re=440$ and (b) $380$. Flow at $t=0$ is initiated from uniform turbulence at $Re=500$. Colors show local cross-flow energy $(v^2 + u_{\spwise}^2)/2$ at $x=L_x/2$, $y=0$ (white: 0, dark orange: 0.02). (c) 
Intermittent alternation between uniform and patterned turbulence at $Re=430$ in a Minimal Band Unit with $L_z=40$.
}
\label{fig:probes_Lz800}
\end{figure}

Finally, for comparison with studies of uniform turbulence, we introduce the friction Reynolds number:
\begin{align}
Re_\tau = \frac{ u_\tau h}{\nu}, ~~~~\text{with } {u_\tau^2} = \nu 
U^\prime_\text{wall}
= \frac{U_{\rm wall}^2}{Re}\left< 
\frac{\partial{u}_{\rm strm}}{\partial y}(y=1)
\right>_{x,z,t}
\label{eq:Retaudef}
\end{align}
Note that $Re_\tau = \sqrt{Re}$ in the laminar state.
\resub{The values of $Re_\tau$ computed throughout this study are given in Appendix \ref{app:Retau}}.

\label{sec:balance}
\section{Spectra in different configurations}
\label{sec:spectra_box_size}

We have carried out simulations in a Long Slender Box of size $L_z=800$ for various $Re$, with the uniform state at $Re=500$ as an initial condition. Two such simulations are shown via spatio-temporal diagrams in  figure~\ref{fig:probes_Lz800} at $Re=440$ and $Re=380$.
With decreasing $Re$, the flow shows intermittent gaps (white spots in the figure) that emerge from the turbulent field at seemingly random locations. 
A \emph{gap} is defined as a weakened turbulent structure, or a quasi-laminar zone, surrounded by turbulent flow. A gap is the opposite of a \emph{band}, which is a turbulent core surrounded by quasi-laminar flow. In plane Couette flow, bands are observed at $Re\in[300, 440]$ \citep{prigent2003long, barkley2007mean,duguet2010formation,shi}. 
Gaps and bands self-organize into \emph{patterns} as $Re$ is decreased. This is the situation observed in a Long Slender Box in figure~\ref{fig:probes_Lz800_R380} ($Re=380$), where a regular alternation of gaps and turbulent bands is visible. 
In a Minimal Band Unit, the system is constrained and the distinction between gaps and patterns is lost.
While the system cannot exhibit the spatial intermittency seen in figure \ref{fig:probes_Lz800_R440}, temporal intermittency is possible and is seen as alternations between uniform turbulence and patterns, as illustrated in figure \ref{fig:probes_MBU_R430} at $Re = 430$. 
\citet[Part 2]{gome2} investigate extensively the dynamical emergence of gap and patterns out of turbulent flow. 

We define the total physical energy and total spectral energy of the flow as:
$$
E(y,z) \equiv  \frac{1}{2} \overline{ \utot \cdot \utot} ~~~ 
\text{and} ~~~  
\Etot(y,k_z) \equiv  \frac{1}{2} \overline{ \uhat^{*} \cdot \uhat},  $$
where $\overline{(.)}$ denotes time and $x$ averaging and the Fourier transform is taken in the band-orthogonal direction $z$:
\begin{align}\uhat(x, y, k_z) \equiv \frac{1}{L_z}\int_{0}^{L_z}  \boldsymbol{u}(x,y,z) e^{- i k_z z } \text{d}z.
\label{eq:fft}
\end{align}
\resub{We will also use the turbulent kinetic energy (TKE) in both physical and spectral space,
$$
K(y,z) \equiv  \frac{1}{2} \overline{ \utot^\prime \cdot \utot^\prime} ~~~ 
\text{and} ~~~  
\tke(y,k_z) \equiv  \frac{1}{2} \overline{ \uprimehat^{*} \cdot \uprimehat },
$$
where the flow has been decomposed into its mean and fluctuating components $\utot = \ubar + \utot^\prime$.}

\begin{figure}
    \centering
\subfloat[Long Slender Box]{ \includegraphics[width=0.5\columnwidth]{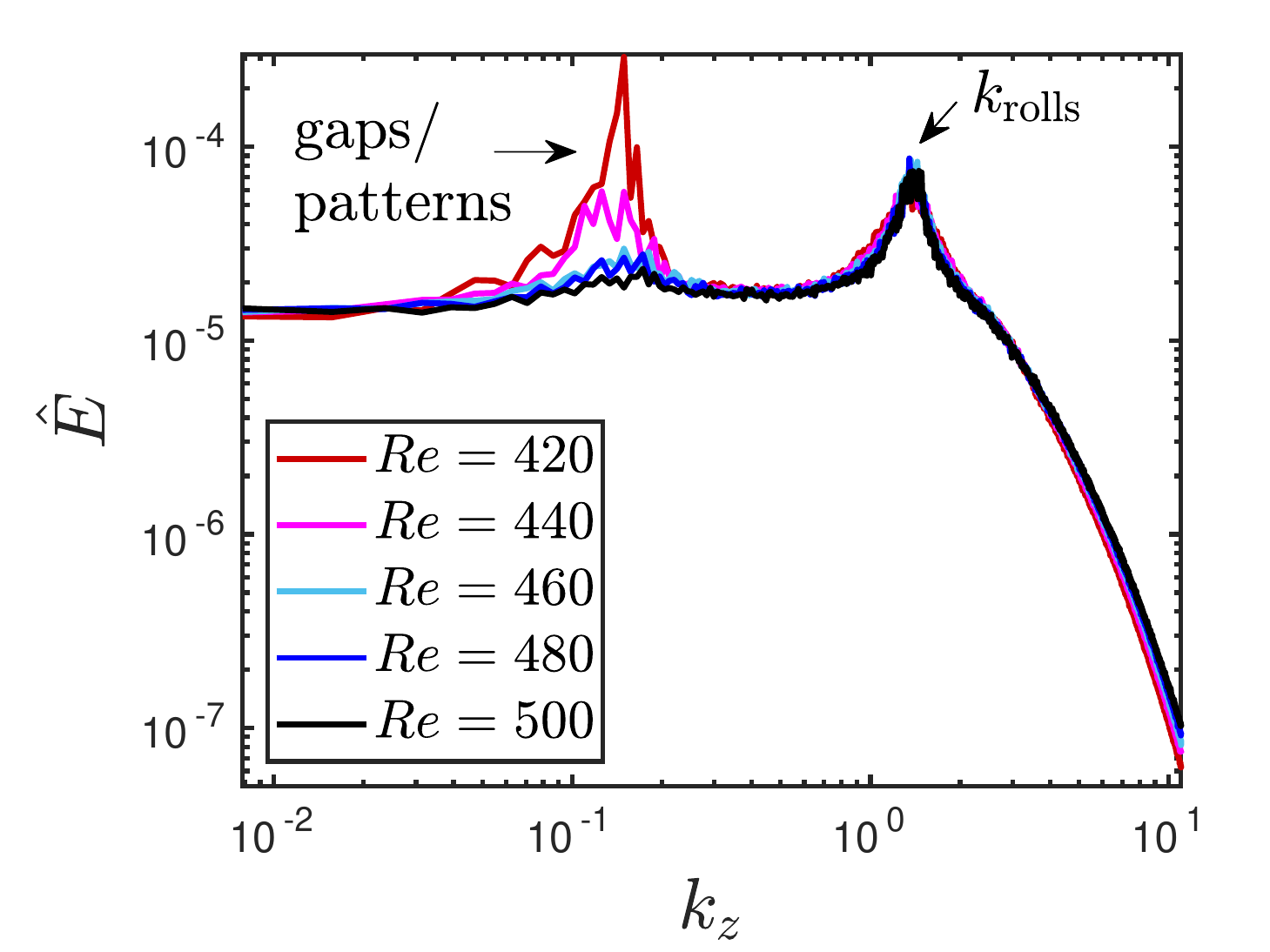}
\label{fig:spec_Lz800_Re}} 
\subfloat[Minimal Band Unit at $Re=430$]{ \includegraphics[width=0.5\columnwidth]{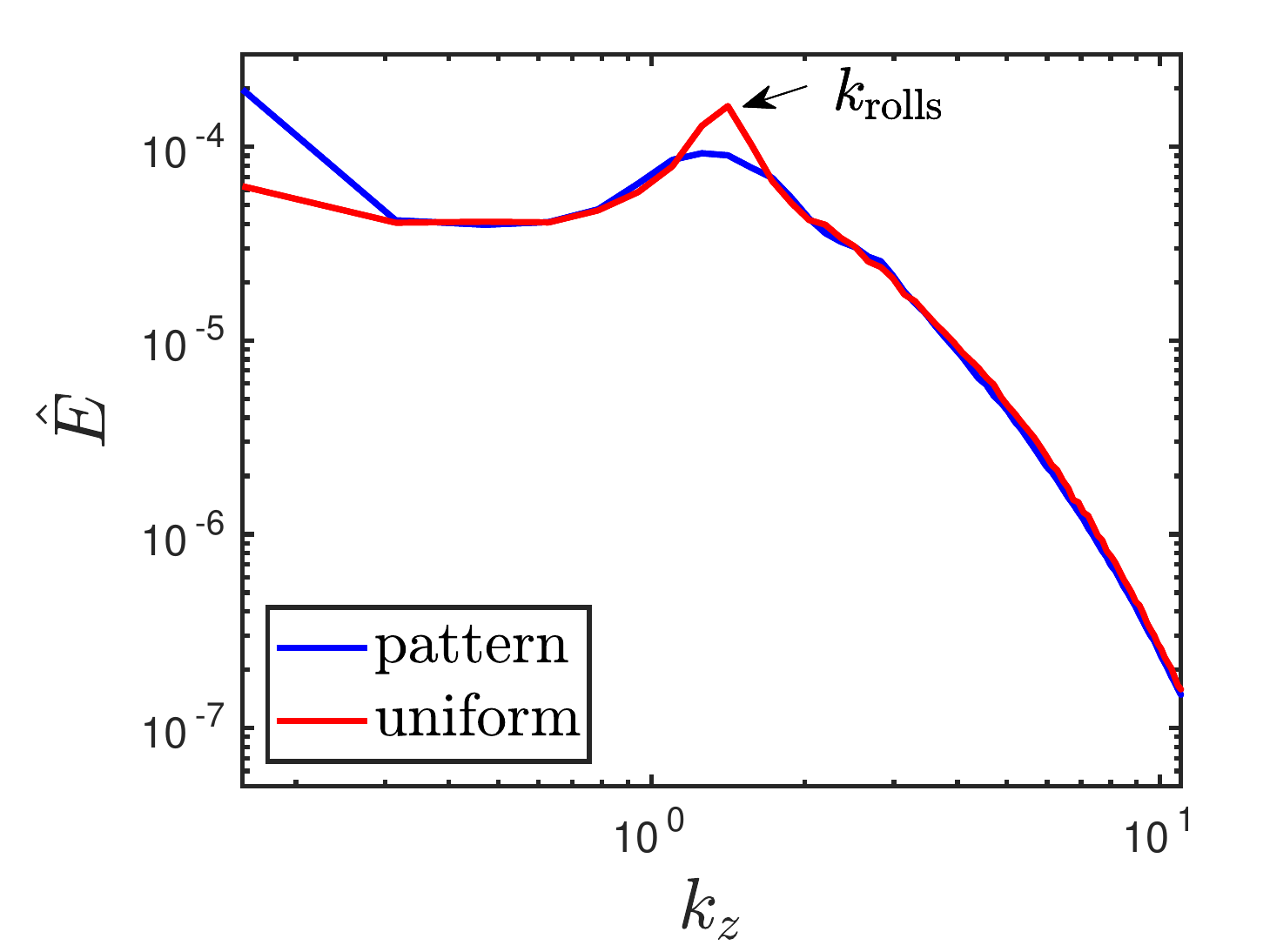}
\label{fig:spec_MBU}} \\
\subfloat[Patterned mean flow in a Minimal Band Unit]{ \includegraphics[width=\columnwidth]{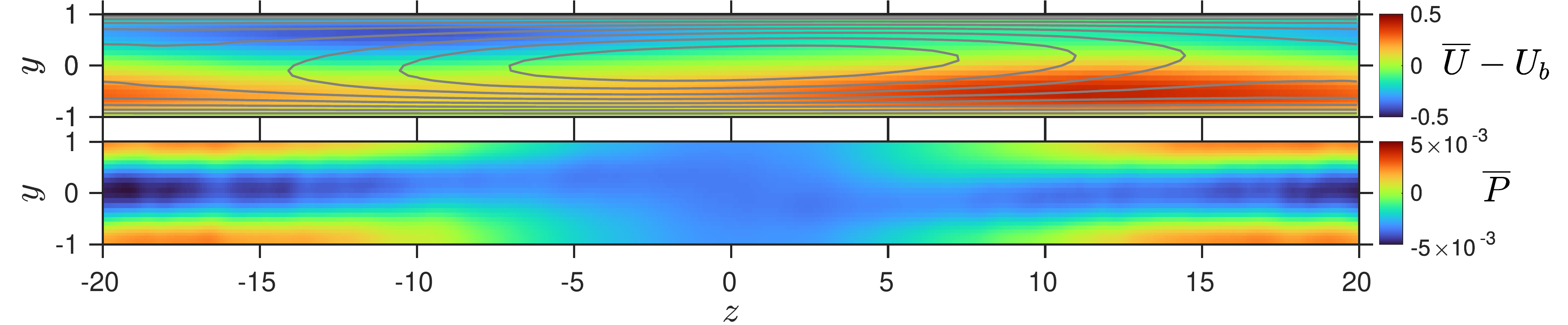} 
\label{fig:mf_u_P}} 
\caption{(a) Total energy spectra in a Long Slender Box $L_z=800$ (black line) at $y=0$, for different $Re$. The spectra are averaged irrespective of the nature of the state (patterned or uniform).
(b) Total energy spectra (continuous lines) in a Minimal Band Unit of size $L_z=40$. The spectra are individually computed in the uniform turbulence (red) and patterned states (blue), at $Re=430$. 
(c) Visualisation of the mean flow: colors show the deviation of the along-band velocity from the laminar base flow $\overline{U}-U_b$ (top) and the pressure $\overline{P}$ (bottom). Streamlines of the mean flow perpendicular to the bands are shown as grey curves. The laminar region is centred at $z=0$.
}
\label{fig:spec}
\end{figure}

Figure~\ref{fig:spec_Lz800_Re} shows $\Etot(y=0, k_z)$ for simulations in a Long Slender Box at different values of $Re$. 
%Here, the average is carried out over a long period of time ($t\in [100, 5000])$.
The average has been carried out over a long period of time ($t\in [100, 5000])$.
The total energy spectra show two prominent energy-containing scales: one at small wavenumbers (around $k_z=0.15$, i.e.\ $\lambda_z \simeq 42$) corresponding to the 
%regular alternation of bands and gaps, 
alternation of turbulent bands and quasi-laminar gaps, and a second one at large wavenumbers ($k_z \simeq 1.41$, $\lambda_z\simeq 4.45$), which we will denote $\ksmax$. This small wavelength corresponds to a spanwise spacing of $\lambda_{\spwise} = 2\pi\cos{\theta}/\ksmax  = 4.06$, which is approximately the idealised periodicity of pairs of streaks and rolls in Couette flow \citep{waleffe1997self}, with individual rolls occupying the height $L_y=2$ of the shear layer. 
In wall units, this peak corresponds to $\lambda_{\spwise}^+ = Re_\tau \lambda_{\spwise} = 130$ at $Re=430$ ($Re_\tau = 31.9$). This is not far from the streak spacing of $\lambda_{\spwise}^+=136$ measured by \citet[]{komminaho1996very} in plane Couette flow at $Re_\tau = 52$. 
For $k_z>\ksmax$, the energy 
%decreases 
falls off rapidly 
with $k_z$ up to the resolution scale.
The scale separation between the large-scale turbulent-laminar patterns
%gaps and bands 
and the small-scale streaks and rolls was already observed in the transitional regime by many authors \citep[]{tsukahara2005dns,tuckerman2011patterns, ishida2016transitional}.
The spectrum varies with $Re$, but mostly at large scales (low $k_z$): 
the large-scale peak is barely visible at $Re=500$ and grows in intensity with decreasing $Re$, becoming dominant for $Re<440$. Meanwhile, the small-scale spectrum is only very weakly affected by the change in $Re$.

We now turn to the Minimal Band Unit, which has exactly the periodicity of a single wavelength of the pattern. %\citep{barkley2007mean}.
The flow in this configuration does not have localised gaps like those which appear in figure \ref{fig:probes_Lz800_R440}.
The system instead 
%reduced to a bistable situation, fluctuating 
fluctuates between patterned and uniform states as seen in figure \ref{fig:probes_MBU_R430}, and each of the two states can be distinguished and consequently analysed separately. In particular, we can take means for patterned and uniform states independently.

The total energy spectrum in a Minimal Band Unit 
%of size $L_z=40$ at a fixed 
at $Re=430$ is presented in figure~\ref{fig:spec_MBU}. Contrary to figure~\ref{fig:spec_Lz800_Re}, where unconditional averaging mixes uniform turbulence and localised gaps in the spectrum, here we have conditionally computed separately the spectrum for the patterned state (blue line) and the uniform state (red line).
As expected, the spectrum for the uniform state lacks the peak at the pattern scale. 
%However, 
The energy of the streak-roll structures $\Etot(\ksmax)$ is higher in the uniform case than in the patterned case. This hints at a redistribution of the energy from small scales (near $\ksmax$) to large scales ($\ll \ksmax$) when the flow changes from uniform to patterned turbulence. For $k_z > 2$, both spectra appear to collapse, suggesting that the small-scale 
%turbulent cascade 
\resub{energy distribution}
is the same in both cases.

%We now introduce the mean-fluctuation decomposition: $\utot = \ubar + \uprime$, 

We now consider the mean flow $\ubar(y,z)$, computed from an $(x,t)$ average over long time intervals in either the patterned or the uniform state in the Minimal Band Unit. %of size $L_z=40$. 
The mean flow in this configuration was studied by \citet{barkley2007mean}. 
We visualise $\ubar = (U(y,z), V(y,z), W(y,z))$ in figure \ref{fig:mf_u_P}, by showing $U-U_b$ and $\overline{P}$ (colors) and plotting the streamlines of $(V,W)$ (grey lines).
%\footnote{
(Figure~\ref{fig:mf_u_P} corrects the erroneous pressure displayed in \citet[figure 5]{barkley2007mean}.)
%}
The flow is centered around the quasi-laminar region, and the total in-plane velocity $(V,W)$ shows a circulation around this region of the flow. $U-U_b$ shows two centro-symmetrically related zones of flow parallel to the band, localised in the upper layer (blue zone) and in the bottom layer (red zone). 

The mean flow $\ubar$ can be decomposed into Fourier modes:
\vspace*{-0.2cm}
\begin{equation}
    \label{eq:trig}
    \ubar(y, z) = \ubar_0(y) + 2\mathcal{R} \left(\ubar_1(y) e^{2\pi i z/L_z} \right) + \ubar_{>1} (y,z)\\[-0.2cm]
\end{equation}
%NB: real part is because of complex conjugates
where $\mathcal{R}$ denotes real part, $\ubar_0 \equiv \widehat{\ubar}(y, k_z=0)= (U_0(y), 0 , W_0(y))$ is the $z$-independent (uniform) component of the mean flow, $\ubar_1 = \widehat{\ubar}(y, k_z=2\pi/L_z)$ is the Fourier coefficient corresponding to wavelength $\lambda_z = L_z$, and $\ubar_{>1} \equiv \sum_{k_z>2\pi/L_z} \widehat{\ubar}(y, k_z) $ is the remainder of the decomposition. 
(To lighten the notation, we omit the hats on $\ubar$ when subscripts $0$, $1$, or $>1$ are used to indicate the corresponding Fourier coefficients.)
%\citet{barkley2007mean} noted that most 
Most of the mean-flow energy lies in the uniform mode $\ubar_0$, with a few percent in the trigonometric component $\ubar_1$. The energy in the remaining terms ($\ubar_{>1}$)
is at least two orders of magnitude lower than that of $\ubar_1$ \citep[]{barkley2007mean}.

\begin{figure}
    \centering
\includegraphics[width=0.9\columnwidth]{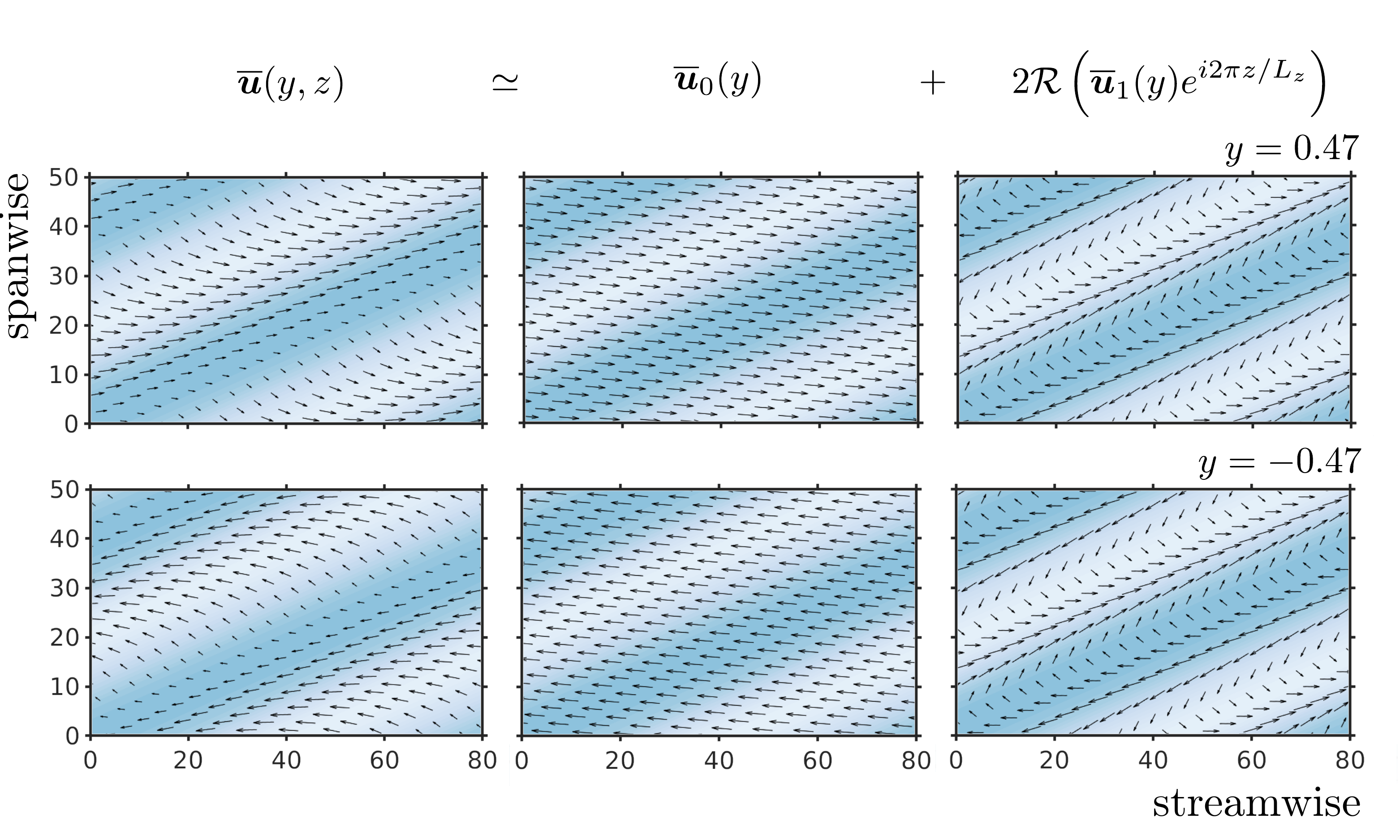}
\caption{Mean-flow decomposition in two Fourier modes $\ubar_0$ and $\ubar_1$ \eqref{eq:trig}, visualised in the planes $y=\pm0.47$ at $Re=400$. Colors show TKE: turbulent and laminar zones are respectively in blue and white.
}
\label{fig:mf_trig}
%\bigskip
 %   \centering
\subfloat{ \includegraphics[width=0.9\columnwidth]{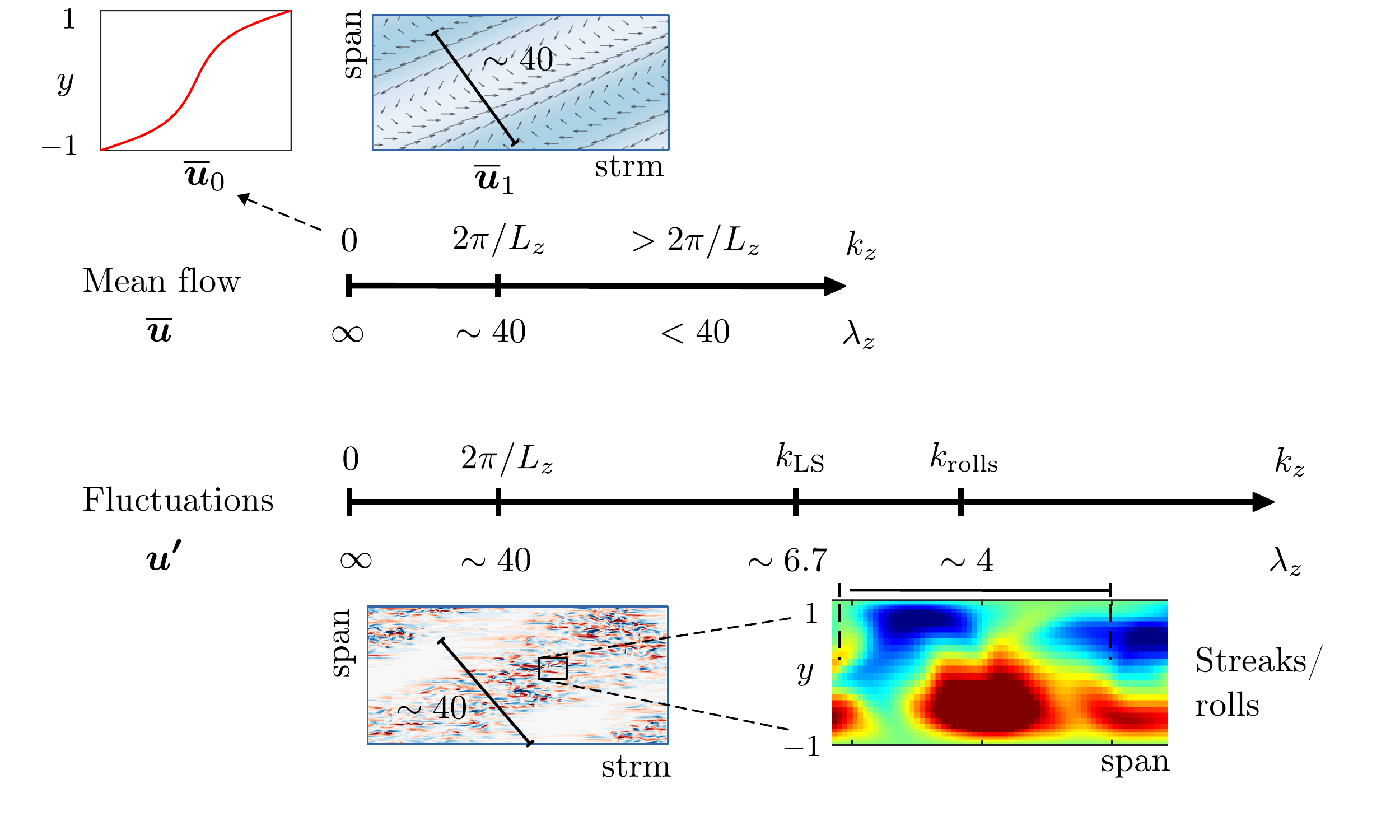}}
\caption{Illustration of the most-relevant scales in transitional patterns. The upper part depicts the mean flow: $\ubar_0 $ is the uniform ($z$-independent) mean shear, illustrated by the mean streamwise velocity profile, while $\ubar_1$ corresponds to scale $\lambda_z \sim 40$ and is dominated by the flow along the laminar-turbulent interfaces as illustrated in the streamwise-spanwise plane. The lower part depicts the fluctuations: scale $\lambda_z \sim 40$ is illustrated by the periodic presence and absence of fluctuating cross-flow velocity. The scale $\kLS$ is that below which fluctuations receive energy by nonlinear interactions (via $\Transnls$, eq. \eqref{eq:tke_bal_spec}), and $\ksmax$ is the scale of rolls and streaks, illustrated in the spanwise-$y$ plane.
}
\label{fig:scales}
\end{figure}

%\begin{samepage}
%\nopagebreak

\newpage 

The decomposition of $\ubar$ into $\ubar_0$ and $\ubar_1$ is illustrated in figure \ref{fig:mf_trig}. The mean flow and the turbulent kinetic energy 
%$K(y,z)$
%\equiv  \frac{1}{2} \overline{\uprime \cdot \uprime}$ 
are visualised in the planes $y=\pm0.47$.
The most relevant scales involved in the mean flow and the fluctuations are illustrated in figure \ref{fig:scales}. Mode $\ubar_0$ has a S-shape profile in $y$ with small spanwise component.
Mode $\ubar_1$ contains the large-scale flow along laminar-turbulent interfaces. 

 \nopagebreak[4]

\section{Physical balance in a Minimal Band Unit}
\label{sec:phys_balance}
\nopagebreak

The remainder of this article will focus on the Minimal Band Unit with a fixed length of $L_z=40$.

Before turning to the energy balance in spectral space, we first consider the traditional turbulent energy decomposition in the physical-space representation \citep{pope2000turbulent}, as carried out in transitional pipe flow by \cite{wygnanski1973transition} and \cite{song2017speed} and in bent pipe flow by \cite{rinaldi2019vanishing}. 
We write the balance equation for the turbulent kinetic energy, $K(y,z)$, in the physical representation:
\begin{equation}
    \label{eq:tke_bal_phys}
    \frac{\partial K}{\partial t} + \ubar \cdot \nabla K = P - \epsilon  + \Transnl + \Transp + \Transv
\end{equation}
where the production term, dissipation term, and rate of strain are:
\begin{equation}
    \label{eq:P_eps}
    P \equiv - \overline{u_i^\prime u_j^\prime} \frac{\partial \overline{u}_i}{\partial x_j} \text{,} ~~~~~ 
    \epsilon \equiv \frac{2}{Re} \overline{s_{ij}^\prime s_{ij}^\prime}
    \text{,} ~~~~~ 
s_{ij}^\prime \equiv \frac{1}{2}\left( \frac{\partial u_i^\prime}{ \partial x_j} + \frac{\partial u_j^\prime}{\partial x_i}  \right).
\end{equation}
Subscripts $i$ and $j$ range over $\{1,2,3\}$ (or equivalently $\{x,y,z\}$) and we use the Einstein summation convention. The transfer terms read:
\begin{align}
    \label{eq:T}
    \Transnl \equiv -\frac{1}{2} \frac{\partial}{\partial x_i}\overline{u_i^\prime u_j^\prime u_j^\prime}, ~~~~~
    \Transp \equiv - \frac{\partial}{\partial x_i} \overline{u_i^\prime p^\prime } , ~~~~~
    \Transv \equiv \frac{2}{Re} \frac{\partial}{\partial x_i}\overline{u_j^\prime s_{ij}^\prime}
\end{align}
which account respectively for nonlinear interactions, work by pressure and viscous diffusion.  We also introduce the total transfer $T\equiv \Transnl + \Transp + \Transv$. This TKE balance is accompanied by the energy balance of the mean flow, $\Emean = \frac{1}{2}\ubar \cdot \ubar = E - K$ \citep[eq. 5.131]{pope2000turbulent}:
\begin{equation}
    \label{eq:mean_bal_phys}
    \frac{\partial \Emean}{\partial t} + \ubar \cdot \nabla \Emean = - \Prodmp - \Dissipmp  + \Transnlm + \Transpm + \Transvm
\end{equation}
where 
\begin{equation}
    \label{eq:epsm_Tm}
    \Dissipmp \equiv \frac{2}{Re} \overline{s}_{ij} \overline{s}_{ij}, \qquad 
    \overline{s}_{ij} \equiv
\frac{1}{2}
\left(\frac{\partial \overline{u}_i}{\partial x_j} + \frac{\partial \overline{u}_j}{\partial x_i} \right)
    \end{equation}
    and
\begin{equation}    
    \Transnlm \equiv -\frac{\partial}{\partial x_i} \overline{u}_j \overline{u_i^\prime u_j^\prime}, ~~~~~
    \Transpm \equiv   - \frac{\partial}{\partial x_i}\overline{u}_i \overline{p} ~~~~ \text{and} ~~~~
    \Transvm \equiv  \frac{2}{Re}\frac{\partial}{\partial x_i} \overline{u}_j   \overline{s}_{ij}
\end{equation}

In order to emphasise the derivation of
\eqref{eq:tke_bal_phys} and 
\eqref{eq:mean_bal_phys} from the Navier-Stokes equations, we have retained temporal derivatives, despite the fact that these equations describe $t$ and $x$ averaged quantities. 
While turbulent-laminar banded patterns are statistically steady in plane Couette flow, there is in fact some slight motion of the band position. To gain in precision, we position the pattern at each time based on the phase of the $z$-trigonometric Fourier coefficient of the along-band flow at the mid-plane: $z_{\text{loc}}(t) = - \phi(t) L_z/2\pi$, where $\phi(t) = \arg \left<\widehat{u}(x, 0, 2\pi/L_z,t) \right>_x$. Temporal averages are computed with this phase alignment and we consider $\partial K/\partial t = 0$ and $\partial \overline{E}/\partial t = 0$. The results in this section are all presented in a frame centered around the quasi-laminar zone, \resub{as was done in \citet{barkley2007mean}.} 

\begin{figure}
    \centering
\subfloat[]{ \includegraphics[width=\columnwidth]{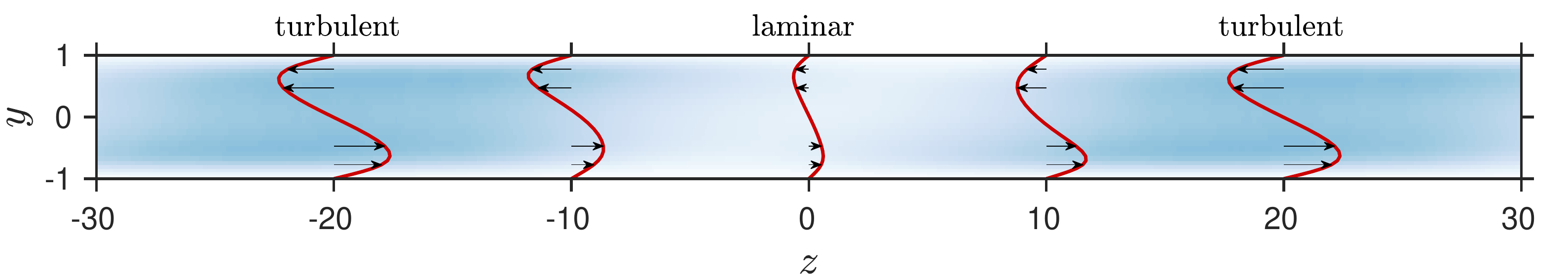} 
\label{fig:schema_flux}}\\
\vspace{-0.9em}
\subfloat[]{ \includegraphics[width=0.5\columnwidth]{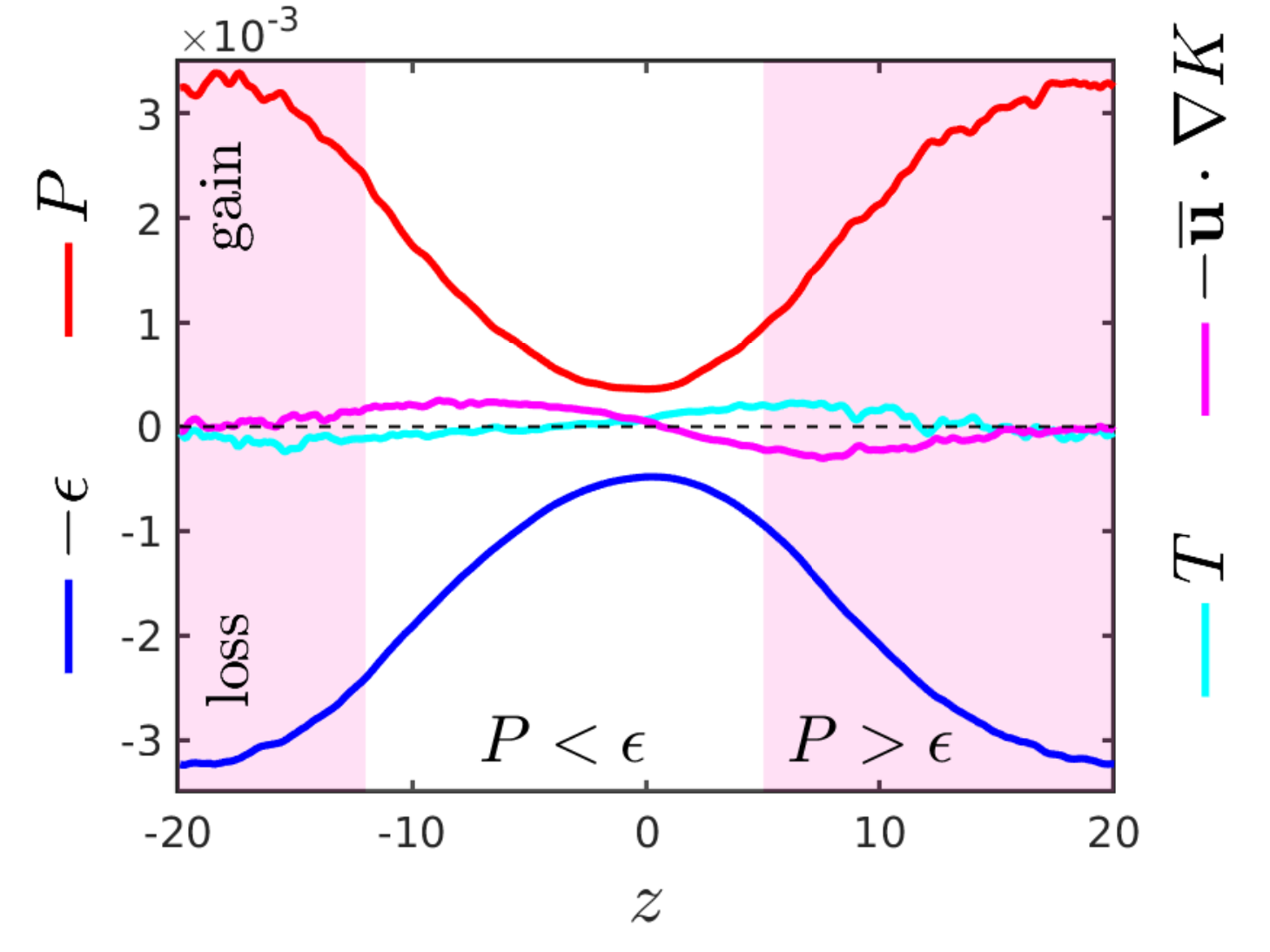} 
\label{fig:budget}}~
\subfloat[]{ \includegraphics[width=0.5\columnwidth]{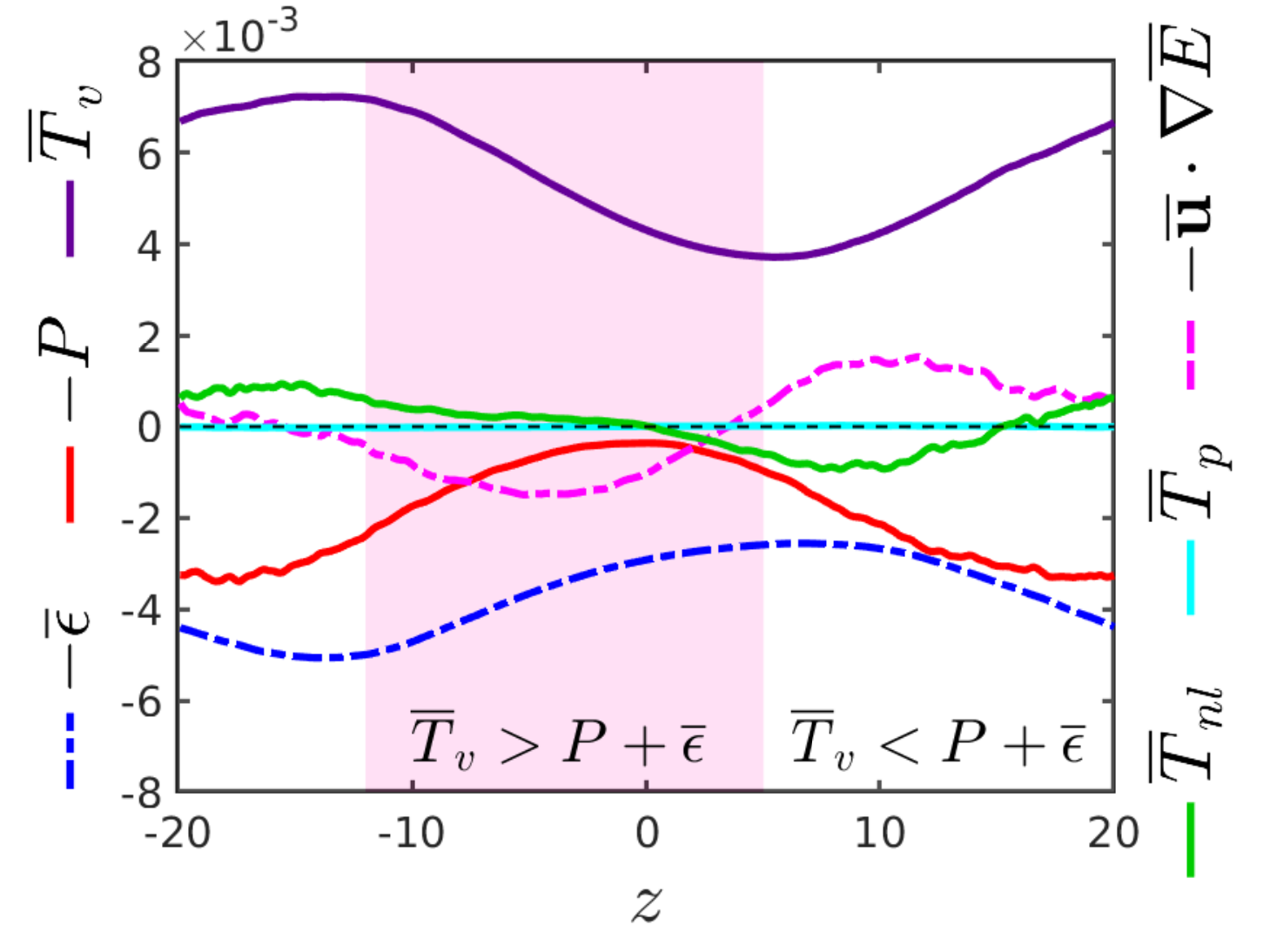} \label{fig:mean_bal_phys}}
\label{fig:mean_bal_phys_all}
\caption{(a) Mean profiles of the deviation from laminar flow ($\overline{W} - W_b$) and the TKE (white: 0, blue: 0.08), in a turbulent-laminar pattern at $Re=400$ centered around the laminar region at $z=0$. 
(b) TKE and (c) mean-flow energy budgets for $Re=400$.
%at $L_z=40$. 
Each term is integrated over the upper half of the domain, $y\in [0, 1]$, where advection by the mean flow is towards the right. \resub{In (b), the regions where $P>\epsilon$ are shaded.} There is a transfer of turbulent energy from shaded to non-shaded regions.
\resub{In (c), the regions where $\Transvm > P + \Dissipmp$ are shaded. There is a transfer of mean-flow energy from shaded to non-shaded regions.} 
}
\end{figure}

In figure \ref{fig:schema_flux} we represent the streamwise mean flow with arrows and the turbulent kinetic energy $K(y,z)$ by colors.
The centre of the tubulent region is at $z\pm 20$, while locations $z\pm10$ correspond to overhang regions \citep{lundbladh1991direct, duguet2013oblique}, in which turbulence extends further towards positive z on top and towards negative z on bottom, and where the along-band large-scale flow is strongest (see figure \ref{fig:mf_u_P}).
%The center of the turbulent region is at $z\pm20$, while \emph{overhang regions} \citep{lundbladh1991direct, duguet2013oblique} are located around $z\pm10$, where the along-band large-scale flow is strongest (see figure \ref{fig:mf_u_P}).
%
Figures \ref{fig:budget} and \ref{fig:mean_bal_phys}
display the terms in the energy budgets of equations \eqref{eq:tke_bal_phys} and \eqref{eq:mean_bal_phys}. 
To better relate these results to those from pipe flow, we integrate the energy budgets over the upper half of the domain, where the $z$ component of the mean flow is from left to right. We use the same symbols $P$, $\epsilon$, etc.\ to denote these half-height averages. 
(The lower half can be obtained from the upper half by symmetry and should be compared to pipe flow with the opposite streamwise direction.) All quantities depend strongly on $z$ and it is this dependence on which we will focus.

Figure~\ref{fig:budget} shows the TKE budget. 
The energy balance is dominated by production and dissipation. Unsurprisingly, production is minimal in the quasi-laminar region where the fluctuations, and hence the Reynolds stresses, are small. The regions where production is larger than and smaller than dissipation are indicated in the figure with shading.
\resub{This local disequilibrium between production and dissipation is accounted for by the transfers from advection and fluctuations, $\ubar \cdot \nabla K$ and $T$ (the former being of larger amplitude than the latter).}
\resub{The spatial transfer of energy goes from the shaded region ($z\simeq 5$ to $z\simeq -12$, taking into account periodicity), to the unshaded region ($z\simeq -12$ to $z\simeq 5$) in figure \ref{fig:budget}.}
\resub{Turbulent energy therefore goes from the turbulent to the quasi-laminar zone.}
%, as indicated by the shaded area in figure \ref{fig:budget}.}
%
These results are consistent with those in a band in plane Poiseuille flow \citep[Fig. 5]{brethouwer2012turbulent} and in a puff in pipe flow \citep[]{song2017speed}: when entering the turbulent region from upstream to downstream, $P>\epsilon$ first, and then $P<\epsilon$, which signifies a spatial flux of energy from upstream to downstream. (In the upper half of our Couette domain, increasing $z$ corresponds to going downstream in a pipe.)

We now look at the energy budget of the mean flow, presented in figure~\ref{fig:mean_bal_phys}, again centered around the laminar region and integrated over the upper half of the domain. 
Unlike pressure-driven channel or pipe flows, the energy is injected into the plane Couette flow by the imposed motion of the wall, and this is captured by the viscous diffusion term $\Transvm$ in the mean-flow energy equation.
\resub{The injected energy is mostly lost to mean-flow dissipation $\Dissipmp$ and TKE production $P$, which extracts energy from the mean flow to fuel fluctuations. The remaining transport terms $-\ubar\cdot \nabla\Emean + \Transnlm + \Transpm$ account for the imbalance between injection, production and mean-flow dissipation: $\Transvm - P - \Dissipmp \ne 0$. We find that the overall transfer term appearing in the mean-flow equation behaves in the opposite way as the total transfer appearing in the TKE balance: mean-flow energy is transferred from the laminar region to the turbulent region, as illustrated by the shaded area in figure \ref{fig:mean_bal_phys}.}

In pipe flow, \citet{song2017speed} reported that the peak in TKE dissipation $\epsilon$ is shifted downstream from the peak in the production $P$.
Our data for the upper-half of plane Couette flow does not support such a $z$-shift in the peaks in $\Dissipp$ and $\Prodp$. 
Meanwhile, unrelated to these considerations, we observe a considerable
$z$-shift between the peaks in mean-flow dissipation $\Dissipmp$ at $z=-15$ and production $\Prodmp$ at $z=-20$, as shown in figure~\ref{fig:mean_bal_phys}. This shift between $\Prodp$ and $\Dissipmp$ is consistent with overhangs in the mean flow located at the sides of turbulent regions where the TKE is maximal. 
\resub{We presume that some shift between $\Prodp$ and $\Dissipmp$ is also present in the pipe flow case, but this remains to be seen. }

\section{Spectral decomposition}
\label{sec:spec_balance}

We now analyse the spectral balance of kinetic energy. In shear flows at higher $Re$, this analysis leads to a detailed understanding of the energy sources and transfers between scales. We refer the reader to \citet{bolotnov2010spectral, lee2015direct, mizuno2016spectra, cho2018scale} for studies at higher $Re_\tau$, and to \citet[]{symon2021energy} for a minimal channel study at $Re_\tau=180$.
In a similar vein, \citet{lee2019spectral} recently computed two-point correlations in channel flow. 

\subsection{Notation and governing equations}
We begin by writing the Reynolds-averaged Navier-Stokes equations and the equation for fluctuations from the mean:

\begin{align}
    \label{eq:RANS}
    \frac{\partial \overline{u}_j}{\partial t} + \overline{u}_i \frac{\partial \overline{u}_j}{\partial x_i} 
    +  \frac{\partial}{\partial x_i} \overline{ u_i^\prime u_j^\prime  }
    &=   - \frac{\partial \overline{p}}{\partial x_j}
    + \frac{2}{Re} \frac{\partial \overline{s}_{ij}}{\partial x_i} 
\end{align}

\begin{align}
    \label{eq:devNS}
    \frac{\partial u_j^\prime}{\partial t} + \overline{u}_i \frac{\partial u_j^\prime}{\partial x_i} 
    +  u_i^\prime \frac{\partial u_j^\prime}{\partial x_i}
    &= - u^\prime_i \frac{\partial \overline{u}_j}{\partial x_i} 
    + \frac{\partial}{\partial x_i}  \overline{ u_i^\prime u_j^\prime }
        - \frac{\partial p^\prime}{\partial x_j}
    + \frac{2}{Re} \frac{\partial s_{ij}^\prime}{\partial x_i} 
\end{align}
By taking the $z$ Fourier transform of (\ref{eq:devNS}) and multiplying by $\ujphatc$, followed by averaging over $x$ and $t$, we obtain a balance equation for the spectral kinetic energy
$\tke(y,k_z) \equiv  \frac{1}{2} \overline{ \uprimehat^{*} \cdot \uprimehat } $:

\begin{align}
    \underbrace{ \vphantom{\mathcal{R}\left\{ \overline{\ujphat \widehat{u^\prime_i \frac{\partial \overline{u}_j}{\partial x_i}} } \right\}} 
    \frac{\partial \tke(y, k_z)}{\partial t}}_0
&+ \underbrace{\mathcal{R}\left\{ \overline{\ujphatc  \widehat{ \overline{u}_i\frac{\partial u_j^\prime}{\partial x_i} }} \right\} }_{-\Advs(y,k_z)}
    =   \underbrace{- \mathcal{R}\left\{ \overline{\ujphatc \widehat{u^\prime_i \frac{\partial \overline{u}_j}{\partial x_i}} } \right\}}_{\Prods(y, k_z)} ~
   - \underbrace{ \vphantom{\mathcal{R}\left\{ \overline{\ujphat \widehat{u^\prime_i \frac{\partial \overline{u}_j}{\partial x_i}} } \right\}}
  \frac{2}{Re} \overline{ \widehat{s_{ij}^\prime} \widehat{s_{ij}^\prime}^*}}_{\Dissips(y,k_z)}\nonumber\\
    &+ \underbrace{\vphantom{- \mathcal{R}\left\{  \overline{\ujphatc \widehat{  u_i^\prime \frac{\partial  u_j^\prime}{\partial x_i}  }} \right\}}
    \frac{2}{Re} \mathcal{R} \left\{  \overline{\frac{\partial}{\partial y} (\ujphatc \widehat{s_{yj}^\prime}) } \right\}}_{\Transvs(y,k_z)}
~  \underbrace{\vphantom{\mathcal{R}\left\{  \overline{\ujphatc \widehat{  u_i^\prime \frac{\partial  u_j^\prime}{\partial x_i}  }} \right\}} 
 ~ - \mathcal{R}\left\{ \overline{\frac{\partial}{\partial y}  (\widehat{u_y^{\prime}}^*\widehat{p^{\prime}})} \right\} }_{\Transps(y,k_z)}
~
    \underbrace{- \mathcal{R}\left\{  \overline{\ujphatc \widehat{  u_i^\prime \frac{\partial  u_j^\prime}{\partial x_i}  }} \right\} }_{\Transnls (y,k_z)} 
     \label{eq:tke_bal_spec}
\end{align}
%vphantom is to vertically align the underbrace 
where we revert from the general partial derivative $\partial/\partial x_i$ or subscript $i$ to the wall-normal coordinate $y$ when this is the only non-zero term. 
\begin{itemize}[leftmargin=*,labelindent=8mm,labelsep=3mm, itemindent=0mm]
    \item[-] $\Advs$ is the interaction between mean velocity and gradient of fluctuations, corresponding to the spectral version of the advection term $\ubar \cdot \nabla K$;
    \item[-] $\Prods$ is the spectral production term, which is an interaction between the mean gradient and fluctuations at scale $k_z$;
    \item[-] $\Dissips$ is the viscous dissipation at mode $k_z$;
    \item[-] $\Transvs$, $\hat{T}_p$ are transfer terms to mode $k_z$ due to strain-velocity and pressure-velocity correlations;
    \item[-] $\Transnls$ is an inter-scale transfer to mode $k_z$ due to triad interactions. \\
    % cancles out
\end{itemize}

When summed over $k_z$ and integrated over $y$, $\Transvs$, $\Transps$ and $\Transnls$ are zero. 

The forms of the pressure, viscous diffusion, dissipation and triadic terms are the same as they would be if the flow were uniform in $z$.
Only advection and production terms, which contain the inhomogeneous mean flow, do not simplify as in the uniform case, and instead require a convolution over wavenumbers.
In the usual analysis of uniform turbulence in a non-tilted box \citep{bolotnov2010spectral, cho2018scale, lee2019spectral}, 
 $\ubar$ reduces to $(U(y),0,0)$ and $\widehat{\ubar}= (U(y), 0, 0)$ for $k_z=0$ and is otherwise 0,
which simplifies the spectral balance. In particular, the advection term $\Advs$ vanishes, because in such cases: 
\begin{align}
    \Advs (y, k_z) =  -\mathcal{R}\left\{ \overline{\ujphatc U(y) \widehat{ \frac{\partial u_j^\prime}{\partial x} }} \right\} = -\frac{1}{2} \mathcal{R}\left\{ U(y) \overline{ \frac{\partial}{\partial x} \ujphatc \ujphat} \right\} = 0
    \label{eq:adv_unif}
    \end{align}
(due to averaging over the periodic $x$ direction). This is also true in the case of tilted uniform turbulence $\ubar=(U(y), 0 , W(y))$.
%
%:
%\begin{align}
%    \Advs (y, k_z) = -\mathcal{R}\left\{W(y) i k_z  %\overline{\ujphatc \ujphat} \right\} = 0
%    
%\end{align}
%
However, this is not true for a patterned mean flow $\ubar=(U(y,z), V(y,z), W(y,z))$ like the one shown in figure \ref{fig:mf_u_P}.

We furthermore introduce the balance equation for the spectral energy of the mean flow $\Esmean \equiv \frac{1}{2} \widehat{\ubar}^* \cdot \widehat{\ubar}$ at wavenumber $k_z$:

\begin{align}
  \underbrace{\vphantom{\mathcal{R}\left\{ \frac{\partial \ujmhatc}{\partial x_i} \widehat{\overline{u_i^\prime u_j^\prime}} \right\} } \frac{\partial \Esmean(y, k_z)}{\partial t} }_0
&+ \underbrace{\mathcal{R}\left\{ \ujmhatc  \widehat{ \overline{u}_i \frac{\partial \overline{u}_j}{\partial x_i} } \right\}  }_{-\Advms(y,k_z)}
       =\underbrace{ \mathcal{R}\left\{ \frac{\partial \ujmhatc}{\partial x_i} \widehat{\overline{u_i^\prime u_j^\prime}} \right\} }_{-\Prodms(y, k_z)}
   - \underbrace{\vphantom{\mathcal{R}\left\{ \frac{\partial \ujmhatc}{\partial x_i} \widehat{\overline{u_i^\prime u_j^\prime}} \right\} } ~
   \frac{2}{Re} \widehat{\overline{s}}_{ij} \widehat{\overline{s}}_{ij}^*}_{\Dissipms(y,k_z)}
   \nonumber\\
    &+  \underbrace{\frac{2}{Re} \mathcal{R}\left\{\frac{\partial}{\partial y} (\ujmhatc \widehat{\overline{s}}_{yj} ) \right\}}_{\Transvms(y,k_z)}
~  \underbrace{ - \mathcal{R}\left\{ \frac{\partial}{\partial y} \widehat{\overline{u}}_y^* \widehat{\overline{p}}  \right\} }_{\Transpms(y,k_z)} 
~     \underbrace{- \mathcal{R}\left\{ \frac{\partial}{\partial y}(\ujmhatc \widehat{\overline{u_y^\prime u_j^\prime}} ) \right\}  }_{\Transnlms (y,k_z)} 
     \label{eq:mean_bal_spec}
\end{align}

where:
\begin{itemize}[leftmargin=*,labelindent=8mm,labelsep=3mm, itemindent=0mm]
    \item[-] $\Advms$ is a nonlinear transfer term for the mean flow. This is a spectral version of the advection term $\ubar \cdot \nabla \Emean$ appearing in the mean-flow balance equation \eqref{eq:mean_bal_phys}.  
    \item[-] $\Prodms$ is the interaction between Reynolds stress at scale $k_z$ and the mean gradient at scale $k_z$, and hence is a production term.
    \item[-] $\Dissipms$ is a dissipation term for the mean-flow energy;
    \item[-] $\Transvms$, $\Transpms$ are transfer terms due to correlations between mean strain and velocity, and mean pressure and velocity;
    \item[-] $\Transnlms$ is a flux term due to the interactions between the Reynolds stress and the mean flow. \\
\end{itemize}

We have presented equations \eqref{eq:tke_bal_spec} and \eqref{eq:mean_bal_spec} with $y$ dependence to facilitate understanding the origin of the various terms. However, in the rest of this section, we will focus on $y-$integrated TKE and mean-flow balance to characterise the %global 
spectral distribution as a function of $k_z$.
As the mean flow $\ubar$ is dominated by $\ubar_0$ and $\ubar_1$, 
we write (\ref{eq:mean_bal_spec}) in $y$-integrated form for $k_z=0$ and $k_z=2\pi/L_z$ and obtain:
\begin{align}
    I + \Advms_0  -\Prodms_0 - \Dissipms_0 = 0  ~~~ \text{and} ~~~   \Advms_1 - \Prodms_1 - \Dissipms_1 = 0
    \label{eq:mf_bal}
\end{align}
where we have introduced
\begin{align}
    \widehat{\overline{\Pi}}_0 \equiv \int_{-1}^{1}\Prodms(y, 0) ~\text{d} y  ~~~~ \text{and} ~~~~ \widehat{\overline{\Pi}}_1 \equiv \int_{-1}^{1}\Prodms\left(y, \frac{2\pi}{L_z}\right) ~\text{d} y 
    \label{eq:pi0_pi1}
\end{align}
with similar definitions for $\Advms_0$, $\Dissipms_0$, $\Advms_1$ and $\Dissipms_1$.
We have also introduced the total energy injection due to the action of the walls:
\begin{align}
    I &\equiv \sum_{k_z} \int_{-1}^1 \Transvms(y, k_z) ~\text{d} y  =  \frac{2}{Re} \sum_{k_z}  \ujmhatc (k_z) ~\widehat{\overline{s}}_{yj} (k_z) \bigg \rvert_{-1}^1 
    \label{eq:injection}
\end{align}
%
%This term is non-zero only for mode $k_z=0$ because the applied wall velocity is uniform, so that
The only non-zero term in the final expression \eqref{eq:injection} is mode $k_z=0$ (because the boundary condition dictates a fixed velocity everywhere on the wall), so that
\begin{align}
    I &= \frac{2}{Re} \ujmhatc (k_z=0) ~\widehat{\overline{s}}_{yj} (k_z=0) \bigg \rvert_{-1}^1 
    =  2 \frac{u_\tau^2}{U_{\rm wall}^2}
\end{align}
Note that $\Transpms$ and $\Transnlms$ integrate to zero, since both $\overline{u}_y$ and the Reynolds stress vanish at the walls.

Two important comments can be made at this stage.
The first one starts from a 
%seemingly trivial
word of caution: all terms in \eqref{eq:mean_bal_spec} are not the Fourier transforms of those in \eqref{eq:mean_bal_phys}. 
(This is a generalisation of the fact that $\Etot (k_z)$ is defined to be $\uhat (k_z)\cdot \uhat (k_z) /2$ and not $\widehat{\utot \cdot \utot}(k_z)/2$.)
This means in particular that although energy is injected only in the balance of $\ubar_0$ via $I$, the energy is not injected uniformly within the flow, as $\Transvm$ is not uniform in $z$ (see figure~\ref{fig:mean_bal_phys}). The connection with the physical injection of energy is indeed only through $z$ averaging:
\begin{equation}
    I= \frac{1}{L_z}\int_0^{L_z} \int_{-1}^1 \Transvm (y, z) ~\text{d}y ~ \text{d} z  \end{equation}

The second comment is about the way in which this injected energy is communicated to the TKE spectral balance. Contrary to the physical-space version of the energy balance, where the same production $\Prodp$ appears in the TKE (\ref{eq:tke_bal_phys}) and the mean flow (\ref{eq:mean_bal_phys})  equations, the spectral production terms appearing in  (\ref{eq:tke_bal_spec}) and (\ref{eq:mean_bal_spec}), $\Prods$ and $\Prodms$, are different. However, the sum over $k_z$ of
these two terms agree, so we can write the total ($y$-integrated) production $\Pi$ as:
\begin{align}
    \Pi \equiv \sum_{k_z} \int_{-1}^1 \Prods(y, k_z) ~\text{d} y =  \sum_{k_z} \int_{-1}^1 \Prodms(y, k_z) ~\text{d} y 
    \label{eq:total_prod}
\end{align}
Furthermore, in the physical-space representation, 
\begin{align}
\Pi = \frac{1}{L_z}\int_{0}^{L_z} \int_{-1}^{1} \Prodp(y,z) ~\text{d}z ~\text{d}y = \frac{1}{L_z} \int_{0}^{L_z} \int_{-1}^{1} \Dissipp (y,z) ~\text{d}z ~\text{d}y
\label{eq:Pi_equals_eps}
\end{align}
%$\Pi = \int_{-1}^{1} \left<P\right>_z \text{d}y = \int_{-1}^{1} \left<\epsilon\right>_z \text{d}y$ 
where the last equality follows since all transfer terms integrate to zero. 
The equivalence (\ref{eq:total_prod}) is key to understanding how TKE and mean-flow energy are connected. This will be further developed in section \ref{sec:spec_results}.

\subsection{Results for the spectral energy balance}
\label{sec:spec_results}

\subsubsection{TKE balance}

\begin{figure}
    \centering
\subfloat[Pattern ($Re=400$), TKE]{ \includegraphics[width=0.5\columnwidth]{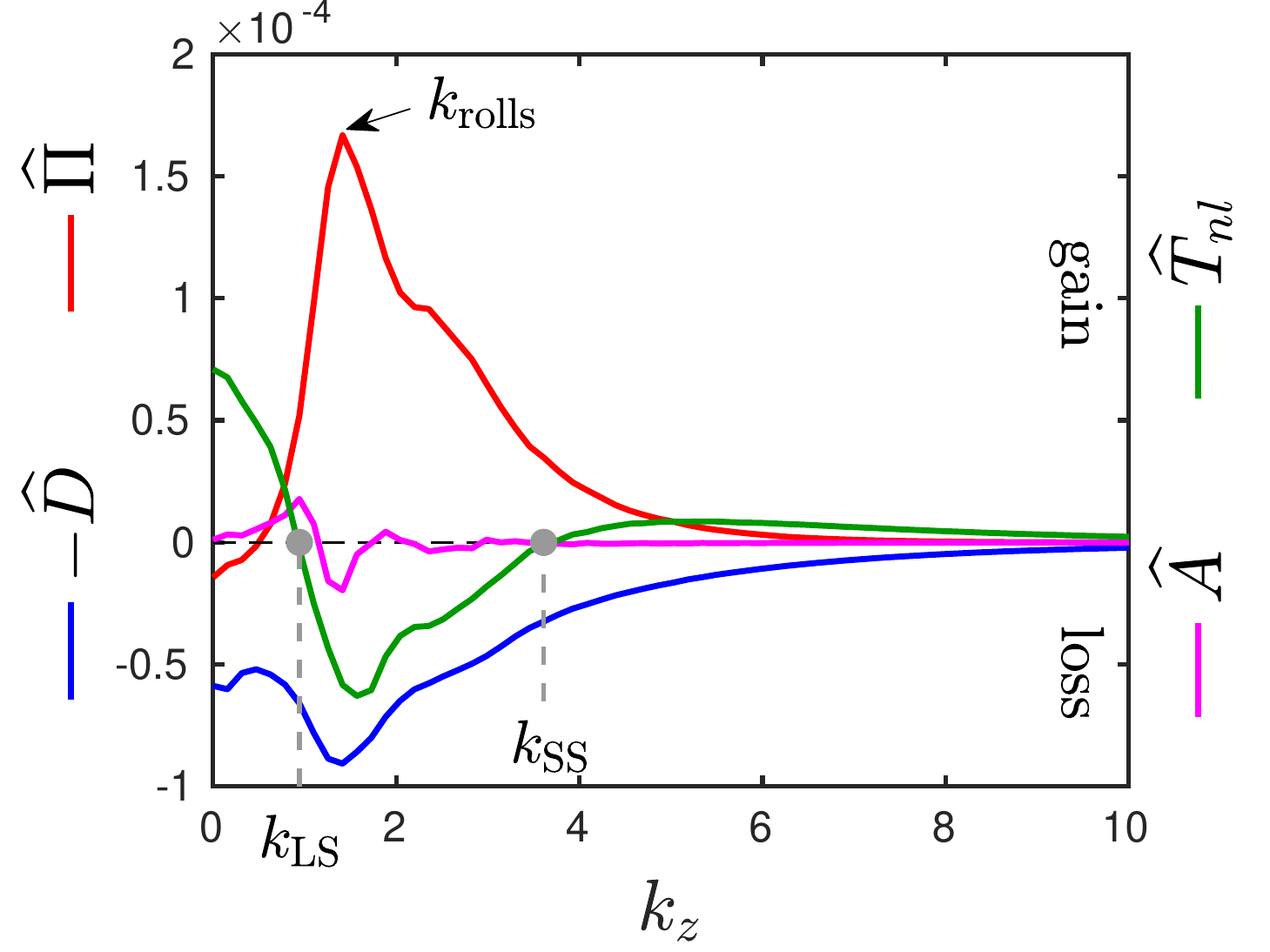} 
\label{fig:tke_bal_spec_pattern}} ~
\subfloat[Pattern ($Re=400$), mean flow]{ \includegraphics[width=0.5\columnwidth]{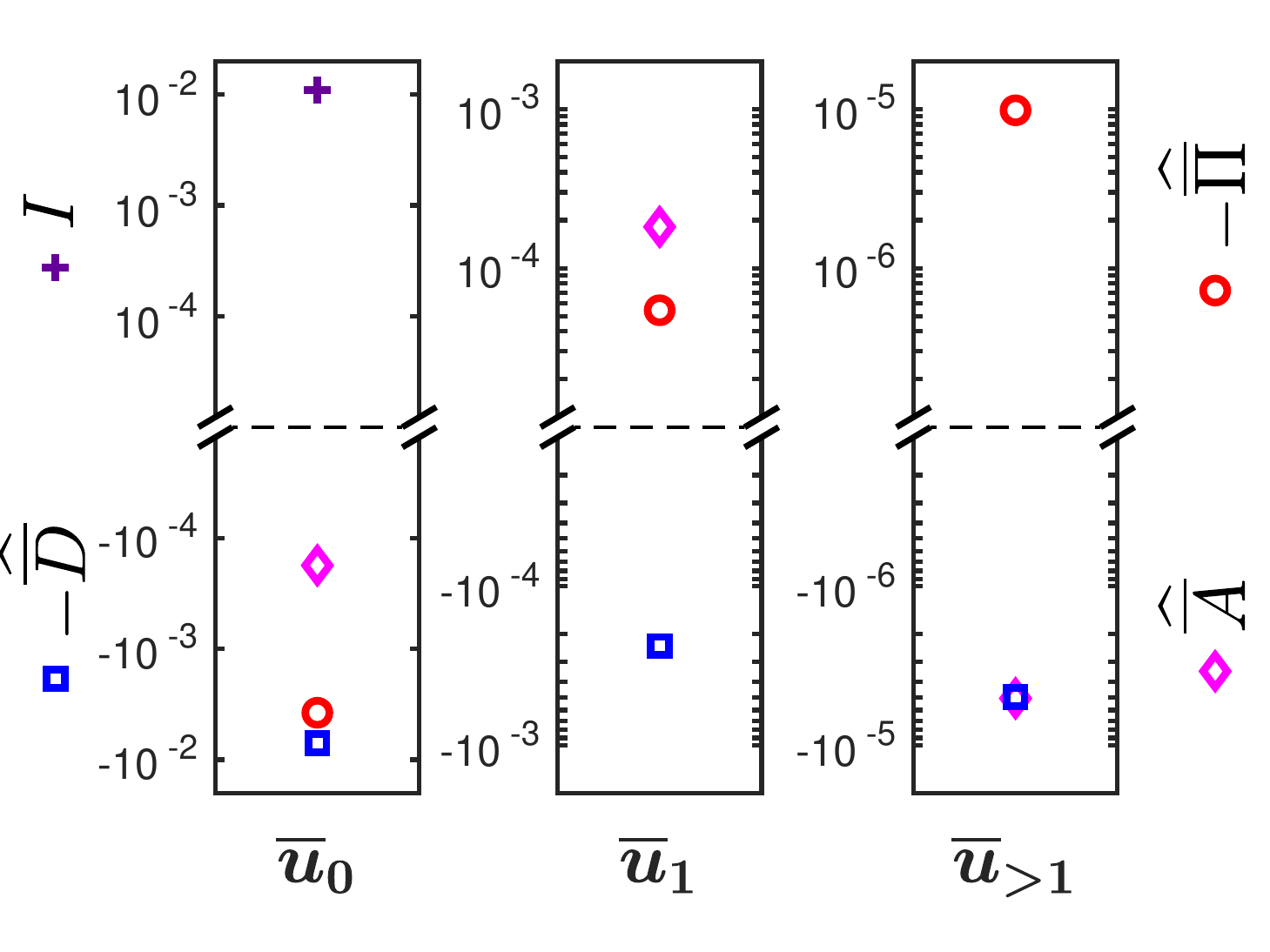} \label{fig:mean_bal_spec_pattern}} \\
\subfloat[Uniform ($Re=500$), TKE]{ \includegraphics[width=0.5\columnwidth]{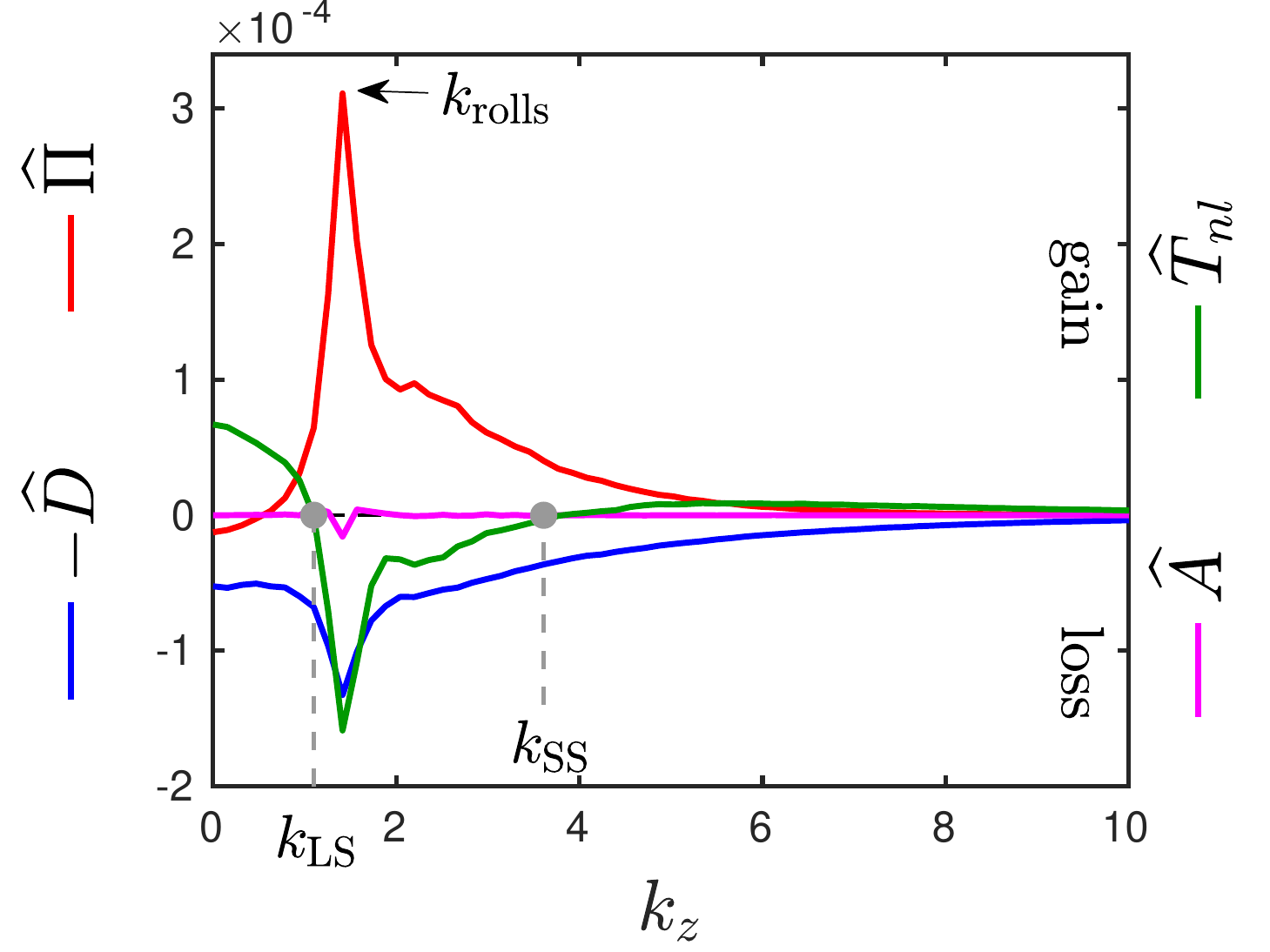} 
\label{fig:tke_bal_spec_unif}} ~
\subfloat[Uniform ($Re=500$), mean flow]{ \includegraphics[width=0.5\columnwidth]{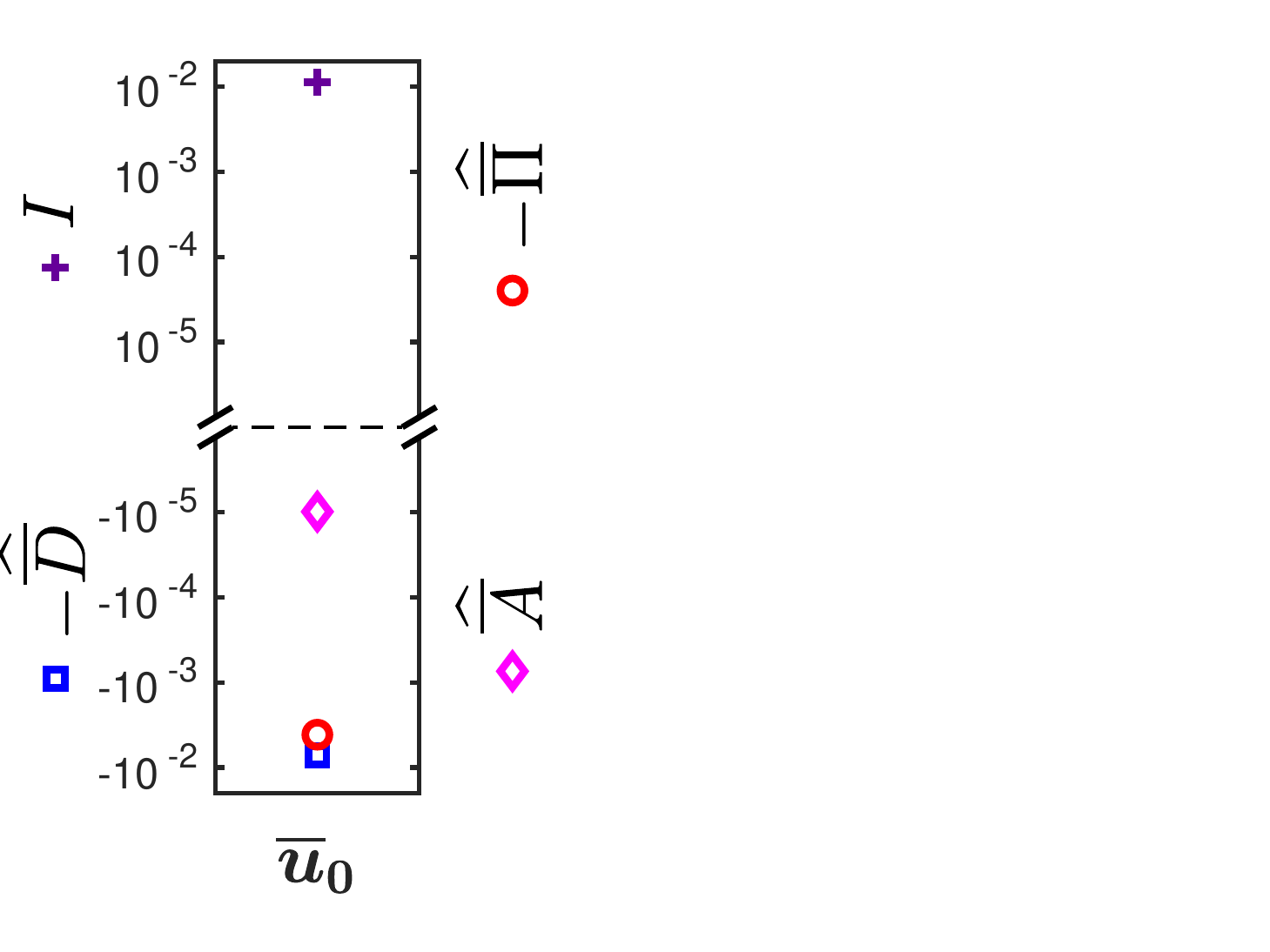} \label{fig:mean_bal_spec_unif}
} 
\caption{(a) Spectral energy budget (\ref{eq:tke_bal_spec}) for a pattern at $Re=400$, integrated over $y\in[-1, 1]$. Viscous and pressure transfers are not shown as they integrate to 0. The grey circles indicate $\kLS$ and $\kSS$, which delimit the spectral region where transfer $\Transnls$ is negative.
(b) Spectral energy budget of the mean flow \eqref{eq:mean_bal_spec} integrated over $y\in[-1, 1]$, shown for $k_z=0$, $k_z=2\pi/L_z$ and summed over $k_z>2\pi/L_z$. 
(c, d) show the same as (a, b) for a uniform state at $Re=500$. \resub{To convey both sign and order of magnitude, panels (b) and (d) show ${\rm sgn}(Q)\log|Q|$ for each quantity $Q$.}
}
\label{fig:tke_mean_bal}
\end{figure} 

We examine the spectral balance of the TKE \eqref{eq:tke_bal_spec}, integrated over the cross-channel direction. This balance is presented for the patterned state in figure \ref{fig:tke_bal_spec_pattern} ($Re=400$) and for the uniform state on figure \ref{fig:tke_bal_spec_unif} ($Re=500$). 
The transfer terms $\Transvs$ and $\Transps$ are not shown as they integrate to zero. 
(The $y$ dependence of energy transfer will be discussed in Appendix \ref{app:y_spec}.)

We first focus on the similarities between patterned and uniform states. We observe a peak in the production and dissipation terms near the energy-containing scale $\ksmax \simeq 1.41$, as we saw for the spectral energy in figures \ref{fig:spec_Lz800_Re} 
and \ref{fig:spec_MBU}.
At this scale, the nonlinear transfer $\Transnls$ is negative and of large amplitude: scale $\ksmax$ produces much more than it dissipates, and the remainder is transferred away from scale $k_z$ to other scales. 
% to other $k_z$. 
%For $k_z>\ksmax$, 
The nonlinear transfer becomes positive above a small-scale wavenumber that we denote $\kSS$. (In both the patterned state at $Re=400$ and in the uniform state at $Re=500$, we have $\kSS \simeq3.6$.)
%
%figure~\ref{fig:tke_bal_spec_pattern} and $k_z\simeq 4.0$ in figure~\ref{fig:tke_bal_spec_unif}).
%
This positive transfer at small scales is indicative of a direct energy cascade to small dissipative scales.

The TKE balance for $k_z<\ksmax$ contrasts with that at large $k_z$. First, production 
%decreases with decreasing $k_z$ and 
becomes negative for $k_z \lesssim 0.47$. This negative production at large scales appears in both patterned and uniform states. It corresponds to energy transfer from the fluctuations to the mean flow.
%
%The zone of negative production spans from $k_z=0$ to $k_z \simeq 0.47$ in both the patterned and uniform cases presented here. 
We note that this unusual sign of part of the production term has been also reported by \citet[]{symon2021energy} in spanwise-constant modes of channel flow in a minimal domain that is too small to support laminar-turbulent patterns.

Second, energy in the range $k_z < 0.94$ is fuelled by a positive nonlinear transfer $\Transnls$, which signifies a transfer from small to large scales. This is present in both patterned and uniform states. We denote the (large) scale at which this transfer becomes positive by $\kLS$ as seen in figures \ref{fig:scales}, 
\ref{fig:tke_bal_spec_pattern}, and \ref{fig:tke_bal_spec_unif}.
In the part of the spectrum $k_z<\kLS$, the influx of energy from smaller scales is mostly balanced by dissipation, while only a relatively small amount of energy is lost to the mean flow via negative production. 

Now considering the differences between the patterned (figure \ref{fig:tke_bal_spec_pattern}) and uniform states (figure \ref{fig:tke_bal_spec_unif}), the advection term $\Advs$ plays a more significant role in redistributing energy between scales in the patterned state: it is positive for $k_z<1.1 < \ksmax$, negative near $\ksmax$, and negligible for $k_z>3$. This role is very similar to that of nonlinear transfers $\Transnls$, but with weaker amplitude.
In the uniform state, $\Advs$ is nearly zero 
and would vanish if the mean flow were strictly uniform in $z$, see \eqref{eq:adv_unif}. 
This is not exactly the case here, especially at $k_z \simeq \ksmax$, probably due to insufficiently long averaging. 
Other differences are visible between the uniform and patterned states, especially regarding the shape and intensity of each individual curve. 
%For instance, the behaviour of $\Transnls$ and $\Dissips$ is changed near $\ksmax$: these are almost equal in the uniform case while $\Dissips$ exceeds $\Transnls$ in the patterned case.
\resub{
For instance, near $\ksmax$ one sees that $\Dissips \simeq \Transnls$ in the uniform case while $\Dissips$ exceeds $\Transnls$ in the patterned case.}

\subsubsection{Mean-flow balance}

The spectral energy balance of the mean flow \eqref{eq:mf_bal} is presented in figure \ref{fig:mean_bal_spec_pattern} and  \ref{fig:mean_bal_spec_unif} for both patterned and uniform states. In the patterned case, the three panels correspond to modes $\ubar_0$, $\ubar_1$ and $\ubar_{>1}$. \resub{For the uniform state, the terms in the balance of $\ubar_1$ and $\ubar_{>1}$ are zero up to numerical error, hence it is not meaningful to show them.} In both the patterned and uniform cases, $\ubar_0$ is fueled by the mean strain via injection term $I$ (purple cross). This energy is dissipated (blue square) and also transferred to the fluctuations via the production $\Prodms_0$ (red circle). Note that $\Prodms > 0$ corresponds to usual positive TKE production and hence a sink of energy with respect to the mean flow: production appears as $-\Prodms$ in the mean balance equation \eqref{eq:mean_bal_spec}.

For $\ubar_1$ in the patterned state (middle panel of figure \ref{fig:mean_bal_spec_pattern}), the main source of energy is the advective term $\Advms_1$, with some energy coming from the negative production $\Prodms_1 < 0$. %This component of the mean flow is therefore fuelled to some extent by interactions between the Reynolds stress and the mean shear.
Thus, the $\ubar_1$ component of the mean flow is fuelled to some extent by a negative transfer from fluctuations back to mean flow, but the advective contribution dominates.
The two sources are balanced by dissipation. 
The remaining scales in the mean spectral balance $k_z>2\pi/L_z$ (right panel of figure \ref{fig:mean_bal_spec_pattern}) are very weak compared to the first two components. 

Our results show that the advection term $\Advms$ plays a crucial role in the mean-flow balance in the patterned state. 
Since this term represents a transfer due to nonlinearities, its sum over $k_z$ and $y$ vanishes.
\resub{At $Re=400$, we find that $\Advms_0\simeq -1.77\times10^{-4}$, $\Advms_1\simeq1.82\times10^{-4}$, and $\sum_{k_z>2\pi/L_z} \int_{-1}^1 \Advms(y, k_z) ~\text{d}y \simeq -5\times10^{-6}$.
}
Hence we have the following approximate equality:
\begin{align}
    \Advms_0 \simeq - \Advms_1.
    \label{eq:adv_mf}
\end{align}
\resub{which holds for other values of $Re$ in the patterned regime.}
Even though the advection is negligible compared with the dominant terms in the $\ubar_0$ balance, it 
is the dominant source of energy
at the pattern scale. In a perfectly uniform case, $\Advms_0$ would be zero.

%In the uniform case, $ \Advms_0  \simeq -\Advms_1 \simeq 0$.

\resub{We refer the reader to Appendix \ref{app:y_spec} for the $y$ dependence of the terms in the mean-flow budget \eqref{eq:mean_bal_spec}.}

\subsubsection{Connection between TKE and mean flow}

We now investigate the connection between the TKE and mean flow, focusing particularly on the spectral production terms $\Prodms$ and $\Prods$.
%which appear differently in the TKE and mean-flow spectral balances (eq. \ref{eq:tke_bal_spec} and \ref{eq:mean_bal_spec}). We recall that these terms are equal when summed over $k_z$ (see, e.g.\ \ref{eq:total_prod}).
Recall that while these production terms take different forms in the TKE and mean-flow spectral balances (eq. \eqref{eq:tke_bal_spec} and \eqref{eq:mean_bal_spec}), upon integration over $y$ and summation over $k_z$ (equation \eqref{eq:total_prod}), they give the same total production $\Pi$.

We decompose the total production in two ways: first by writing the total TKE production $\Pi$ as a sum of its positive and negative parts, and second by considering the dominant contributions from $\ubar_0$ and $\ubar_1$ in the mean-flow production $\Prodms$:
\begin{align}
     \Pi = \widehat{\Pi}^{<0} + \widehat{\Pi}^{>0} \simeq \Prodms_0 + \Prodms_1 
     \label{eq:total_prod_appro}
\end{align}
where:
\begin{align}
    \widehat{\Pi}^{>0} \equiv \sum_{k_z=0}^{\infty} \int_{-1}^1\Prods (y,k_z) \text{ d}y ~ \Theta \left(  \int_{-1}^1 \Prods (y, k_z)  \text{ d}y \right) \\
    ~~~\text{and}~~~ 
    \widehat{\Pi}^{<0} \equiv \sum_{k_z=0}^{\infty} \int_{-1}^1 \Prods (y, k_z) \text{ d}y~ \Theta \left(- \int_{-1}^1 \Prods (k_z) \text{ d}y \right),
\end{align}
where $\Theta$ is the Heaviside function. 
We recall that figure \ref{fig:tke_bal_spec_pattern} shows that $\int_{-1}^1 \Prods(y, k_z) \text{ d}y <0$ occurs mostly at large scales.
%
\begin{comment}
\begin{align}
    \widehat{\Pi}^{>0} \equiv \sum_{k_z=0}^{\infty} \Prods_y (k_z)  ~ \Theta (  \Prods_y (k_z)) ~~~\text{and}~~~  \widehat{\Pi}^{<0} \equiv \sum_{k_z=0}^{\infty} \Prods_y (k_z)~ \Theta (-\Prods_y (k_z)),
\end{align}
with $\Prods_y = \int_{-1}^1\Prods (y,k_z) ~\text{d}y $ the $y$-integrated production and $\Theta$ the Heaviside function. 
We recall that figure \ref{fig:tke_bal_spec_pattern} shows that $\Prods_y<0$ occurs mostly at large scales.
\end{comment}
Each term in (\ref{eq:total_prod_appro}) in the patterned and uniform states is displayed in table \ref{tab:total_prod} for various values of $Re$. 

\begin{table}
  \begin{center}
\def~{\hphantom{0}}
    \begin{tabular}{ccccccc}
       State~ & $Re$~~ & $\Pi$ & $\Prodms_0$ &  $\Prodms_1$ & $\widehat{\Pi}^{>0}$ & $\widehat{\Pi}^{<0}$
       \\[3pt]
           Pattern~ & 400~~ & $3.71\times10^{-3}$ & $3.77\times10^{-3}$ & $-5.43 \times 10^{-5}$ & $3.76\times10^{-3}$ & $-5.34\times10^{-5}$ \\ 
           Pattern~ & 430~~ & $3.82\times10^{-3}$ & $3.87\times10^{-3}$ & $-4.10\times10^{-5}$ & $3.87\times10^{-3}$ & $-5.36\times10^{-5}$ \\
            Uniform~ & 430~~ & $4.14\times10^{-3}$ & $4.14\times10^{-3}$ & 
            $O(10^{-6})$ &$4.20\times10^{-3}$ & $-6.30\times10^{-5}$ \\
            Uniform~ & 500~~ & $4.12\times10^{-3}$ & $4.11\times10^{-3}$ & $O(10^{-6})$ & $4.17\times10^{-3}$ & $-5.64\times10^{-5}$ \\
  
     \end{tabular}
   \caption{Production terms appearing in the mean flow (\ref{eq:mean_bal_spec}) and the TKE (\ref{eq:tke_bal_spec}) balance, as decomposed in (\ref{eq:total_prod_appro}), for the flow regimes presented in figure \ref{fig:tke_mean_bal}.}
  \label{tab:total_prod}
  \end{center}
\end{table}

We observe that in the patterned case the positive production is very close to $\Prodms_0$ and the negative production is very close to 
$\Prodms_1$, {\em i.e.}
$\Prodms_0  \simeq \widehat{\Pi}^{>0}$ and
$\Prodms_1 \simeq \widehat{\Pi}^{<0}$.
In the uniform case, $\Prodms_1$ is very small and $\Prodms_0$ accounts for essentially all the production, so it is the sum of the positive and negative parts. 
In other words:
\begin{align}
\Prodms_0  ~
\begin{dcases}
    \simeq \widehat{\Pi}^{>0}  & \text{in patterned state}\\
    \simeq \widehat{\Pi}^{>0} + \Pi^{<0} & \text{in uniform state  }
 \end{dcases}
 ~~~~~ \Prodms_1 ~
 \begin{dcases}
    \simeq \Pi^{<0}   & \text{in patterned state} \\
    \ll \Pi^{<0}& \text{in uniform state  }
 \end{dcases}
\end{align}
This supports an essential connection between the TKE and the mean-flow production terms: in the patterned state, almost all negative TKE production goes to $\ubar_1$, and almost all positive TKE production comes from $\ubar_0$; in the uniform state, the negative TKE production is absorbed by $\ubar=\ubar_0$.
(In all cases, the negative production, $\Pi^{<0}$, represents less than $1.5\%$ of $\Pi$: $-\Pi^{<0}/\Pi \simeq 1.46\% $ at $Re=400$ and $1.37\%$ at $Re=500$.) 

\begin{figure}
    \centering
\subfloat{ \includegraphics[width=0.5\columnwidth]{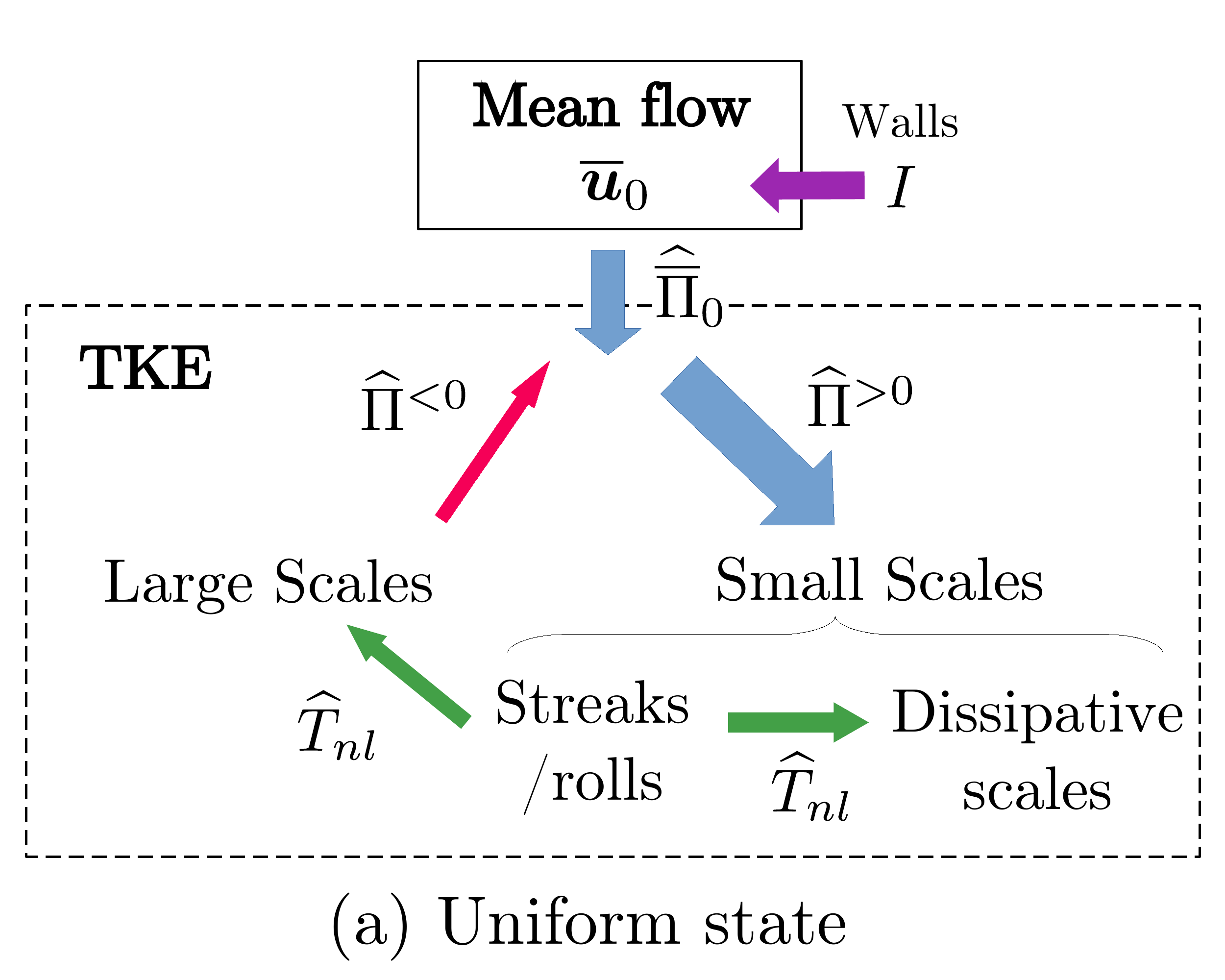}     \label{fig:sketch_unif}}~
\subfloat{ \includegraphics[width=0.5\columnwidth]{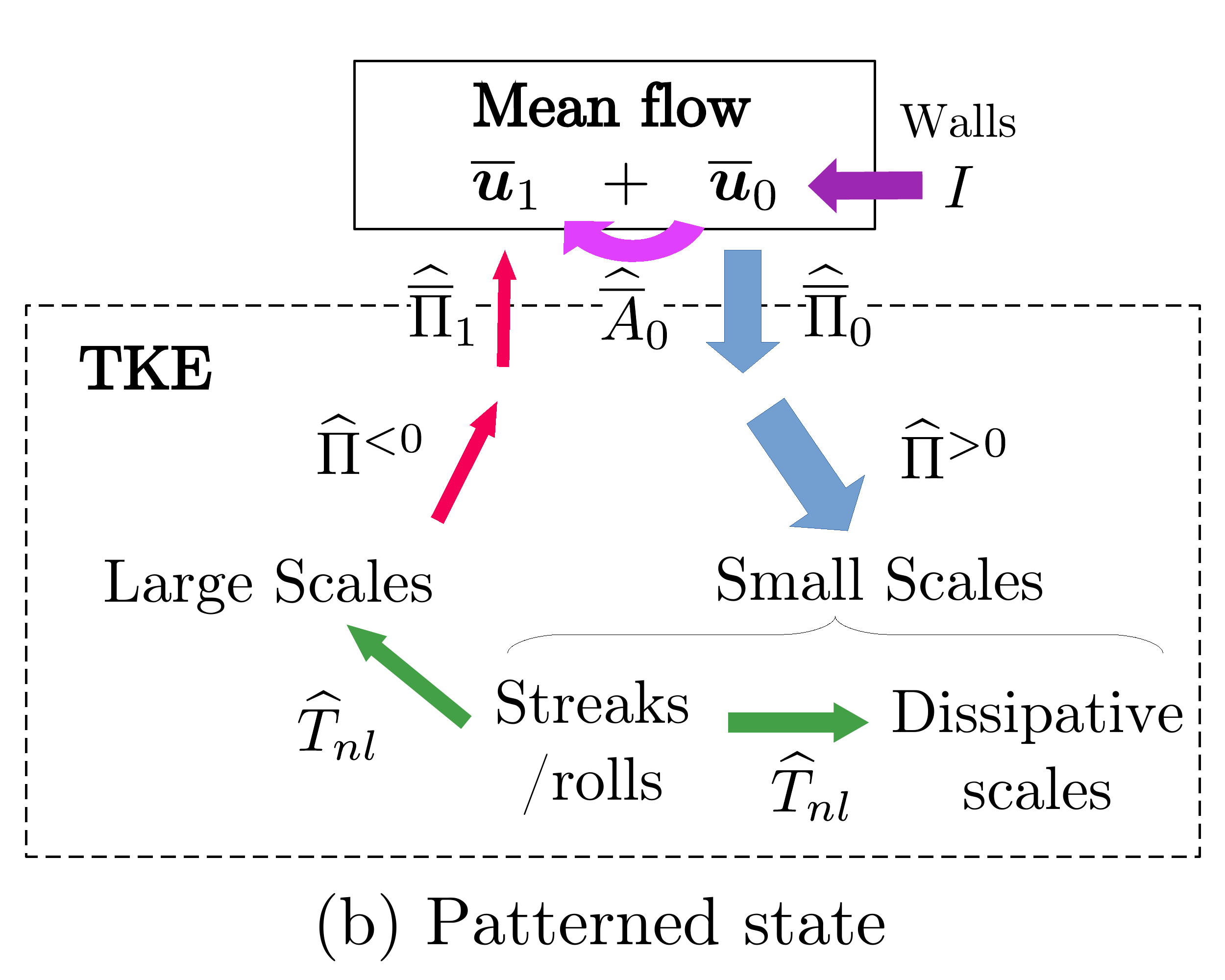}    \label{fig:sketch_pattern}} \\
\caption{Illustration of the mean flow - TKE interaction for (a) the uniformly turbulent state 
and (b) the patterned state. 
In the uniform state, $\Prodms_0 \simeq \Prods^{<0} + \Prods^{>0}$, while in the patterned state, $\Prodms_0 \simeq  \Prods^{>0}$ and $ \Prodms_1 \simeq \Prods^{<0} $.}
\label{fig:sketch}
\end{figure}

At this stage, we can draw the following conclusions, illustrated in figures \ref{fig:sketch_unif} and \ref{fig:sketch_pattern}:

\begin{enumerate}[leftmargin=*,labelindent=8mm,labelsep=3mm, itemindent=0mm, label=(\arabic*)]
    \item Most of the energy flows into the mean flow and then to TKE according to the usual picture from developed shear flows: energy is injected to $\ubar_0$ by viscous stress, and is transferred to fluctuations via positive production. TKE is mostly produced at the scale of the energy-containing eddies (here, streaks and rolls) and is transferred to smaller scales where it is dissipated. 
    
   \item An important modification to this usual picture is the presence of an inverse transfer of some TKE to large scales via triad interactions $\Transnls$.
   This energy is not entirely dissipated and instead feeds back to the mean flow via negative production $\Prods^{<0}$.
   
   \item Although weak compared to total production $\Pi$, this negative production $\Prods^{<0}$ fuels $\ubar_1$ in the patterned state. 
   
   \item In the patterned state, $\Advms_1$ is the main source of energy of $\ubar_1$: nonlinearities of the mean flow play a stronger role than negative production. 
\end{enumerate}

We have defined large scales as those for which the nonlinear transfer is negative:
$k_z < \kLS$ in figures \ref{fig:tke_bal_spec_pattern} and \ref{fig:tke_bal_spec_unif}.
This separates the large and small scales in figure \ref{fig:sketch}. Note, however, that the scales at which production becomes negative are even larger: $k_z \lesssim 0.5 < \kLS$ in figures \ref{fig:tke_bal_spec_pattern} and \ref{fig:tke_bal_spec_unif}. We do not distinguish these different notions of large scales in figure \ref{fig:sketch}.

\begin{figure}
    \centering
\includegraphics[width=0.5\columnwidth]{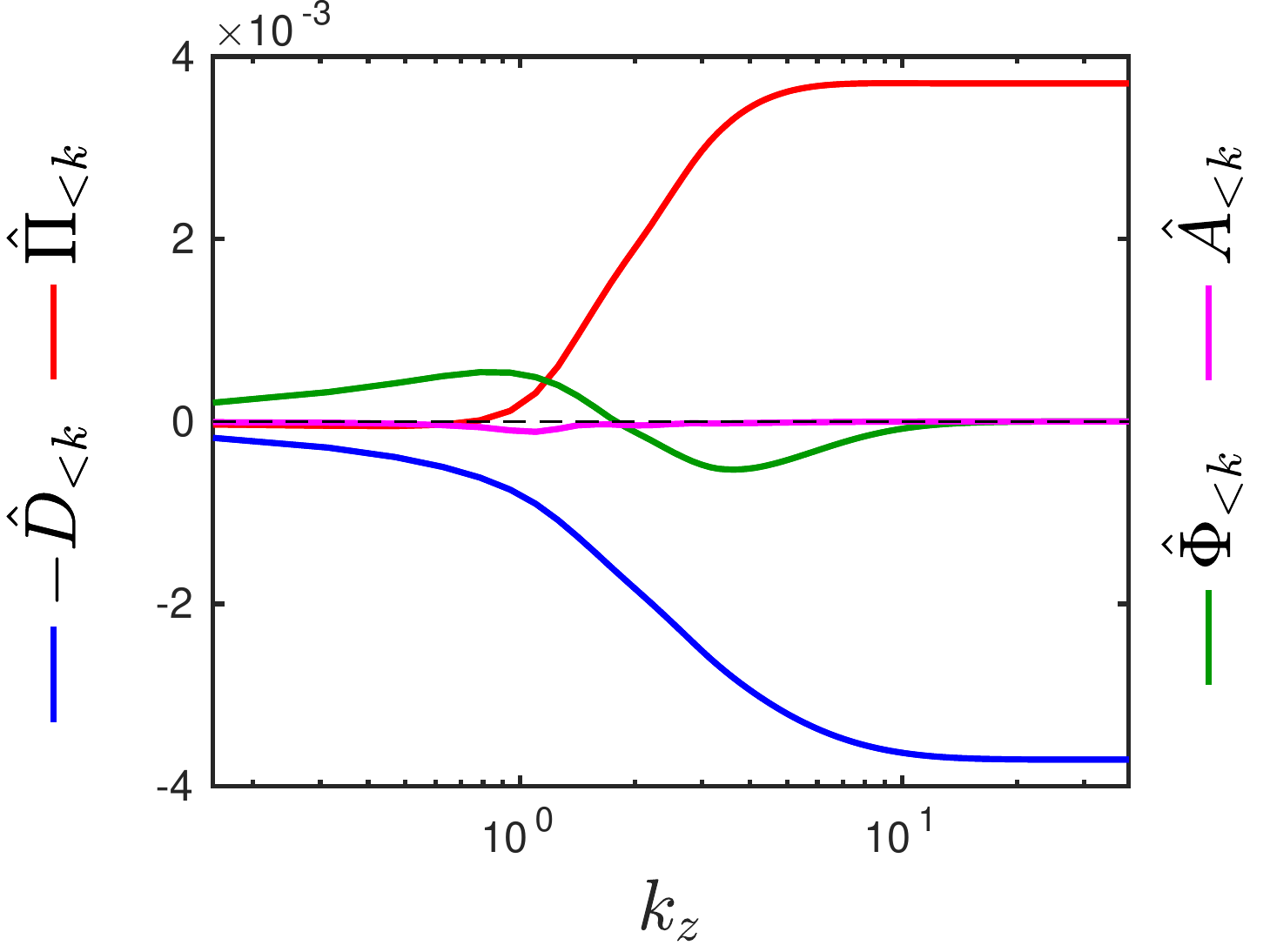}
 \caption{%Scale-by-scale 
 Cumulative energy balance \eqref{eq:k_spec_bal} integrated over $y\in[-1, 1]$ in the patterned case at $Re=400$.}
\label{fig:k_spec_bal}
\end{figure}

We extend these considerations of transfers across scales by considering the quantities
\begin{align}
    \widehat{\Pi}_{<k}(k_z) \equiv \sum_{k_z^\prime < k_z }\int_{-1}^1 \Prods (y, k_z^\prime) ~ \text{d}y, ~~~
    \widehat{D}_{<k}(k_z) \equiv \sum_{k_z^\prime < k_z }\int_{-1}^1 \Dissips (y, k_z^\prime) ~ \text{d}y \nonumber \\
    \widehat{\Phi}_{<k}(k_z) \equiv \sum_{k_z^\prime < k_z }\int_{-1}^1 \Transnls (y, k_z^\prime) ~ \text{d}y, ~~~
    \widehat{A}_{<k}(k_z) \equiv \sum_{k_z^\prime < k_z }\int_{-1}^1 \Advs (y, k_z^\prime) ~ \text{d}y 
    \label{eq:k_spec_bal}
\end{align}
These scale-to-scale quantities 
are shown in figure~\ref{fig:k_spec_bal}. $\widehat{\Phi}_{<k}$ is the nonlinear energy flux across a wavenumber $k_z$.
This integrated picture 
%conveys the presence of 
reveals a zone of inverse flux of energy to large scales ($\widehat{\Phi}_{<k} >0$ for $k_z<1.88$). For $k_z < O(1)$, this inverse transfer is the dominant source and is mostly balanced by dissipation. Starting at $k_z > O(1)$, production comes into play and eventually is the only source.

We emphasise that this strong inverse transfer does not correspond to an inverse cascade {\em per se}, because it does not lead to an accumulation of energy towards the largest available scale in the system. 
%
%Indeed, simulations in Large Slender Boxes have emphasized the presence of a small range of energetic large scales, around $\lambda_z \simeq 40$ (figure~\ref{fig:spec_Lz800_Re}). It is expected that 
%scales larger than 40 are less energetic and that 
%there is a build-up making energy condense around this finite scale.
\resub{Rather, as presented in figure~\ref{fig:spec_Lz800_Re}, 
simulations in Large Slender Boxes show that energetic large scales are concentrated around $\lambda_z \simeq 40$, the scale of the turbulent-laminar patterns and not the domain scale.}

%It is expected that 
%scales larger than 40 are less energetic and that there is a build-up making energy condense around this finite scale.

\section{Evolution with Reynolds number}
\label{sec:bal_Re}

%We now address the dependence of the global energy balance on $Re$. Unlike previous studies \citep[]{tuckerman2011patterns, rolland2011ginzburg}, we do not focus on an order parameter for the transition between uniform turbulence and patterns, but rather compute the Reynolds decomposition for each of the two states through the transition. 
%\resub{Note that as patterned and uniform states coexist over a finite range of Reynolds number in the Minimal Band Unit (as previously illustrated in figure \ref{fig:probes_MBU_R430}), the transition between them is discontinuous with $Re$, which explains the discontinuities in the global quantities that we will compute.}
%
%\resub{We will show in \citet{gome2} that in a Long Slender Box where the flow is less spatially constrained, the transition from uniform to patterned turbulence is in fact smooth.} 

%Figure \ref{fig:Re_evol} presents the evolution of several quantities computed in a Minimal Band Unit of $L_z=40$ for the uniform states at higher $Re$ and for the patterned states at lower $Re$.  For fluctuating bistable cases at intermediate $Re$ (e.g $Re=430$, as shown in figure \ref{fig:probes_MBU_R430}), conditional averaging has been carried out over selected time windows during which the state is either patterned or uniform. 
%To compare across different values of $Re$, we will normalise these quantities either by viscous wall units or by global quantities (total injection $I$ or total production $\Pi$).

We now address the dependence of the global energy balance on $Re$. Unlike previous studies \citep[]{tuckerman2011patterns, rolland2011ginzburg}, we do not focus on an order parameter for the transition between uniform turbulence and patterns, but rather compute the Reynolds decomposition for each of the two states throughout the transition. \resub{Recall that at intermediate $Re$ (e.g.\ $Re=430$, as shown in figure \ref{fig:probes_MBU_R430}), the flow fluctuates between patterned and uniform states, similar to the dynamics of a fluctuating bistable system. With this in mind, conditional averaging has been carried out over selected time windows during which the state is either patterned or uniform. This results in discontinuities with $Re$ in most global measures, because the patterned and uniform states are different. Again, here we are not investigating the nature of the transition, rather we are seeking to quantify properties of the patterned and uniform flows through it.} 
\resub{In our companion paper \citet[Part 2]{gome2} we investigate 
%SG: the nature of
the transition in a Long Slender Box where the flow is not tightly constrained spatially as here, and show that the transition from uniform to patterned turbulence is in fact smooth.} 

%For fluctuating bistable cases at intermediate $Re$ (e.g $Re=430$, as shown in figure \ref{fig:probes_MBU_R430}), conditional averaging has been carried out over selected time windows during which the state is either patterned or uniform. 
%To compare across different values of $Re$, we will normalise these quantities either by viscous wall units or by global quantities (total injection $I$ or total production $\Pi$).

\begin{figure}
    \centering
\includegraphics[width=\columnwidth]{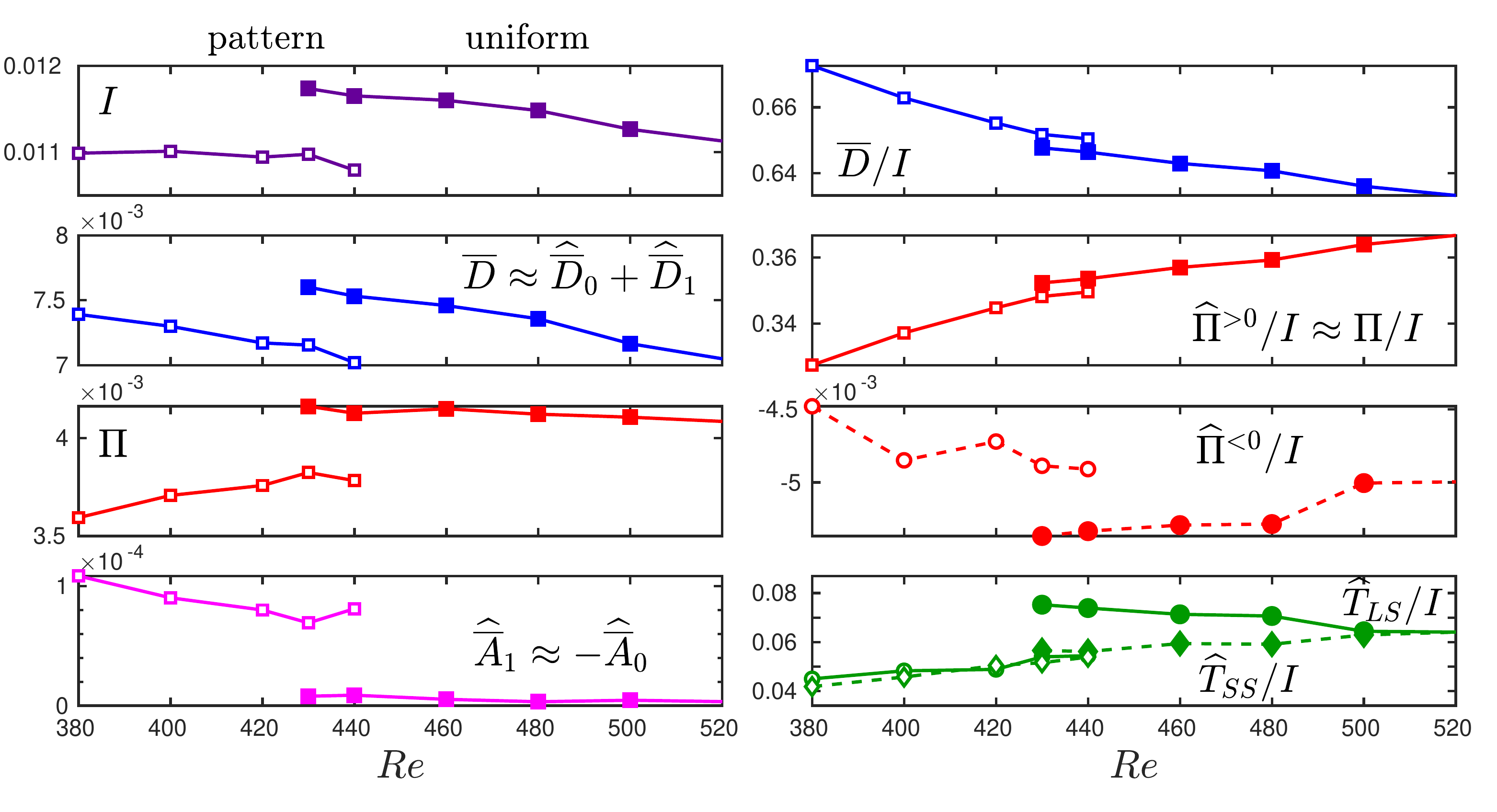}  
\caption{Evolution with $Re$ of various energetic quantities defined throughout the text (equations \eqref{eq:mf_bal},  \eqref{eq:pi0_pi1}, and \eqref{eq:injection}). Open and filled symbols are used for the patterned and uniform states respectively.
In the bottom-right panel, circles and diamonds stand respectively for $\widehat{T}_{LS}/I$ and $\widehat{T}_{SS}/I$.}
    \label{fig:Re_evol}
\end{figure}

Figure \ref{fig:Re_evol} presents the evolution of several quantities computed in a Minimal Band Unit of $L_z=40$ for the uniform states at higher $Re$ and for the patterned states at lower $Re$.  
We first consider the terms appearing in the mean-flow balance and show their evolution with $Re$. 
\resub{
We plot the injection $I$ along with total mean-flow dissipation $\overline{D}$ and the total TKE production $\Pi$, where $\overline{D} \simeq \Dissipms_0 + \Dissipms_1$ since these terms dominate the dissipation (see figure~\ref{fig:mean_bal_spec_pattern}). Recall also that the total TKE production
and dissipation are equal; equation~\eqref{eq:Pi_equals_eps}.}
As seen in figure \ref{fig:Re_evol}, all quantities reach a maximum in the uniform state at $Re=430$, before dropping discontinuously to the patterned state as $Re$ is decreased. We also show the main source of energy of the large-scale flow, $\Advms_1\simeq -\Advms_0$, which directs energy from $\ubar_0$ to $\ubar_1$ via advection.
$\Advms_1$ undergoes an especially dramatic increase %by a factor of nearly ten,
when going from uniform turbulence to the patterned state. \resub{(We recall that $\Advms_1$ vanishes in the uniform state up to numerical error.)} 

\resub{The right panels of figure \ref{fig:Re_evol} show quantities normalised by the injection rate $I$.
$\overline{D}/I$ is seen to increase with decreasing $Re$ in figure \ref{fig:Re_evol}, signifying that the mean-flow dissipation $\overline{D}$ increases more rapidly than the injection rate $I$ with decreasing $Re$. 
As a consequence, relatively less energy is transferred to turbulence with decreasing $Re$. This is confirmed by the plot of normalised TKE production $\widehat{\Pi}^{>0}/I \simeq \Pi/I$, which shows a decrease with decreasing $Re$.
Meanwhile, while considerably smaller in magnitude, the normalised negative production, $\widehat{\Pi}^{<0}/I$, decreases with decreasing $Re$ in the uniform regime, before switching to a value of lower intensity in the patterned state.
Altogether, this shows that a larger fraction of the total energy is retained by the mean flow at lower $Re$. }

We finally turn to the evolution of transfer terms with $Re$. For this purpose, we focus only on the nonlinear transfers into large scales at $k_z < \kLS$, and into small dissipative scales at $k_z > \kSS$. (See figure \ref{fig:tke_mean_bal}.) We define the total nonlinear transfer to large scales $\widehat{T}_{LS}$ and to small scales $\widehat{T}_{SS}$ by
\begin{align}
\widehat{T}_{LS} =  \sum_{k_z \leq \kLS} \int_{-1}^1 \widehat{T}_{nl}~\text{d}y \qquad
\widehat{T}_{SS} =  \sum_{k_z> k_{SS}} \int_{-1}^1 \widehat{T}_{nl}~\text{d}y ,
\label{eq:Tss}
\end{align}
We plot both $\widehat{T}_{LS}/I$ (circles) and $\widehat{T}_{SS}/I$ (diamonds) in figure \ref{fig:Re_evol}. 
For the uniform state, slightly more energy is transferred to large scales as $Re$ decreases.  $\widehat{T}_{LS}/I$ undergoes a discontinuous drop at the transition to patterns, where relatively less transfer goes to large scales. 
%On the other hand, the small-scale transfers are barely impacted by decreasing $Re$ and by the change in state. 
On the other hand, the small-scale transfers decrease monotonically with decreasing $Re$ and 
are barely impacted by the change in state.
We find that $\widehat{T}_{LS} \sim \widehat{T}_{SS} $ in the patterned state, for reasons unknown.

\resub{Interestingly, the normalised quantities $\overline{D}/I$, $\Pi/I$ and $\overline{T}_{SS}/I$ show a less abrupt transition from the uniform to the patterned state than their non-normalised counterparts, and could even be approximated as continuous as the system transitions from uniform to patterned flow
(e.g.\ we find a difference of order $0.5\%$ in $\overline{D}/I$ in the uniform and patterned states at $Re=430$, while it is of $7\%$ in $\overline{D}$). This signifies that the relative turbulent dissipation is approximately the same in the patterned and the uniform states at fixed value of $Re$. }

\resub{In summary, as $Re$ decreases, the mean flow dissipates its energy more rapidly than the fluctuations do,
i.e.\ the flow is less turbulent and the mean flow retains more energy (this is mostly due to small turbulent scales dissipating less of the total energy,
and secondly, because of greater fuelling of the mean flow by negative production $\widehat{\Pi}^{<0}$).
It seems that there is a point at which the mean-flow undergoes a sort of \emph{dissipation crisis}, and diverts some of its energy from the uniform mode $\ubar_0$ to the large-scale mode $\ubar_1$, via the mean advection term.
Therefore, the patterned state can be seen as more adapted to an increasingly dissipative environment when $Re$ decreases.
}

\section{Conclusion}

Wall-bounded turbulence at low Reynolds numbers is marked by a strong scale separation between the streak/roll scale of the self-sustaining process that comprises the turbulence, and the large-scale flow associated with oblique laminar-turbulent patterns. In this article, we have computed the spectral energy balances for both the mean flow and the turbulent fluctuations in a Minimal Band Unit, thus revealing the energy transfers connecting the different scales in transitional plane Couette flow. 

As expected, TKE production is maximal at the scale of streaks and rolls, and a direct cascade sends energy to smaller dissipative scales. 
However, part of the TKE is also transferred to large scales via nonlinear interaction. At large scales, this energy is partly sent to the mean flow, via negative production. 

The intense large-scale flow along laminar-turbulent bands appears in the trigonometric component of the mean flow $\ubar_1$.  The main energy source for $\ubar_1$ is its nonlinear interaction with the uniform component $\ubar_0$ (via the term called $\Advms_1$ in this article). This interaction is due to the mean advection, which plays a significant role in both spatial and spectral transfers of mean-flow energy. Interestingly, the $\ubar_1$ component of the mean flow is also fueled by negative production transferring energy from fluctuations to 
mean flow. However, this is only a secondary driver of $\ubar_1$, as negative production accounts for only approximately $20 \%$ of its energy sources (see figure \ref{fig:mean_bal_spec_pattern}). 

Negative production has not received much attention although it has been reported for spanwise-constant modes at $Re_\tau=180$ in a minimal channel by \citet[]{symon2021energy}. 
%We have found negative production at large scales for $Re_\tau\lesssim 66$, which, although of weak intensity, plays a role in feeding the inhomogeneous mean flow in transitional patterns.
%
\resub{We have found negative production at large scales in both patterned and uniform turbulence in plane Couette, in the region $Re_\tau \lesssim 66$ studied here.}
%as low as $Re_\tau \simeq 30$, corresponding to $Re = 400$. (DB: haven't we stated this next sentence already? Maybe cut.) Although of weak intensity, plays a role in feeding the inhomogeneous mean flow in transitional patterns.}
%
\resub{Altogether, the processes energising large-scale motions (inverse transfers and negative production) described here in both patterned and uniform turbulence at low $Re_\tau$ seem to be different from those reported in fully developed wall-bounded turbulence \citep[]{cimarelli2013paths, mizuno2016spectra,  aulery2017spectral, cho2018scale, lee2019spectral, kawata2021scale, andreolli2021global}. 
%they indeed dominate the large-scale TKE balance and occupy the whole shear layer
See Appendix \ref{sec:y_spec} for a more detailed analysis.}

\resub{Our results indicate that 
as the environment becomes more dissipative with decreasing $Re$,
the mean flow in the uniform regime absorbs more and more energy, up to a most dissipative point where the flow transitions to the patterned state.
The patterned state reorganises this energy between the uniform mean flow $\ubar_0$ and the large-scale flow $\ubar_1$ through advection, in such a way that negative production is directed into the large-scale flow.
%the energy is reorganised such as to balance large-scale and small-scale transfers.
%Note however that the role of these energy processes in the origin of the transition from uniform to patterned state is only speculative at this stage. 
}

\resub{In physical space, a possible equivalent picture is that nucleation of quasi-laminar gaps becomes necessary for turbulence to be sustained: with decreasing $Re$, the turbulent mean flow dissipates too much energy compared to what is injected in the flow, such that it needs additional fluxes from quasi-laminar regions, as those reported in \S \ref{sec:phys_balance}.
\resub{This is essentially the physical argument put forward by \cite{barkley2016theoretical} for the localisation of turbulent puffs in pipe flow.}
%
%This scenario will be partly supported in Long Slender Boxes
\resub{Further support for this effect can be found} in our companion paper \citet[Part 2]{gome2}, where we will show that in Long Slender Boxes, as $Re$ is decreased, the total dissipation reaches a maximum just before laminar gaps become steady and patterns emerge.
}

Our analysis of energy budgets
does not directly invoke a dynamical mechanism, such as the self-sustaining process governing wall-bounded turbulence and related autonomous mechanisms describing large scales in developed turbulence \citep[]{hwang2010self, hwang2016self, de2017streak, cho2018scale}.
Further investigations are required to understand whether the %strong 
inverse transfers and negative production that we observe are connected to the nonlinear regeneration of rolls in the self-sustaining process. 
Note that the energetic imprint of the self-sustaining process in developed wall-bounded turbulence was recently analysed by \citet[]{cho2018scale} and \citet{kawata2021scale}, the latter emphasising the role of nonlinear transfers.

Finally, although the oblique simulation domain is very useful for the study of inter-scale distribution of energy in patterned transitional turbulence, further confirmation via simulations 
in large streamwise-spanwise oriented domains is also required: our simulation domain restricts the flow in a number of ways, such as imposing an orientation as well as a mean streak spacing due to the restrained short size $L_x=10$. These features do not seem to alter the robust observations that we have made about mean-turbulent interaction and inverse transfers in uniform turbulence (see Appendix \ref{app:mfu}). 
However, it would be beneficial to disentangle the streamwise and spanwise directions in the energy budget and to compute inter-component transfers, so as to better understand the role of the self-sustaining process in the generation of transitional large-scale structures. 

In \citet[Part 2]{gome2}, the energy processes described above will be essential to understand the selection of a finite wavelength of transitional patterns.

\section*{Acknowledgements}

The calculations for this work were performed using high performance computing resources
provided by the Grand Equipement National de Calcul Intensif at the Institut du D\'eveloppement
et des Ressources en Informatique Scientifique (IDRIS, CNRS) through Grant No. A0102A01119.
This work was supported by a grant from the Simons Foundation (Grant No. 662985).
The authors wish to thank Yohann Duguet, Santiago Benavides, Anna Frishman and Tobias Grafke for fruitful discussions, as well as the referees for their useful suggestions.

\section*{Declaration of Interests}
The authors report no conflict of interest.

\appendix
\section{Wall-normal dependence of spectral balance}
\label{app:y_spec}

\subsection{Values of $Re_\tau$ in a Minimal Band Unit and a Long Slender Box}
\label{app:Retau}

\begin{table}
  \begin{center}
\def~{\hphantom{0}}
  \begin{tabular}{ccccccccccc}
      &$Re$ & 400 & 420 & 440 &460 & 480 & 500 & 550 & 600 & 1000\\ 
        \hline 
      \multirow{2}{*}{MBU} 
      &$Re_\tau^p$ & 29.68 & 31.09 & 32.82 &34.61 & 35.90&37.33 & - & - & -\\
      &$Re_\tau^u$ & 30.65 & 32.24 & 33.69 & 35.08 &36.42 & 37.67 &40.66 & 43.62 & 66.42 \\
  \hline
     LSB &$Re_\tau$ & 29.83 & 31.68 & 33.51 & 35.00 & 36.34 & 37.63 & 40.67 & 43.63 & 66.42
     \end{tabular}
  \caption{Values of $Re_\tau$ for various values of $Re$ in a Minimal Band Unit of size $L_z=40$ and in a Long Slender Box of size $L_z=800$. In the MBU, $Re_\tau^p$ and $Re_\tau^u$ are obtained by averaging over the patterned or uniform state, respectively, while the time-average is unconditional in a LSB.
  For $Re>500$, the patterned state does not occur. 
  }
  \label{tab:Re_tau}
  \end{center}
\end{table}

% Wall units $y^+=(1-y)Re_\tau$ are generalised to define $k_z^+ = k_z / Re_\tau$.
%Later we will need the wall-normal coordinate in wall units $y^+=(1-y)Re_\tau$, and similarly for other quantities.
In a Minimal Band Unit at a transitional Reynolds number, the turbulence may be uniform or patterned during different time periods, i.e. it is temporally as well as spatially intermittent.
For this reason, for each value of $Re$, we take the time average in \eqref{eq:Retaudef} over a period during which the flow retains qualitatively the same state. 
This yields two slightly different values, $Re_\tau^u$ for a uniform state and $Re_\tau^p$  for a patterned state, as presented in table \ref{tab:Re_tau} for $L_z=40$. In the nondimensionalisations carried out in this article, we have used either $Re_\tau^u$ or $Re_\tau^p$, as appropriate for the flow state.

\resub{In Long Slender Boxes, we compute $Re_\tau$ by averaging unconditionally the flow state.}  This procedure does not resolve the local variability of the wall shear stress due to spatial intermittency; for this, we would need to omit $z$-averaging in \eqref{eq:Retaudef} to produce $z$-dependent values of $Re_\tau$; see \citet[]{kashyap2020flow} for a thorough analysis of fluctuations of $Re_\tau$ within and outside of turbulent bands.

\subsection{Energy balance at various $y$ locations}
\label{sec:y_spec}
In the main part of the article, nothing has been said about the location of the energy transfers in the wall-normal direction and no distinction has been made between near-wall and bulk effects on the mean flow and turbulent energies.
In this section, we present results on the TKE balance and subsequently the mean-flow balance for the patterned state at $Re=400$.

\begin{figure}
    \centering
 \includegraphics[width=1\columnwidth]{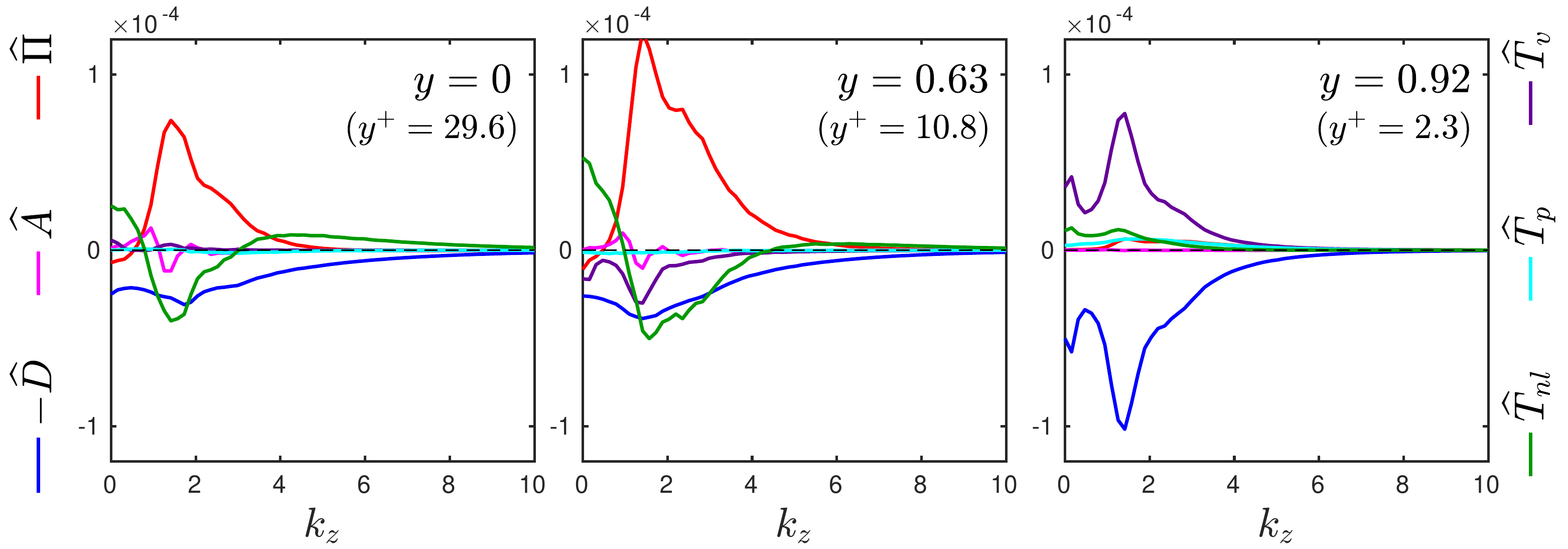} \\
  \caption{TKE spectral balance (\ref{eq:tke_bal_spec}) at different $y$ locations (left: mid-plane, $y=0$; middle: $y=0.63$;  right: near-wall, $y=0.92$). Shown is a patterned case at $Re=400$.}
 \label{fig:tke_bal_spec_y}
\end{figure}
\begin{figure}
    \centering
 \includegraphics[width=1\columnwidth]{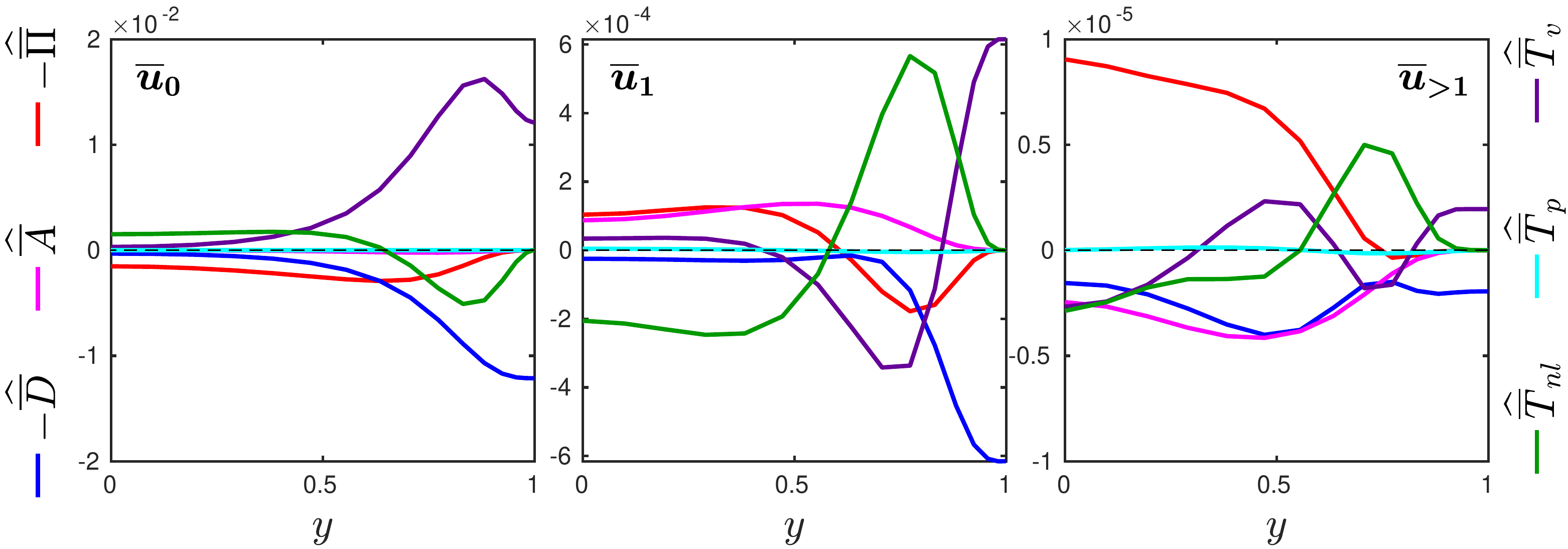}
\caption{Evolution of the mean-flow energy balance (\ref{eq:mean_bal_spec}) with $y$, for $k_z=0$, $k_z=2\pi/L_z$, and summed over $k_z>2\pi/L_z$. (Values for $y<0$ are obtained by reflection in $y=0$.) Shown is a patterned case at $Re=400$.}
\label{fig:mean_bal_spec_y}
\end{figure}

Figure \ref{fig:tke_bal_spec_y} shows the spectral TKE balance at different $y$ locations: the mid-plane ($y=0$, $y^+= 29.6$, left panel), the layer of maximal spectral production $\Prods$ ($y=0.63$, $y^+\simeq 11$, middle panel) and the near-wall region ($y=0.92$, $y^+\simeq 2.4$, right panel).

The balance in the near-wall region is simple because it is dominated by viscous effects, with injection of energy via the rate-of-strain compensated by dissipation.
A small portion of the energy comes from a positive transfer $\Transnls$. 
In the plane $y=0.63$, the production term $\Prods$ is maximal (as will be shown in \S \ref{subsec:y_kz_P_T}.). Production peaks at the roll scale $\ksmax$, while the dissipation, viscous diffusion and nonlinear transfers are all negative with similar magnitudes near this scale. Production becomes negative and nonlinear transfers positive at long length scales (small $k_z$), similar to what we showed for $y$-integrated quantities in \S \ref{sec:spec_results}.
The spectral balance at the mid-plane is qualitatively similar to that at the plane $y=0.63$, with the notable exception that the viscous diffusion $\Transvs$ vanishes due to reflection symmetry about the midplane. 
$\Prods$ and $\Transnl$ are smaller in the mid-plane than in the plane $y=0.63$, while $\Dissips$ and $\Advs$ have nearly the same magnitude in both planes.

The $y$ dependence of the mean-flow energy balance (\ref{eq:mean_bal_spec}) is displayed in figure \ref{fig:mean_bal_spec_y}. In line with our previous observations on $y$-integrated quantities (\S \ref{sec:spec_results}), figure \ref{fig:mean_bal_spec_y} reveals a different phenomenology depending on the wavenumber ($k_z=0$, $k_z=2\pi/L_z$ or $k_z>2\pi/L_z$). The gain in energy in $\ubar_0$ (left panel) due to the viscous transfer term $\Transvms$ is large near the wall where energy is injected into the flow, whereas the two terms involving the Reynolds stress, $\Prodms$ and $\Transnlms$, are dominant and in approximate balance at the mid-plane. Note that while $\Transnlms$ integrates to zero, it has a local influence: flow above $y=0.63$ transfers energy to flow below.

The balance of $\ubar_1$ (middle panel) presents a complex and interesting behaviour. We know from \S \ref{sec:phys_balance} that when integrated over $y$, the balance for mode $\ubar_1$ is such that $\Prodms_1<0$ and that this mode extracts energy from TKE. However, the $y$ dependence of this term shows a change in sign:
the production is only negative (i.e.\ $-\Prodms_1>0$) 
for $-0.6 \lesssim y \lesssim 0.6$. 
\resub{This suggests the importance of turbulence in the bulk region for sustaining the bands.}
$\Transnlms$ undergoes a change in sign at approximately the same $y$ value, with similar behaviour, although their $y$ integrals differ (the integral of $\Transnlms$ vanishes whereas that of $\Prodms$ is negative). $\Transnlms$ dominates the energy source at $y\simeq0.8$. At the wall ($y=1$), the energy balance is between viscous diffusion and dissipation.
The advection term $\Advms$ is always positive.

The situation at $k_z>2\pi/L_z$ is perhaps of negligible importance because of the small amplitude of the energy at this scale. However, we note that the balance near the wall (i.e.\ $0.7 \lesssim y \lesssim 1$) is qualitatively similar to that of mode $\ubar_1$, dominated by viscous diffusion, dissipation, and triad interaction.
In the bulk, energy comes from $\Prodms$ and is diverted towards the other terms.

\subsection{Production and nonlinear transfers in the $(y, k_z)$ plane}
\label{subsec:y_kz_P_T}

\begin{figure}
    \centering
\subfloat[Pattern, $Re=400$ ($Re_\tau = 29.7$)]{ \includegraphics[width=0.5\columnwidth]{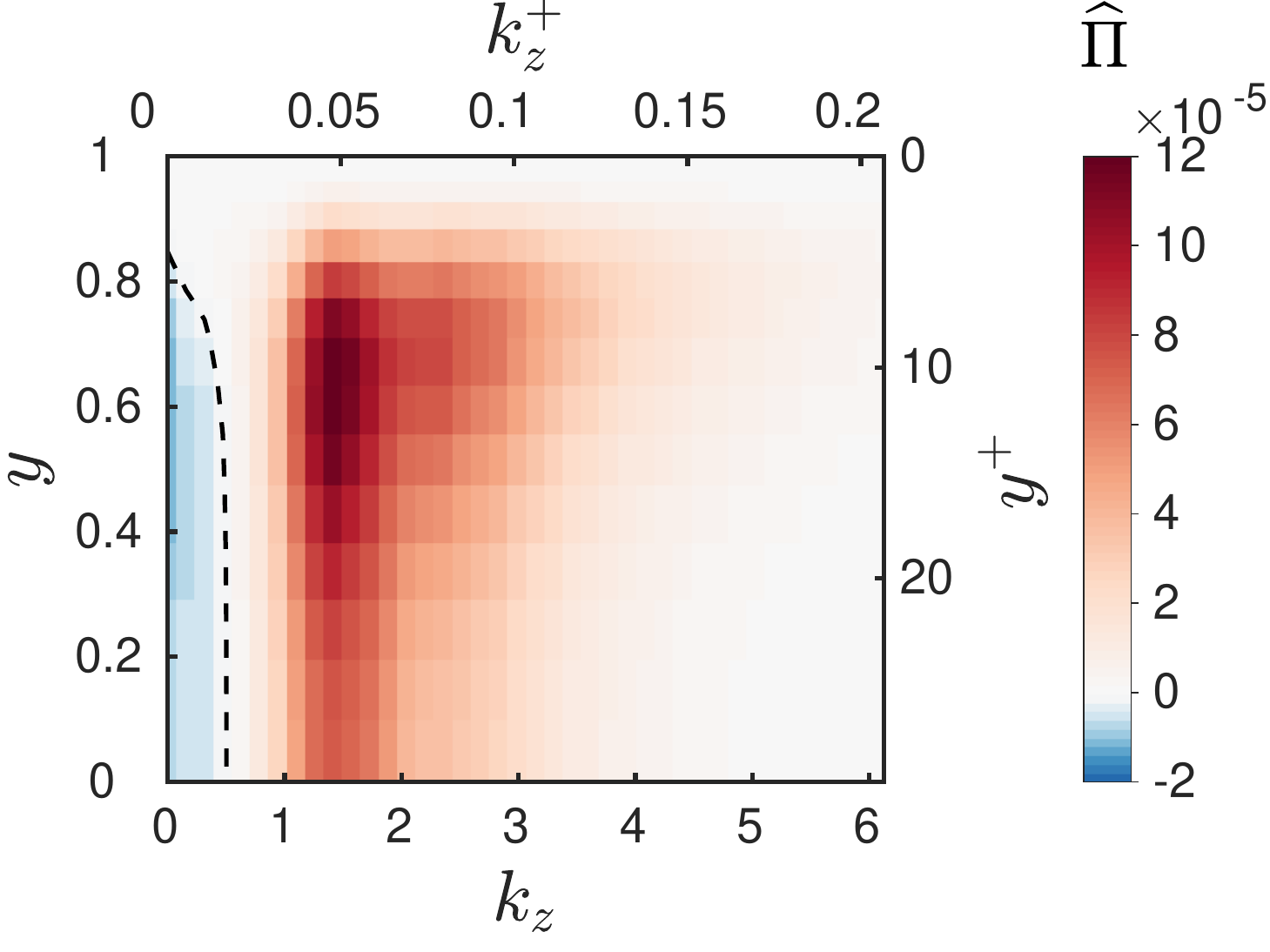}     \label{fig:prod_y_kz_380p}}~
\subfloat[Pattern, $Re=430$ ($Re_\tau = 31.9$)]{ \includegraphics[width=0.5\columnwidth]{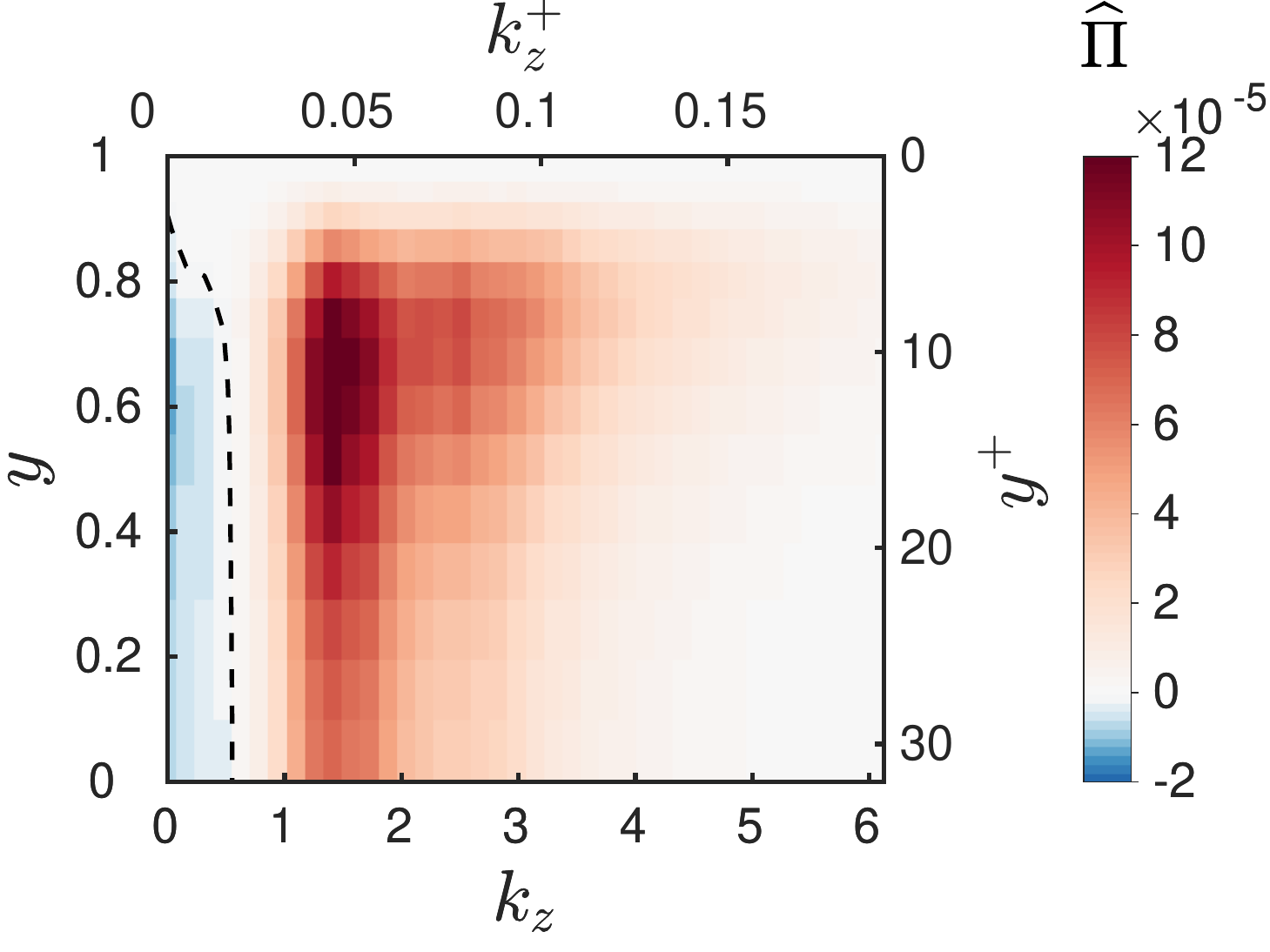}    \label{fig:prod_y_kz_430p}} \\
\subfloat[Uniform, $Re=430$ ($Re_\tau = 33.0$)]{ \includegraphics[width=0.5\columnwidth]{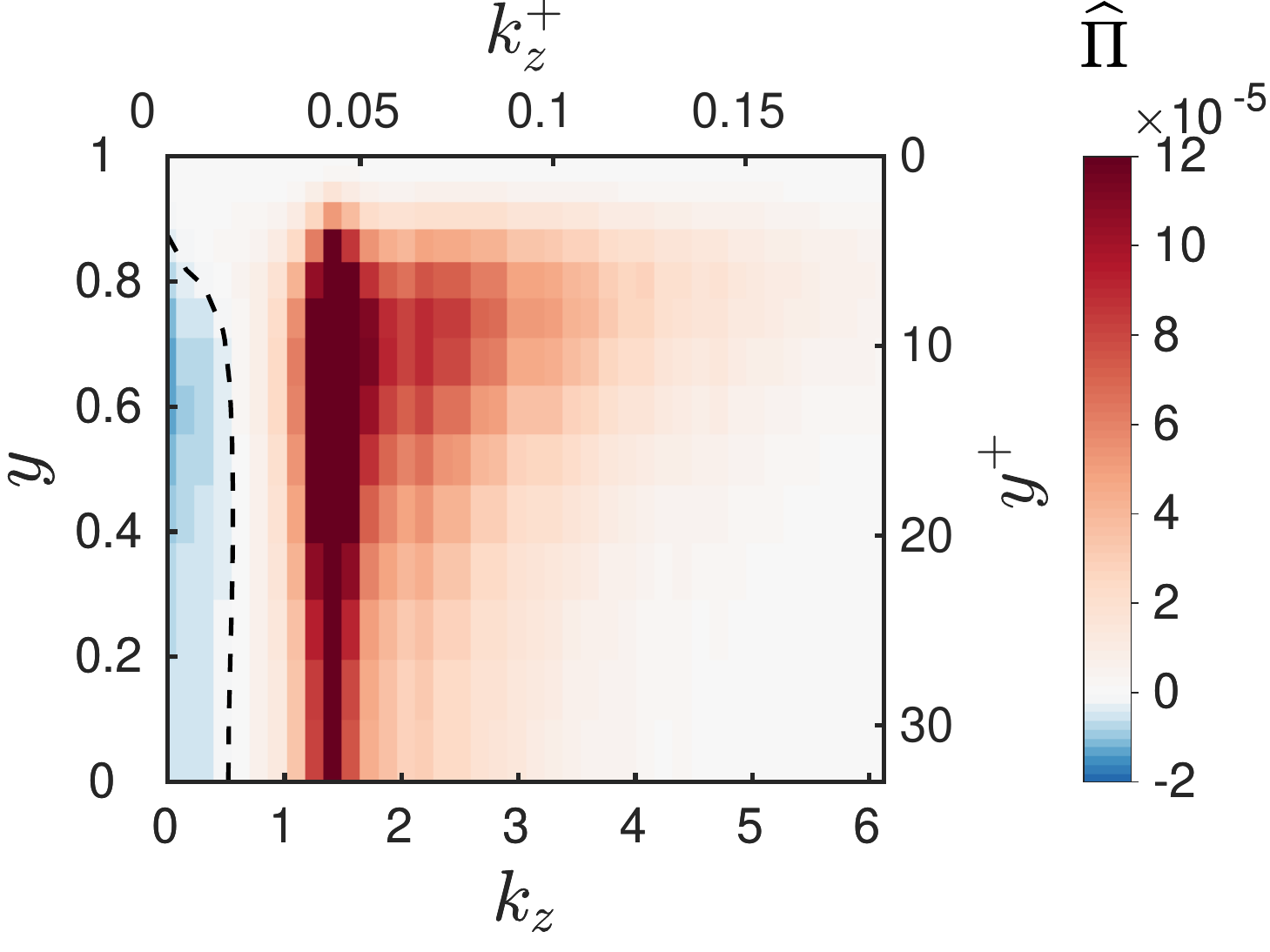} \label{fig:prod_y_kz_430u}} 
~
\subfloat[Uniform, $Re=1000$ ($Re_\tau = 66.4$)]{ \includegraphics[width=0.5\columnwidth]{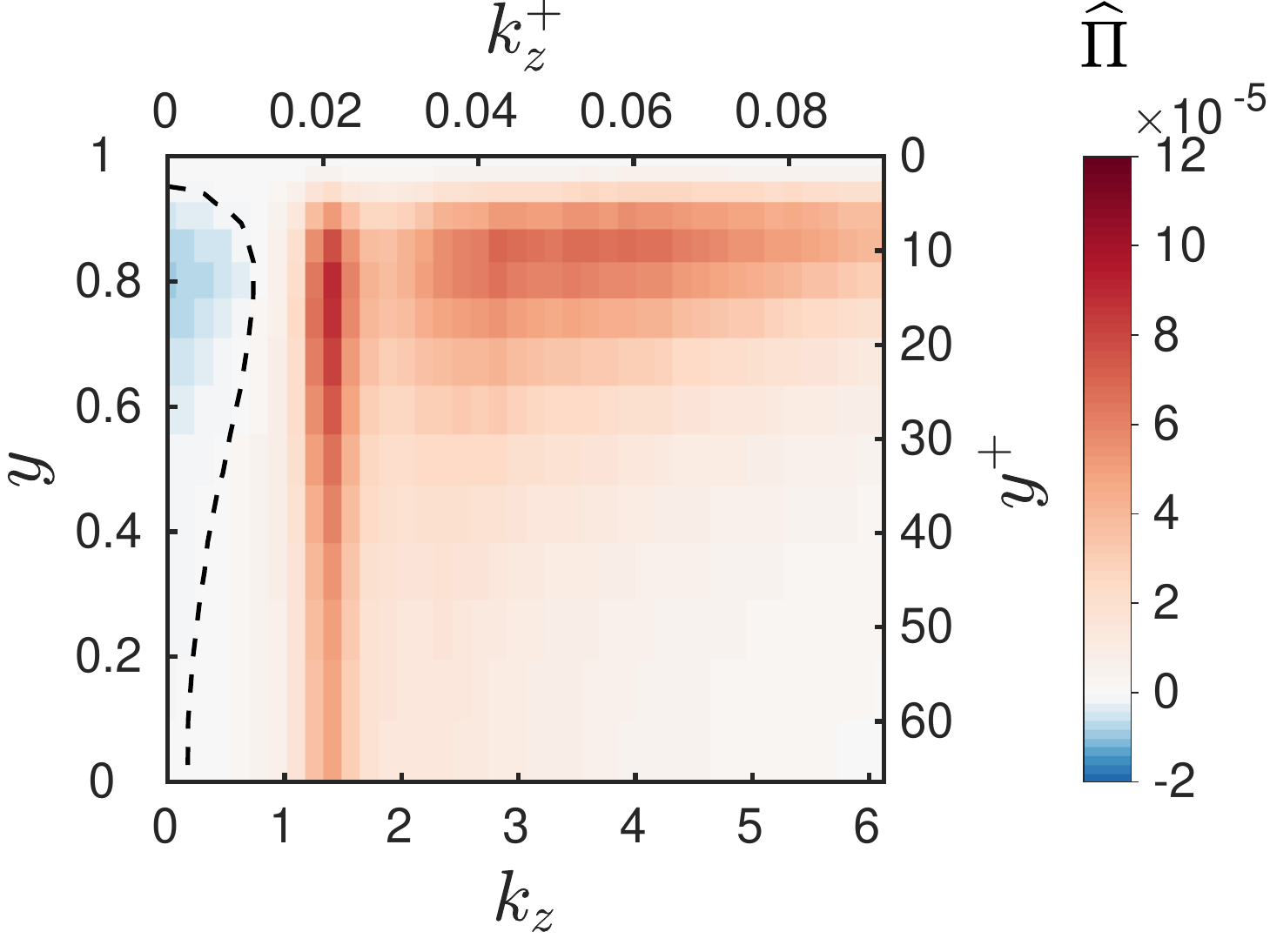}    \label{fig:prod_y_kz_1000u} }
\caption{Visualisations of production $\Prods(y, k_z)$ for different $Re$ and states. The cross-channel range is from the mid-plane ($y=0$, $y^+=Re_\tau$, lower axis) to the wall ($y=1$, $y^+=0$, upper axis). Dashed line separates positive from negative value for small $k_z$.}
 \label{fig:prod_y_kz}
\end{figure}

\begin{figure}
    \centering
\subfloat[Pattern, $Re=400$ ($Re_\tau = 29.7$)]{ \includegraphics[width=0.5\columnwidth]{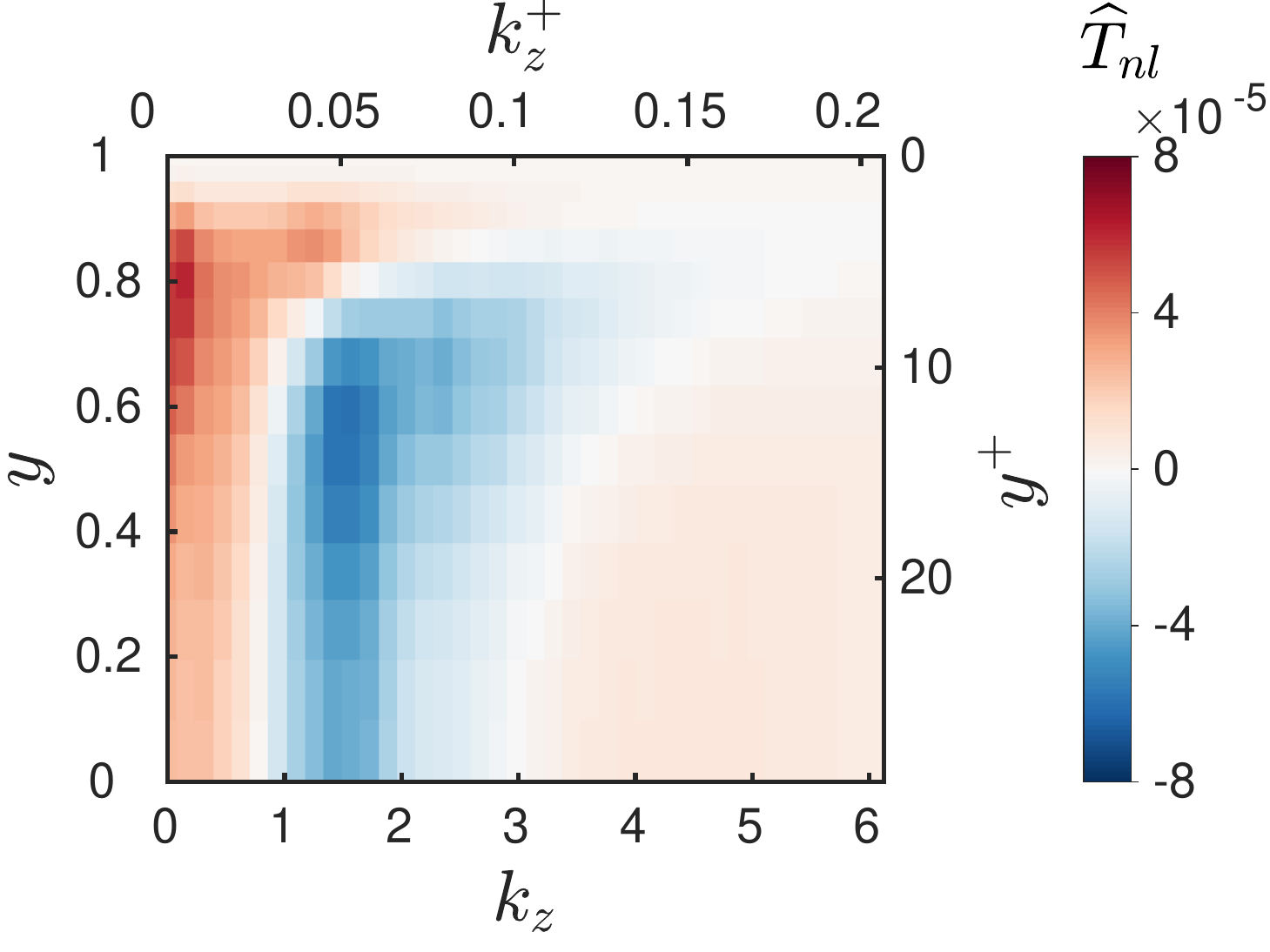}     \label{fig:trans_y_kz_380p}}~
\subfloat[Pattern, $Re=430$ ($Re_\tau = 31.9$)]{ \includegraphics[width=0.5\columnwidth]{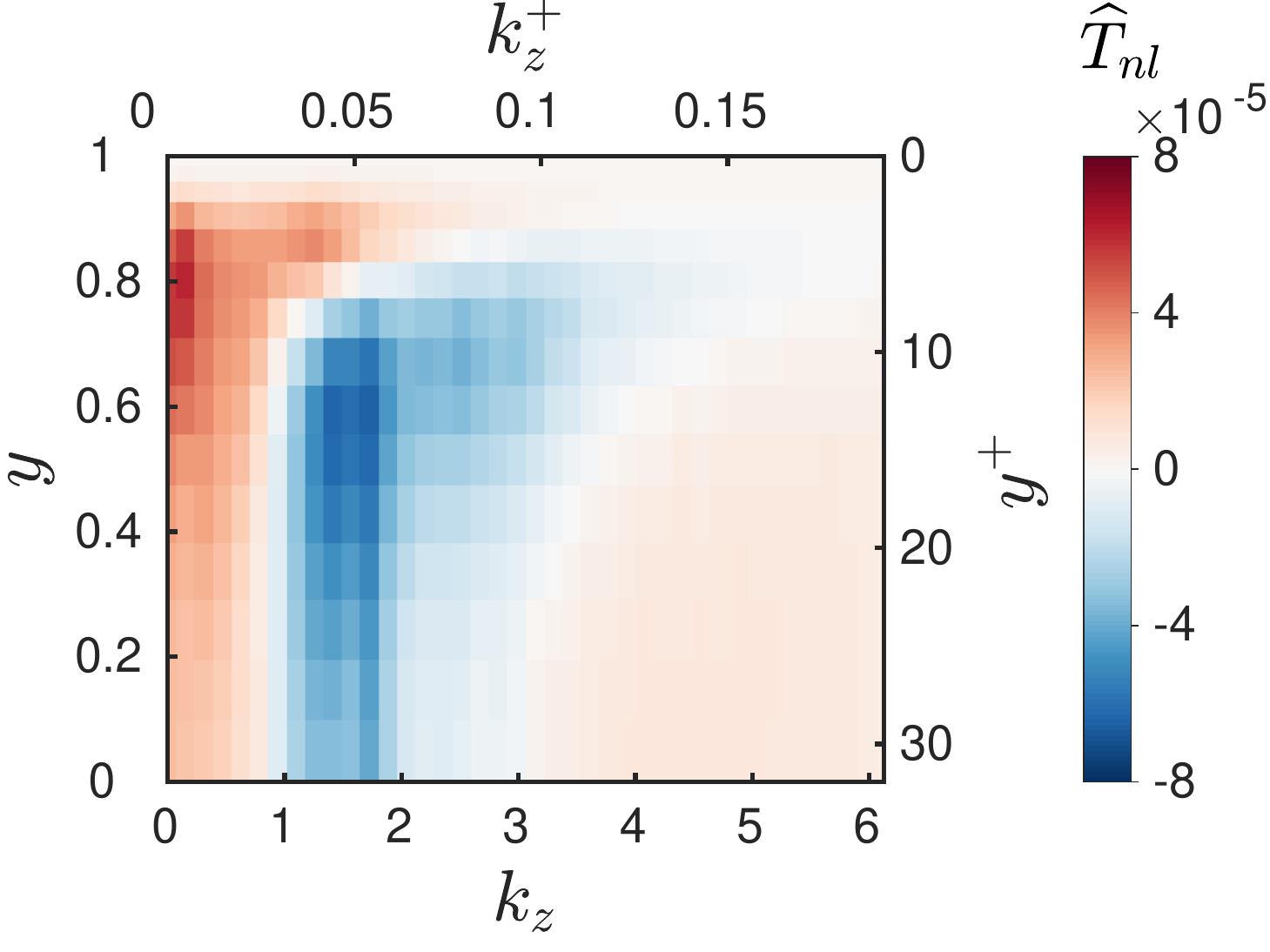}   \label{fig:trans_y_kz_430p}} \\
\subfloat[Uniform, $Re=430$ ($Re_\tau = 33.0$)]{ \includegraphics[width=0.5\columnwidth]{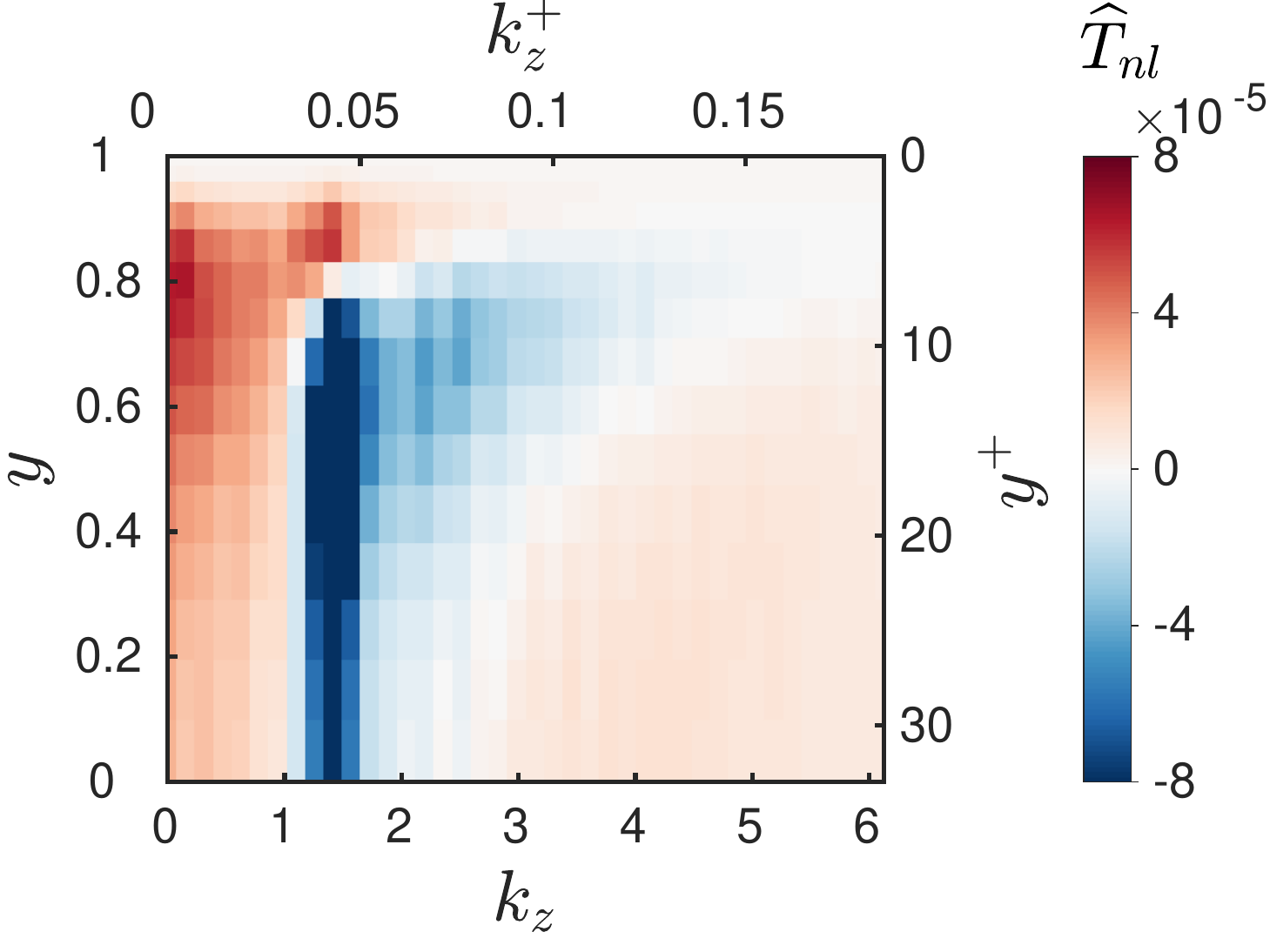} \label{fig:trans_y_kz_430u}} ~
\subfloat[Uniform, $Re=1000$ ($Re_\tau = 66.4$)]{ \includegraphics[width=0.5\columnwidth]{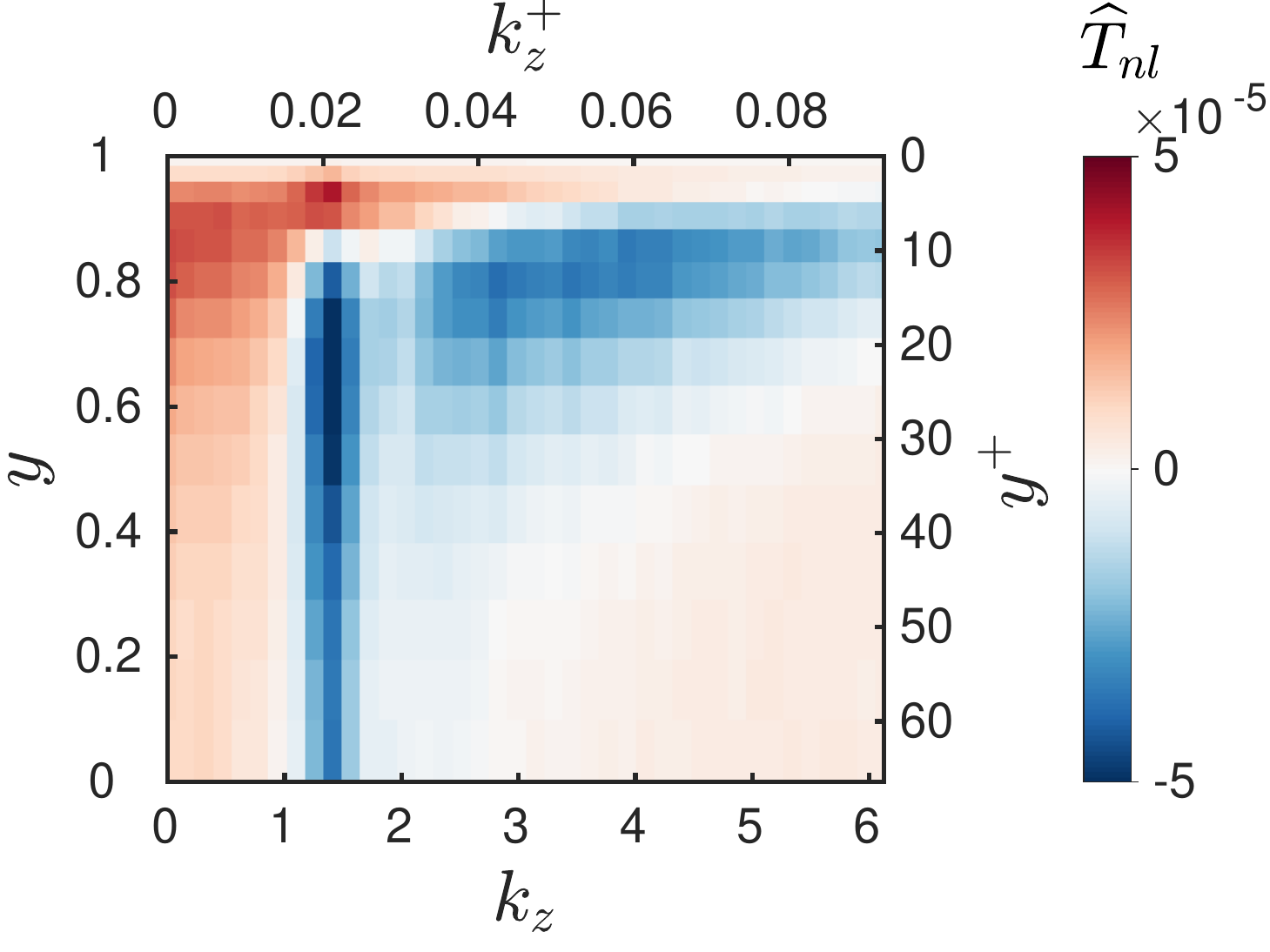} \label{fig:trans_y_kz_1000u} }
\caption{Visualisations of  of nonlinear transfer $\Transnls(y, k_z)$ for different $Re$ and states. The cross-channel range  is from the mid-plane ($y=0$, $y^+=Re_\tau$, lower axis) to the wall ($y=1$, $y^+=0$, upper axis).
%\LT{ Color bar for (c) could be the same as that for (a,b), as is done in Fig 13.}
}
\label{fig:trans_y_kz}
\end{figure}

Figure \ref{fig:prod_y_kz} and \ref{fig:trans_y_kz} show, respectively, $\Prods(y,k_z)$ and $\Transnls(y,k_z)$ 
for different states and $Re$ ranging from 380 to 1000.
We focus on these terms because of their unusual signs in the balance at large scales (small $k_z$).
The zone of negative production at large scales is encircled by the dashed contour. We note that negative production spans the range $y\in[0,  0.8]$ at low $Re$, whereas it is more concentrated between $y=0.6$ and 0.9 at $Re=1000$. In viscous units, it spans approximately from $y^+=5$ to $y^+ \simeq30$ at all $Re$.
\resub{Furthermore, the positive part of the spectrum in marked by a peak at $k_z=\ksmax$ corresponding to the most-producing streaks and rolls, whose wall-normal localisation increases with $Re$, from $y=0.63$ ($y^+\simeq11$) at $Re=400$ to $y=0.8$ ($y^+\simeq13$) at $Re=1000$. }

The triadic interaction term $\Transnls$ is shown on figure \ref{fig:trans_y_kz}.
Inverse transfers are present from $k_z^+=0$ up to $k_z^+ \simeq 0.07$ in the patterned cases, and $k_z^+ \simeq 0.05$ in the uniform case at $Re=1000$ ($Re_\tau = 66.4$), i.e.\ scales smaller than that of rolls and streaks ($\ksmax\simeq1.41$, $\ksmax^+\simeq 0.04$ for $Re\leq430$).
However, this small-scale part of the inverse transfer is localised only near the wall ($y^+<8$), while for $k_z^+< 0.02$, the inverse transfer concerns the whole $y$ domain.

We see two caveats that prevent further quantitative comparisons to other studies in non-tilted domains, for both transitional and non-transitional regimes.
First, the imposition of an angle ($\theta=24^\circ$) 
is completely arbitrary for 
uniform turbulence, and along with the short domain size $L_x$, the streak spacing is imposed in our numerical domain. In Appendix \ref{app:mfu}, we present results in a non-oblique flow unit $(L_{\strm}, L_{\spwise})=(30,20)$ to confirm our observations in the Minimal Band Unit in the non-transitional case $Re=1000$ ($Re_\tau=66.2$).
Second, the reduction to one dimension
can miss the two-dimensionality 
of energy transfers: inter-scale transfers can actually be orientational, i.e.\ they may differ for 
wavenumbers $(k_x, k_z)$ with the same modulus but different orientations. Therefore, inverse transfers in a one-dimensional spectrum can be misleading as they mix transfers between different orientations and transfers between different scales $|\boldsymbol{k}|$.

\resub{These remarks aside, we can draw qualitative comparisons with the energetic large-scales also present in high-$Re$, developed wall-bounded turbulence \citep[and references therein]{jimenez1998largest, smits2011high, lee2018extreme}. The large-scale motions characterising our transitional regime are 
%probably 
of a different nature than those observed in uniform shear flows at higher $Re$,
which are typically streamwise-elongated modes dictated by inertial effects far from the wall (in the outer zone). 
These large scales in
fully-developed turbulence
can also be energised by inverse transfers from small scales
\citep[]{cimarelli2013paths, mizuno2016spectra,  aulery2017spectral, cho2018scale, lee2019spectral, kawata2021scale, andreolli2021global}. However, these inverse transfers are weaker than those reported here in transitional turbulence, and are essentially concentrated near the wall, while we observe inverse transfers over the whole shear layer that dominate the TKE budget at large scales. Furthermore, we recall that our large-scale transfers feed back on the mean flow via negative production, which, to the extent of our knowledge, has never been observed in developed turbulence.
 }

\subsection{Spectral balance in a streamwise-spanwise domain at $Re_\tau = 66$.}
\label{app:mfu}
\begin{figure}
    \centering
\subfloat[Production]{ \includegraphics[width=0.5\columnwidth]{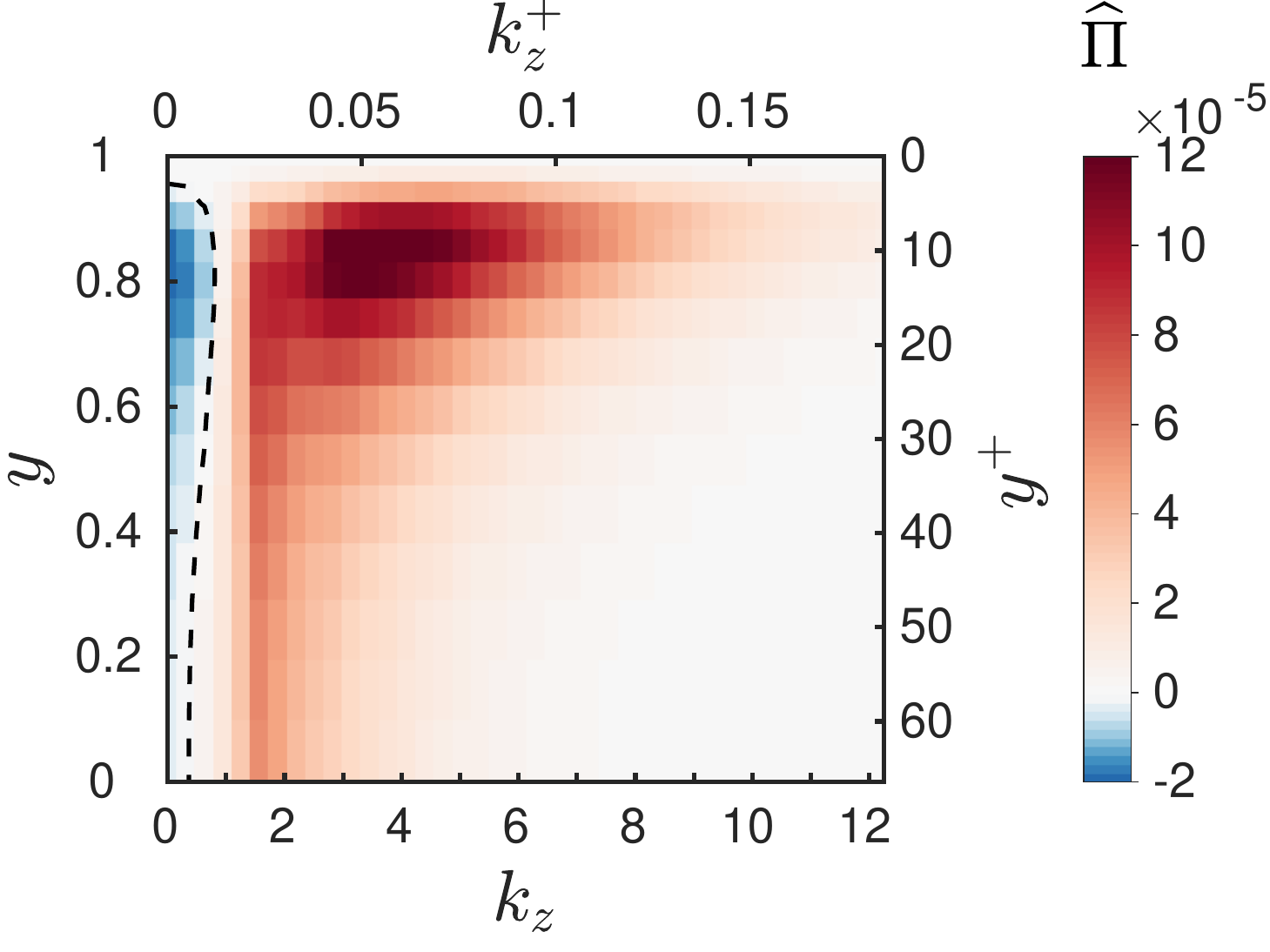}     \label{fig:prod_mfu}}~
\subfloat[nonlinear transfer]{ \includegraphics[width=0.5\columnwidth]{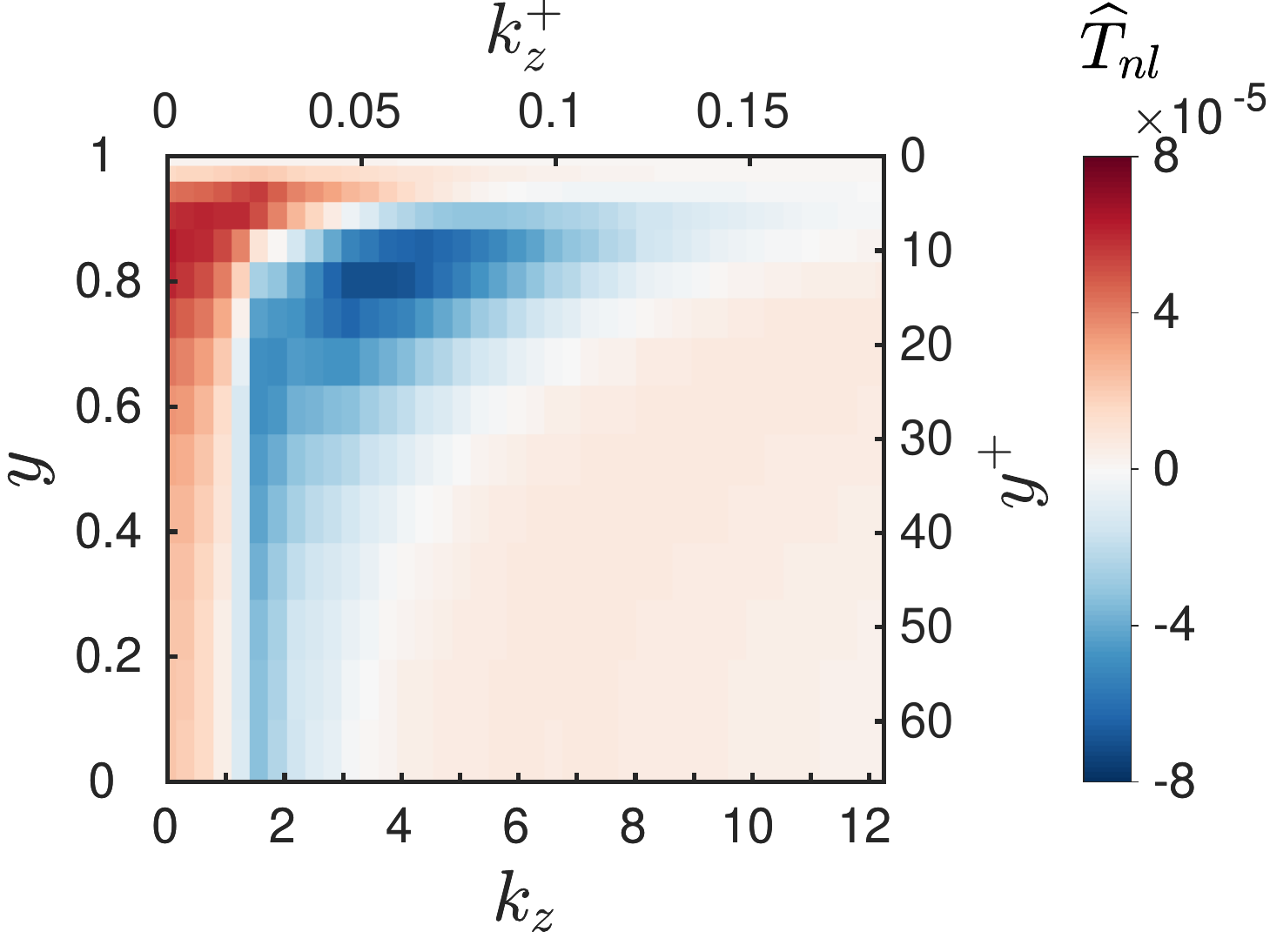}   \label{fig:trans_mfu}} \\
\caption{Production and transfer spectra in a non-tilted domain with $(L_{\strm}, L_{\spwise})=(30,20)$ for $Re=1000$ ($Re_\tau = 66$). }
\label{fig:spec_mfu}
\end{figure}
The use of a Minimal Band Unit of size $(L_x, L_z) = (10,40)$ to study $Re$ outside of 
the transitional regime can be misleading, mainly because the short size and the tilt angle impose a
strict spacing for the streaks. This is certainly why the production and transfer spectra shown at $Re=1000$ (figures \ref{fig:prod_y_kz_1000u} and \ref{fig:trans_y_kz_1000u}) present a sharp peak at $k_z = 1.41$ ($k_z^+ = 0.0214$, $\lambda_z^+ = 290$) along with a tenuous maximum around $k_z^+ = 0.05$ ($\lambda_z^+ = 126$, $\lambda_{\spwise}^+ = 138$). 
In a streamwise-spanwise domain of size $(L_{\strm}, L_{\spwise})=(30,20)$ and number of grid-points $(N_{\strm}, N_{\spwise})=(375,250)$, the streamwise-averaged spectrum is computed as a function of spanwise wavenumber $k_{\spwise}$ on figure \ref{fig:spec_mfu}, and presents a peak located around $k_{\spwise}^+ = 0.05$, $\lambda_{\spwise}^+ \simeq 130$, and no peak below. This is also true for the transfer spectrum.
However, the features observed in a Minimal Band Unit are still present: negative production for $k_{\spwise}^+ < 0.01$ and inverse transfer occupying the whole shear layer for $k_{\spwise}^+ < 0.02$.

\FloatBarrier
\bibliographystyle{jfm}
\bibliography{bib}% Produces the bibliography via BibTeX.

\end{document}